\documentclass[useAMS,fleqn,usenatbib]{mnras}

\usepackage{upgreek}
\usepackage[T1]{fontenc}
\usepackage{ae,aecompl}
\usepackage{amsmath}
\usepackage{amssymb}
\usepackage{graphicx}
\usepackage{natbib}
\usepackage[version=3]{mhchem}
\usepackage{setspace}
\usepackage{float}
\usepackage{mathrsfs}
\usepackage{subfigure}
\usepackage{epstopdf}
\usepackage{ulem}
\usepackage{color}
\defcitealias{maclean2016}{ML16}
\defcitealias{maclean2017}{ML18a}
\defcitealias{marino2017}{Mar17}
\defcitealias{marino2008}{Mar08}
\defcitealias{wang2017}{W17}
\defcitealias{ivans1999}{I99}
\defcitealias{sb2005}{SB05}
\defcitealias{campbell2017}{C17}

\title[On the AGB stars of M\,4]{On the AGB stars of M\,4: A robust disagreement between spectroscopic observations and theory}

\author[B. T. MacLean, et al.]{B. T. MacLean$^{1}$\thanks{E-mail: ben.maclean@monash.edu}, 
S. W. Campbell$^{1}$, 
A. M. Amarsi$^{2}$,
T. Nordlander$^{3,4}$,
\newauthor P. L. Cottrell$^{1,5}$,
G. M. De Silva$^{6,7}$, 
J. Lattanzio$^{1}$, 
T. Constantino$^{8,1}$,
\newauthor V. D'Orazi$^{9,10,11}$ and 
L. Casagrande$^{3,4}$\\
$^{1}$Monash Centre for Astrophysics, School of Physics and Astronomy, Monash University, Victoria 3800, Australia \\ 
$^{2}$Max Planck Institute f\"ur Astronomy, K\"onigstuhl 17, D-69117 Heidelberg, Germany \\
$^{3}$Research School of Astronomy and Astrophysics, Australian National University, Canberra, ACT 2611, Australia \\
$^{4}$ARC Centre of Excellence for All Sky Astrophysics in 3 Dimensions (ASTRO 3D) \\
$^{5}$School of Physical and Chemical Sciences, University of Canterbury, Private Bag 4800, Christchurch 8140, New Zealand \\
$^{6}$Australian Astronomical Observatory, 105 Delhi Rd, North Ryde, NSW 2113, Australia \\
$^{7}$Sydney Institute for Astronomy, School of Physics, The University of Sydney, NSW 2006, Australia \\
$^{8}$Physics and Astronomy, University of Exeter, Exeter, EX4 4QL, United Kingdom \\
$^{9}$INAF Osservatorio Astronomico di Padova, vicolo dell'Osservatorio 5, I-35122, Padova, Italy \\
$^{10}$Research Centre for Astronomy, Astrophysics \& Astrophotonics (MQAAAstro), Macquarie University, Sydney, NSW 2109, Australia \\ 
$^{11}$Department of Physics and Astronomy, Macquarie University, North Ryde, NSW 2109, Australia 
}
\date{Accepted TBC. Received TBC; in original form TBC}

\begin{document}

\pagerange{\pageref{firstpage}--\pageref{lastpage}} \pubyear{2016}

\maketitle

\label{firstpage}

\begin{abstract}

Several recent spectroscopic investigations have presented conflicting results on the existence of Na-rich asymptotic giant branch (AGB) stars in the Galactic globular cluster M\,4 (NGC\,6121). The studies disagree on whether or not Na-rich red giant branch (RGB) stars evolve to the AGB. For a sample of previously published HERMES/AAT AGB and RGB stellar spectra we present a re-analysis of O, Na, and Fe abundances, and a new analysis of Mg and Al abundances; we also present CN band strengths for this sample, derived from low-resolution AAOmega spectra. Following a detailed literature comparison, we find that the AGB samples of all studies consistently show lower abundances of Na and Al, and are weaker in CN, than RGB stars in the cluster. This is similar to recent observations of AGB stars in NGC\,6752 and M\,62. In an attempt to explain this result, we present new theoretical stellar evolutionary models for M\,4; however, these predict that all stars, including Na-rich RGB stars, evolve onto the AGB. We test the robustness of our abundance results using a variety of atmospheric models and spectroscopic methods; however, we do not find evidence that systematic modelling uncertainties can explain the apparent lack of Na-rich AGB stars in M\,4. We conclude that an unexplained, but robust, discordance between observations and theory remains for the AGB stars in M\,4.

\end{abstract}


%%%%%%%%%%%%%%%%%%%%%%%%%%%%%%%%%%%%%%%%%%%%%%%%%%%%%%%%%%%%%%%%

\section{Introduction}
\label{m4_2_intro}

In early GC studies stars were observed at the same evolutionary stage but with different CN strengths, which cannot be explained only with evolutionary effects \citep[e.g.][]{hesser1977,norris19816752}. These and other findings led to the general consensus that Galactic GCs contain multiple populations of stars, identified by variations in light elemental abundances that are \textit{intrinsic} -- inherited at birth -- to the stars. Variations are typically observed in the abundances of C, N, Na, and O, and sometimes Mg and Al (see the extensive reviews of \citealt{sneden1999,gratton2012} and references therein; but see \citealt{bastian2013} for an opposing view). In this paper we designate those GC stars with halo-like abundances (CN-weak, Na poor) as subpopulation one (SP1), and all stars enriched in Na (or that present as CN-strong) as subpopulation two (SP2).

Over the decades since the first spectroscopic studies of Galactic GCs, stars in each evolutionary phase have been targeted to evaluate the consistency of the light-elemental abundance distributions along the stellar evolutionary tracks. While systematic observations of the asymptotic giant branch (AGB, the final phase of nuclear burning) have only been performed relatively recently, AGB stars had previously been included among the GC stellar samples of last century. The literature reviews of \citet{sneden00conf} and \citet{campbell2006} noted that the distribution of CN band strengths of AGB stars in certain globular clusters are very different to those seen in RGB stars -- most strikingly that the AGB stars of NGC\,6752 are exclusively CN-weak. This is in contradiction to the theoretical prediction that the N abundance of a star, which is traced by the CN band strength, should \textit{increase} as a result of `deep mixing' on the RGB \citep{langer1985,henkel2017}. 

Seeking a more reliable diagnostic tool, \citet{campbell2013} measured Na abundances for a sample of 20 AGB and 24 RGB stars in NGC\,6752. Just as in the earlier low-resolution CN studies of the cluster, they found homogeneity in their entire sample of AGB stars: the [Na/Fe] values were all within $\pm 0.1$~dex and very low ([Na/Fe] $\lesssim 0.12$~dex). This contrasted greatly with their RGB sample for which a variation in [Na/Fe] of $\sim 0.9$~dex was reported. While this result was challenged observationally \citep{lapenna2016}, a detailed reanalysis by \citet[hereafter \citetalias{campbell2017}]{campbell2017} supported the original conclusion: that up to 100\% of the Na enhanced stars (SP2; which represent 70\% of the total RGB population) in NGC\,6752 appear to be avoiding the AGB entirely.

It is generally agreed that stars enriched in N and Na are also enriched in He \citep{dupree2011,nardiello2015}. Stars with a He-enhancement evolve faster and thus have lower initial masses than stars of the same age but normal helium. Assuming these stars experience the same amount of mass loss on the RGB, they will retain less envelope on the horizontal branch (HB) and appear bluer \citep{sweigart97,catelan2009}.

The results of \citet{campbell2013} conflict with the prediction of stellar evolutionary theory that only HB stars with extremely thin envelopes avoid the AGB, becoming AGB-manqu\'{e} stars \citep{agb-manque0,agb-manque}. At the metallicities of GCs this only occurs in stellar models with effective temperatures (T$_{\rm eff}$) higher than $\sim 15,000$~K \citep{dorman1993}, corresponding to $\sim$30\% of the most helium enhanced stars in NGC\,6752. Efforts to explain these observations have not been able to reproduce the results -- see for example \citet{cassisi2014}, who could not reproduce the NGC\,6752 observations using population synthesis \citep[also see][]{campbell2013,chantereau2016}.

Adding to the debate on this topic, \citet[hereafter \citetalias{maclean2016}]{maclean2016} reported O, Na, and Fe abundances for a sample of 15 AGB and 106 RGB stars in M\,4 (NGC\,6121), which contains no HB stars predicted to become AGB-manqu\'{e} stars -- M\,4's HB extends only to $\sim 9000$~K in T$_{\rm eff}$. Surprisingly, all 15 AGB stars were found to have SP1-like O and Na abundances despite a significantly larger spread in the RGB abundances. This is the third such finding (after NGC\,6752 and M\,62) of a paucity of SP2 AGB stars in a globular cluster; but the first for a GC without an extended blue HB. While AGB stars have been included within stellar samples of spectroscopic M\,4 studies in the past (\citealt{norris1981m4,suntzeff1991,ivans1999}; and the literature reviews of \citealt{sneden00conf,sb2005}), \citetalias{maclean2016} was the first study that specifically targeted the AGB to investigate stellar evolution using the multiple population phenomenon of M\,4.

Due to the controversial nature of the discovery of \citetalias{maclean2016}, and uncertainties regarding the separation of the subpopulations in [Na/O] space, caveats to the conclusions arising from the study were noted. M\,4 is a moderately metal-poor \citep[{[}Fe/H{]} $= -1.16$;][]{harris1996} cluster that displays a distinctly bimodal HB \citep{marino2011} and a well established Na-O anti-correlation on the RGB and HB. While M\,4 does not exhibit a Mg-Al anti-correlation (Mg has been observed to be homogeneous in M\,4), Al correlates with Na \citep{marino2008}.

The conclusions of \citetalias{maclean2016} motivated the publication of three additional studies (to date) of AGB stars in M\,4 by three separate research groups. Using the photometric index C$_{UBI}$ \citep[which has been shown to correlate with light-elemental abundances in RGB stars;][]{monelli2013}, \citet{lardo2017} determined the spread in C$_{UBI}$ values to be quantitatively similar between the AGB and RGB in M\,4, in contradiction to the spectroscopic findings of \citetalias{maclean2016}. Using high-resolution spectra, \citet[hereafter \citetalias{marino2017}]{marino2017} came to the same conclusion as \citet{lardo2017} by showing that a sample of 17 AGB stars had a similar range in [Na/Fe] values as the RGB sample from \citet[hereafter \citetalias{marino2008}]{marino2008}. However, with similar data, \citet[hereafter \citetalias{wang2017}]{wang2017} found that M\,4 AGB stars have lower [Na/H] values than stars on the RGB, and that the most Na-rich stars did appear to be missing from the AGB, but not to the extreme degree that \citetalias{maclean2016} had concluded. Thus a significant uncertainty exists within the literature with regard to the nature of M\,4's AGB population.

The mixed and contradictory results of recent studies into the light-elemental abundances of M\,4's AGB population call for a detailed, quantitative reinvestigation of the available data in order to idenitify why the results differ. In this paper we adopt the $\mathscr{F}$ parametrisation of SP2 AGB deficits\footnote{$\mathscr{F} = (1-{\frac{\mathscr{R}_{\rm AGB}}{\mathscr{R}_{\rm RGB}}}){\cdot}100$\%, where the percentages of RGB and AGB stars in a GC that are found to be members of SP2 are written as $\mathscr{R}_{RGB}$ and $\mathscr{R}_{AGB}$. For example, \citet{campbell2013} reported $\mathscr{R}_{RGB} = 70$\% and $\mathscr{R}_{AGB} = 0$\% for NGC\,6752.} that was used in \citetalias{maclean2016} and \citet[hereafter \citetalias{maclean2017}]{maclean2017}. 

This paper is structured as follows. In Section~\ref{m4_2_reanalysis} we re-analyse our previously published sample of high-resolution M\,4 stellar spectra in order to test the robustness of our earlier study on M\,4 (\citetalias{maclean2016}). In Section~\ref{m4_2_cn} we calculate CN band strengths from previously unpublished low-resolution spectra of M\,4 stars. In an attempt to resolve the conflicting conclusions in recent (and historical) spectroscopic studies, we compare our abundance and CN results with M\,4 AGB and RGB data from the literature in Section~\ref{m4_2_lit_comp}. In Section~\ref{m4_2_monstar} we use 1D stellar evolution models to establish a precise, quantitative theoretical expectation of the abundance distribution of the AGB of M\,4. In Section~\ref{m4_2_atmos_tests} we investigate possible explanations for the AGB results found in this study (and throughout the literature) including a series of tests utilising a range of stellar atmospheric models. Finally, we summarise our results and conclusions in Section~\ref{m4_2_summary}.

%%%%%%%%%%%%%%%%%%%%%%%%%%%%%%%%%%%%%%%%%%%%%%%%%%%%%%%%%%%%

\section{High-resolution spectra re-analysis}
\label{m4_2_reanalysis}

In order to be confident in our earlier results, which have been challenged in the literature, we re-analysed our sample of M\,4 stellar spectra upon which our \citetalias{maclean2016} results were based. The motivation behind this re-analysis was to i) check the \citetalias{maclean2016} results in light of recent debate on stellar parameter determination for AGB stars in GCs \citep[see][]{lapenna2016,campbell2017}, and ii) increase the number of elements available for use as a diagnostic of multiple populations. Specifically, we redetermined the stellar parameters (T$_{\rm eff}$, $v_t$, log~$g$, and [Fe/H]) and abundances (Na and O) that were published in \citetalias{maclean2016}. We also determined abundances of Mg and Al for our full sample of 15 AGB and 106 RGB stars.

\subsection{Targets and data}
\label{m4_2_data}

The reduced M\,4 high-resolution spectra and photometry used in this study are the same as those used in \citetalias{maclean2016}. M\,4 suffers from significant differential reddening, however constant reddening values were used in \citetalias{maclean2016}. Here we improve upon this, with each star corrected using the reddening map of \citet{hendricks2012}. Individual corrections are included in Table~\ref{tab:m4_2_obs}. We found an average reddening value of $E(B-V)=0.37$ and a 1$\sigma$ star-to-star scatter of $\pm 0.02$. This differential reddening map, however, does not cover our entire sample, and some stars were only adjusted according to the average reddening value.

\begin{figure}
\centering
\includegraphics[width=0.9\linewidth]{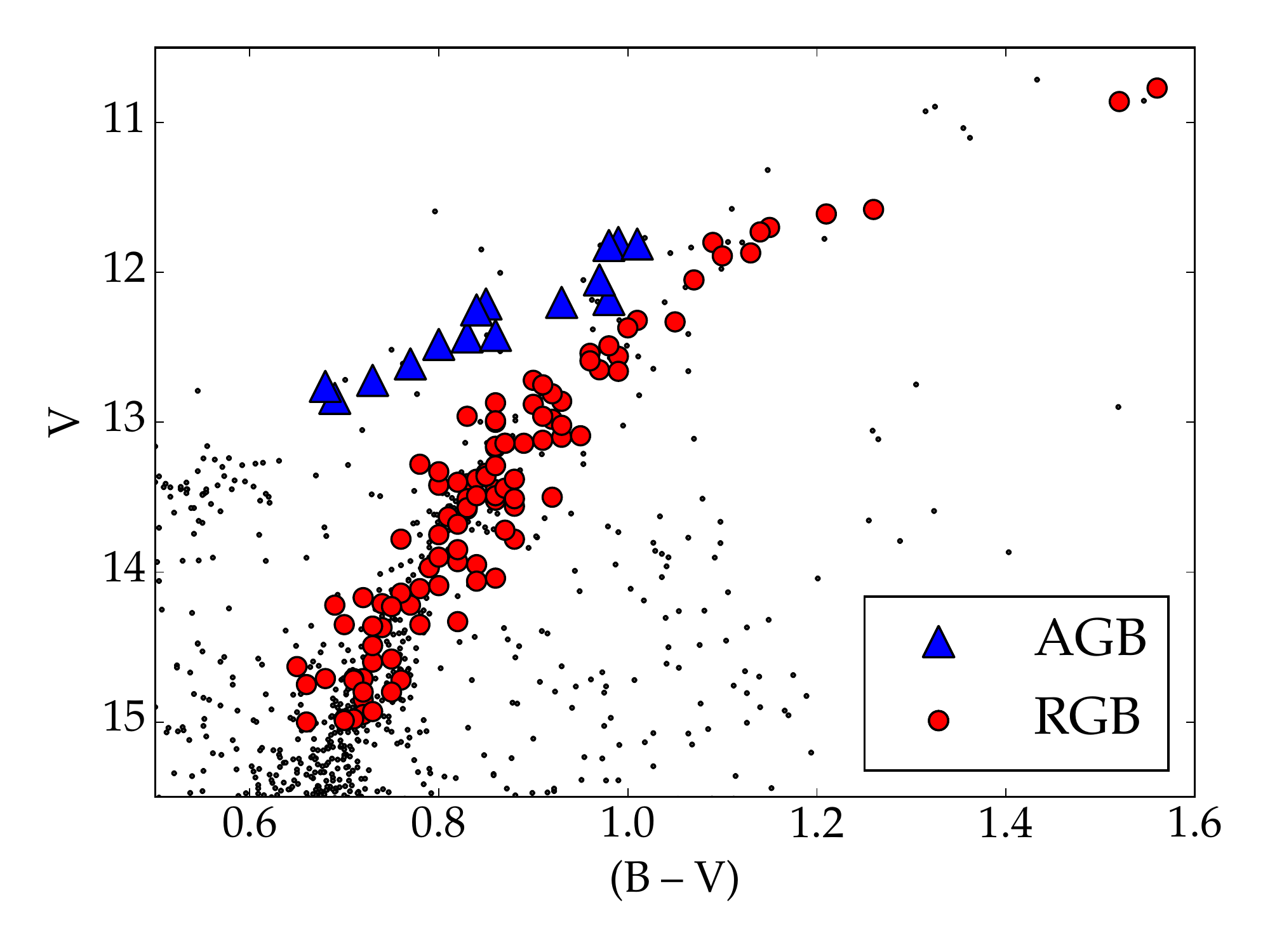}
\includegraphics[width=0.9\linewidth]{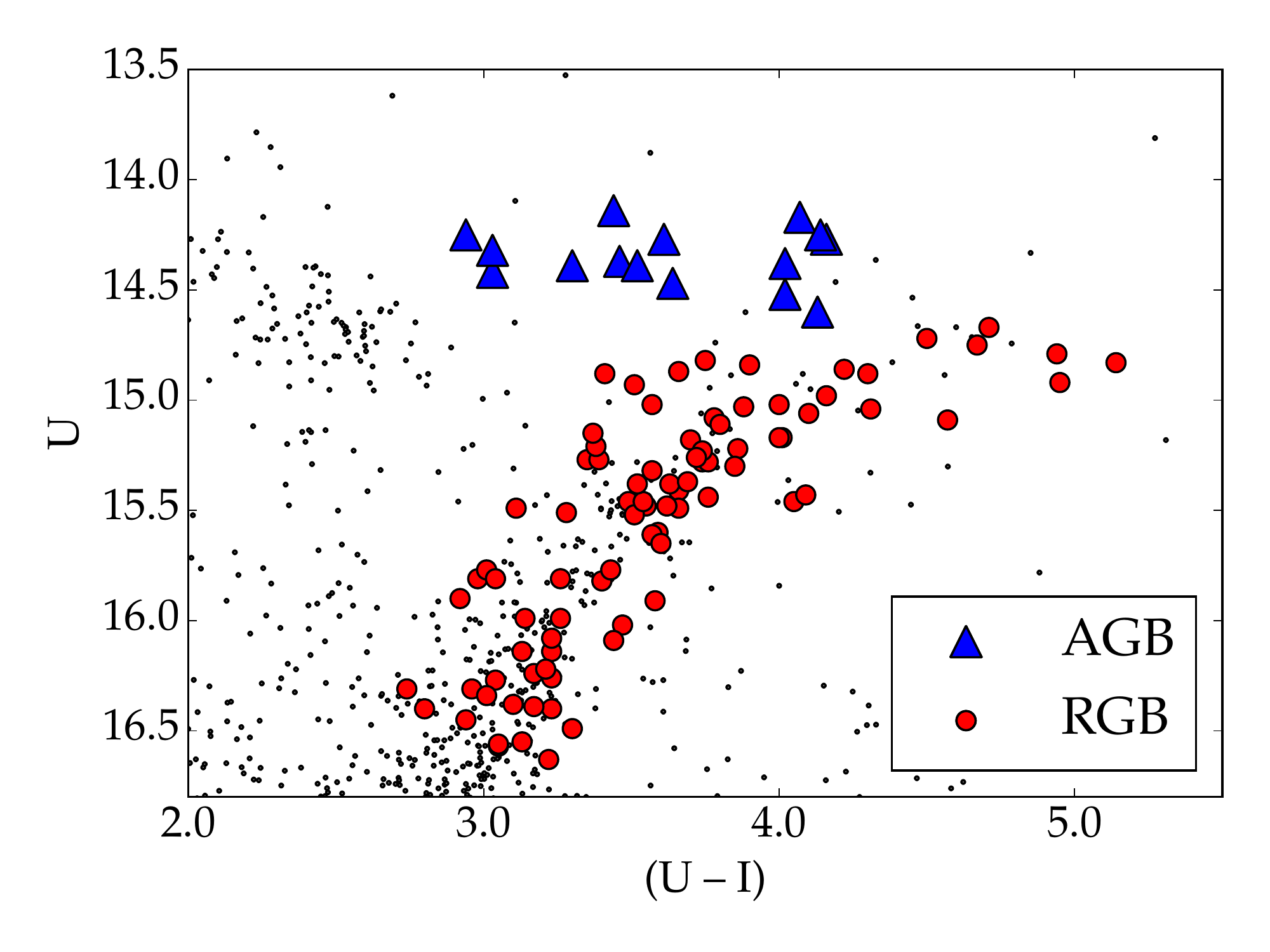}
\caption{$V$ - $(B-V)$ and $U$ - $(U-I)$ colour-magnitude diagrams of M\,4 RGB and AGB target stars, displayed over the full photometric sample of \citet[black points]{momany2003}. In the top panel, targets have been corrected for extinction according to the differential reddening map of \citet{hendricks2012}, and a constant value of $(B-V)$ = $0.37$ was applied to the non-target photometric data. No reddening correction has been applied to the $(U-I)$ photometry in the bottom panel.}
\label{fig:m4_2_cmds}
\end{figure}

The M\,4 targets included in this study are presented in Figure~\ref{fig:m4_2_cmds}. In total, 24 AGB stars were identified in the photometry of \citet{momany2003}. Seven of these were not observable due to 2dF fibre positioning restrictions, and two were found in \citetalias{maclean2016} to be non-members, leaving a final sample of 15. Due to the randomness of stellar astrometry within a GC, we did not identify any sources of selection bias.

\begin{table*}
\centering
\caption{M\,4 target details including data from \protect{\citet[$UBVI$ photometry and target IDs]{momany2003}} and 2MASS \protect{\citep[$JHK$ photometry]{2mass}}, and differential reddening corrections. Gaps in 2MASS data represent targets with low quality flags. Stars for which no reddening value is listed were outside the reddening map of \protect{\citet{hendricks2012}}, and were corrected according to the average reddening value of E$(B-V) =0.37$. Only the first five rows are shown; the full table is available online.}
\label{tab:m4_2_obs}
\begin{tabular}{ccccccccccc}
\hline
ID  & Type & 2MASS ID         & $V$ & $B$ & $U$ & $I$ & $J$ & $H$ & $K$ & E$(B-V)$ \\
\hline\hline
788   & AGB  & 16235772-2622557 & 12.21  & 13.43  & 14.14  & 10.69  & 9.64   & 9.00   & 8.82 & -     \\
3590  & AGB  & 16232184-2630495 & 12.48  & 13.64  & 14.37  & 10.92  & -      & -      & -    & 0.36    \\
10092 & AGB  & 16233067-2629390 & 12.61  & 13.74  & 14.39  & 11.09  & -      & -      & -    & 0.36    \\
11285 & AGB  & 16233195-2631457 & 12.84  & 13.90  & 14.42  & 11.40  & 10.35  & 9.77   & 9.58 & 0.37    \\
13609 & AGB  & 16233477-2631349 & 12.76  & 13.81  & 14.25  & 11.31  & 10.21  & 9.65   & 9.48 & -    \\
\vdots & \vdots & \vdots & \vdots & \vdots & \vdots & \vdots & \vdots & \vdots & \vdots & \vdots \\
\hline
\end{tabular}
\end{table*}

%%%%%%%%%%%%%%%%%%%%%%%%%%%%%%%%%%%%%%%%%%%%%%%%%%%%%%%%%%%%%%%

\subsection{Atmospheric parameters}
\label{m4_2_atmos_method}

For the determination of surface gravity (log~$g$), we did not adopt the standard spectroscopic approach, wherein ionisation balance between abundances determined from neutral and singly-ionised Fe lines is enforced. This is because such an approach can be biased by not accounting for non-LTE effects on Fe\,{\sc i} lines \citep{ivans1999,lapenna2016,sitnova2015}. Therefore, we instead calculated log~$g$ using estimates of T$_{\rm eff}$, luminosity and mass. The luminosity was computed from de-reddened V magnitudes, with bolometric corrections from \citet{alonso1999}. We assumed a mass of 0.8~M$_{\odot}$ and 0.7~M$_{\odot}$ for the RGB and AGB, respectively \citep{miglio2016}.

We investigated different approaches to determining the effective temperatures (T$_{\rm eff}$) of our stars. T$_{\rm eff}$ determinations can be subject to significant uncertainties, both random and systematic. Incorrect modelling assumptions, and degeneracies in the stellar spectra with respect to different stellar parameters, can lead the standard spectroscopic method (requiring a balance of line-by-line Fe\,{\sc i} abundances over a range of excitation potentials) to give unreliable and/or significantly offset T$_{\rm eff}$ values. Similarly, the photometric method (utilising empirical relations between T$_{\rm eff}$ and photometric magnitudes) can potentially produce large uncertainties (up to $\pm 200$~K); for example, see \citet[\citetalias{campbell2017}]{campbell2017} for a detailed investigation of T$_{\rm eff}$ determination using the photometric method, and its effect on Fe and Na abundances determined for AGB and RGB stars. 

Due to i) the high level of differential reddening in M\,4 \citep[and the fact that our sample is not fully covered by the reddening map of][]{hendricks2012}, and ii) the debate within the literature as to appropriate selections of colour-T$_{\rm eff}$ empirical relationships (see \citetalias{campbell2017}), we endeavoured to further improve the spectroscopic T$_{\rm eff}$ determination from our spectroscopic code {\sc phobos}. Version one of this code ({\sc phobos v1}) was used in \citetalias{maclean2016} to determine parameters spectroscopically, but it was dependent on having accurate initial photometric estimates of T$_{\rm eff}$. In \citetalias{campbell2017} we noted that spectroscopic codes and methods appear to give effective temperatures that inherit some of the biases/trends in colour-T$_{\rm eff}$ relations (see \S4 in \citetalias{campbell2017}). We investigated this problem in {\sc phobos v1} and found that, in our case, this bias was due to the choice of the numerical scheme employed to iterate to a solution.

In principle, the choice of photometric estimate should have no bearing on the spectroscopic parameters that the code determines -- that is, the spectroscopic parameters should only be a function of the Fe absorption line-list, and not the initial photometric estimates. We have improved the numerical scheme in {\sc phobos v2} to search for global minima in the stellar parameter space, so that the initial T$_{\rm eff}$ estimates only require an accuracy of $\sim 1000$~K, and so that the code is `agnostic' about the initial T$_{\rm eff}$ estimate. {\sc phobos v2} determines T$_{\rm eff}$ by requiring no trend between the excitation potential of Fe\,{\sc i} absorption lines and the abundances calculated from those lines. Initial microturbulence ($v_t$) estimates were determined using the empirical relation from \citet{gratton1996}, while final spectroscopic values are required to have no trend between the reduced wavelength of Fe\,{\sc i} lines and their associated line-by-line abundances.

To test the efficacy of our improved code ({\sc phobos v2}), we conducted two tests, using our entire M\,4 sample of 121 giant stars, to determine spectroscopic parameters primarily based on two very different sets of photometrically estimated initial-guess T$_{\rm eff}$ values. The first set of initial guesses (T$_{\rm eff, ph}$) are an average of six predictions from the empirical $B-V$ and $V-K$ relations of \citet{ramirez2005}, \citet{gonzalez2009}, and \citet{casagrande2010}, and one direct calculation by implementing the infrared flux method (IRFM) at an estimated log~$g$ of each star using $BVI$ and 2MASS $JHK$ photometry \citep{casagrande2014}. For stars that were flagged for low quality and/or contamination in the 2MASS database, only the $B-V$ relations were used to determine T$_{\rm eff, ph}$, while for all other stars, the mean of the seven estimates was adopted as T$_{\rm eff, ph}$. These methods are mildly dependent on metallicity, for which a value of [Fe/H] $= -1.10$ was assumed (a change in adopted metallicity of 0.1~dex alters T$_{\rm eff, ph}$ values by $\sim 10$~K). Table~\ref{tab:m4_2_teff_diffs} summarises the average difference between the adopted T$_{\rm eff, ph}$ values and those of the individual photometric relations and IRFM -- the systematic differences between the relations highlight that individual photometric relations are often poor choices for determining stellar parameters. Individual T$_{\rm eff, ph}$ values are listed in Table~\ref{tab:m4_2_params}. For the second, and extreme, test of {\sc phobos v2}, the initial T$_{\rm eff}$~guesses of every star (regardless of evolutionary phase) were assumed to be identical: T$_{\rm eff} = 4500$~K, log~$g$ = 2.5, and $v_t$ = 1.5 -- broadly representative of a giant GC star. We designate this second set of initial guesses as T$_{\rm eff, 4500}$. 

\begin{figure}
\centering
\includegraphics[width=0.9\linewidth]{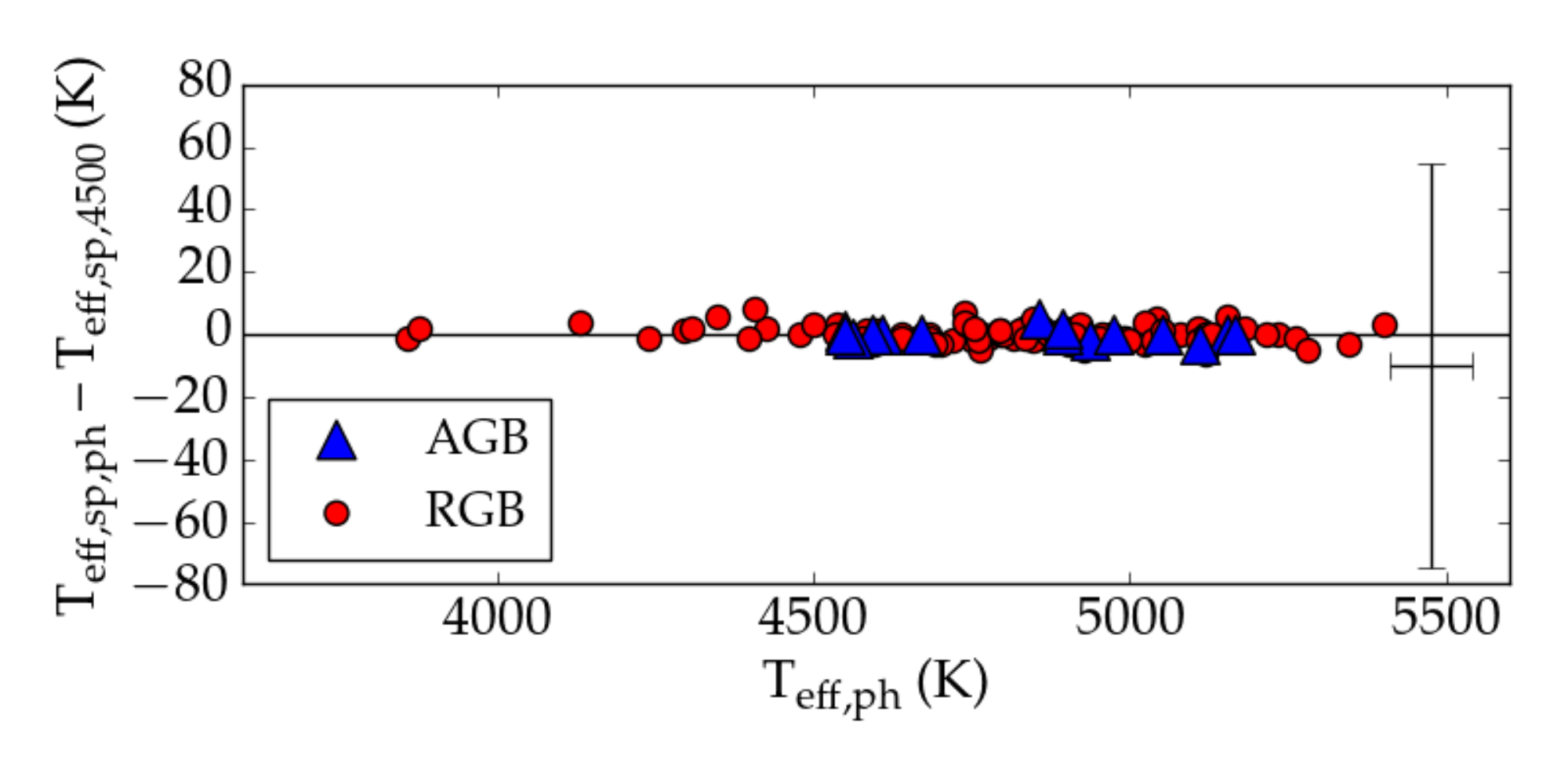}
\includegraphics[width=0.9\linewidth]{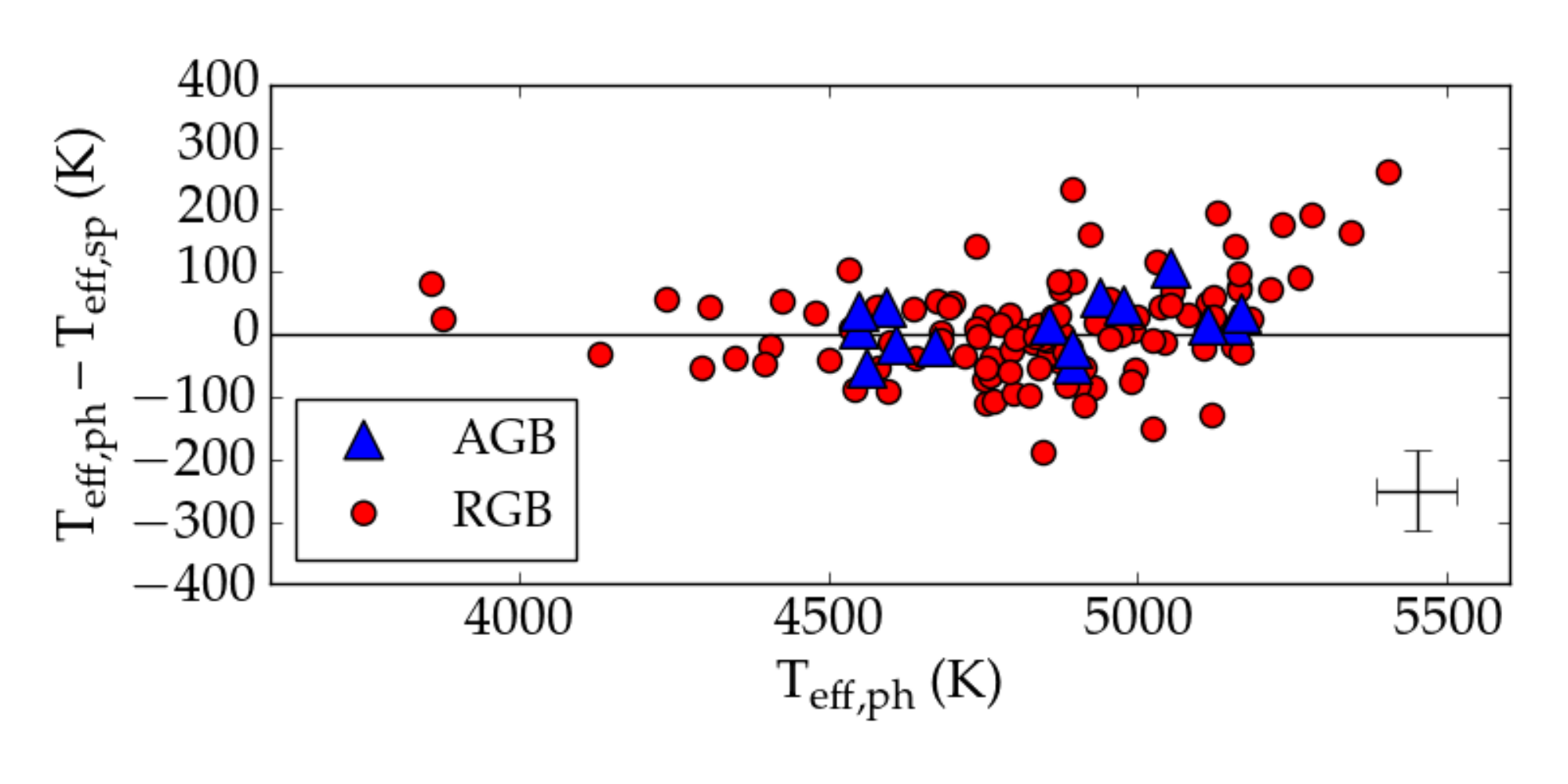}
\caption{\textbf{Top panel:} The star-to-star differences between the spectroscopic T$_{\rm eff, sp}$ values determined (using {\sc phobos v2}) based on initial estimates from i) photometrically estimated stellar parameters (T$_{\rm eff, ph}$), and ii) a single T$_{\rm eff}$ of 4500~K (T$_{\rm eff, 4500}$). The average difference between the spectroscopic values of the two tests is ${\Delta}T_{\rm eff} = 0 \pm 2$~K. \textbf{Bottom panel:} The star-to-star differences between our photometrically estimated T$_{\rm eff, ph}$ values and final adopted spectroscopic T$_{\rm eff, sp}$ values. Error bars in both panels are our typical T$_{\rm eff, sp}$ uncertainties -- $\sim 65$~K, as determined by {\sc phobos v2} (see text for more detail).}
\label{fig:m4_2_bisection_test}
\end{figure}

We used {\sc phobos v2} to determine spectroscopic parameters twice, once using the parameter set T$_{\rm eff, ph}$, and again using the T$_{\rm eff, 4500}$ set of parameters for the initial guess. As seen in Figure~\ref{fig:m4_2_bisection_test}, the differences between the spectroscopically determined effective temperature values using the two different initial estimates (T$_{\rm eff, sp, ph}$ and T$_{\rm eff, sp, 4500)}$) are extremely small, with ${\Delta}T_{\rm eff} = 0 \pm 2$~K, while the the average difference between the photometric (T$_{\rm eff, ph}$) and spectroscopic (T$_{\rm eff, sp}$) values is ${\Delta}T_{\rm eff} = 12 \pm 76$~K. This indicates that no information from the photometric T$_{\rm eff}$ estimates is retained within the spectroscopic results. This is beneficial because the final stellar parameters are independent of the choice of colour-T$_{\rm eff}$ relation, and are therefore reproducible and consistent. 

\begin{figure}
\centering
\includegraphics[width=0.9\linewidth]{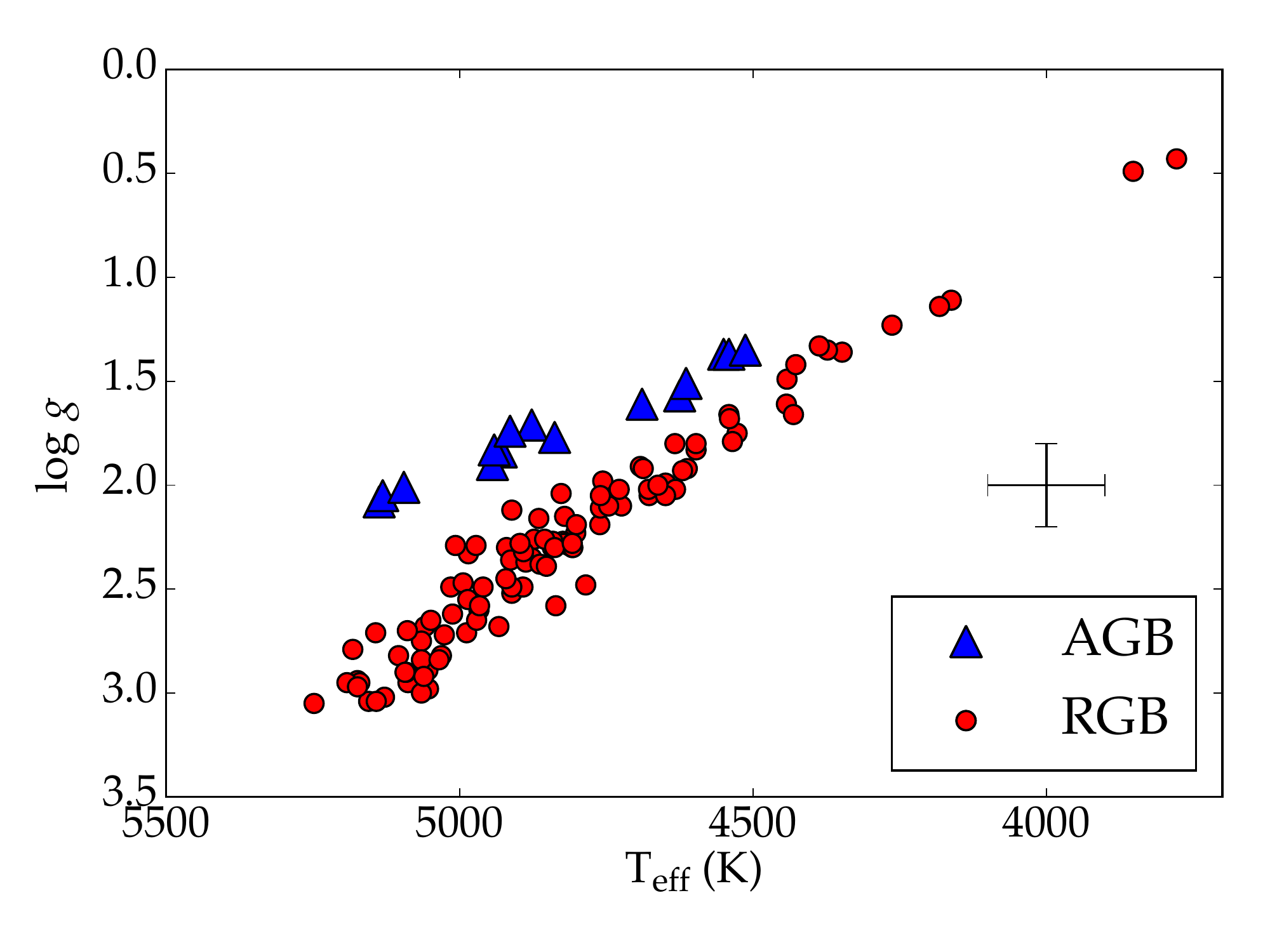}
\caption{Final T$_{\rm eff}$ and log~$g$ values of our M\,4 stellar sample, determined spectroscopically using {\sc phobos v2}. Typical uncertainties are indicated (see Table~\ref{tab:m4_2_atmos}).}
\label{fig:m4_2_hr}
\end{figure}

In summary, we adopt the spectroscopic parameters included in Table~\ref{tab:m4_2_params} and presented in Figure~\ref{fig:m4_2_hr}. The subsequent elemental abundance determinations were based on these 
stellar parameters. {\sc Phobos v2} now also calculates star-to-star T$_{\rm eff}$ and $v_t$ uncertainties based on the standard error of the slope between excitation potential and reduced wavelength, and line-to-line Fe\,{\sc i} abundances. These uncertainties are included in Table~\ref{tab:m4_2_params}. The typical 1$\sigma$ T$_{\rm eff}$ and $v_t$ uncertainties of our sample are 65~K and 0.1~km/s, respectively, and we adopt a 1$\sigma$ log~$g$ uncertainty of 0.2~dex.

%------------------------------------------------------------------------------------

\begin{table}
\centering
\caption{Average differences between the average T$_{\rm eff, ph}$ values and each photometric estimate (T$_{\rm eff, ph}$ -- T$_{\rm eff, estimate}$) for our first {\sc phobos} test. Uncertainties are the standard deviations of the stellar sample, with the quoted uncertainty of each relation in brackets (except for IRFM, which is the average IRFM uncertainty of our sample).}
\label{tab:m4_2_teff_diffs}
\begin{tabular}{lc}
\hline 
Method & $\Delta$T$_{\rm eff}$ (K)  \\
\hline\hline
Ram ($B-V$)$^1$            &  ~~~~$0 \pm 71$ (51) \\
Gonz ($B-V$)$^2$           &  $-49 \pm 70$ (57) \\
Casa ($B-V$)$^3$            &  $-74 \pm 89$ (73) \\
Ram ($V-K$)            &  ~$132 \pm 52$ (28) \\
Gonz ($V-K$)           &  ~~\,$24 \pm 48$ (23) \\
Casa ($V-K$)           & ~~$-2 \pm 54$ (25) \\
IRFM$^3$                 &  ~~$-5 \pm 62$ (33) \\
\hline
Average $\sigma$        & ~~~64          \\
\hline
\multicolumn{2}{l}{\footnotesize{$^1$\protect\shortcite{ramirez2005}}} \\
\multicolumn{2}{l}{\footnotesize{$^2$\protect\citet{gonzalez2009}}} \\
\multicolumn{2}{l}{\footnotesize{$^3$\protect\citet{casagrande2010}}} \\
\end{tabular}
\end{table}

\begin{table*}
\centering
\caption{Stellar parameters for each star in our M\,4 sample. Spectroscopic effective temperatures (T$_{\rm eff, sp}$), microturbulence values ($v_t$), and uncertainties were determined using {\sc phobos v2}, while log~$g$ values were calculated based on the empirical relation from \protect{\citet{alonso1999}}. These were adopted as our final parameters. T$_{\rm eff, ph}$ values are the effective temperatures estimated from photometric colour-T$_{\rm eff}$ relations, were used in the {\sc phobos} test, and are included for comparison (also see Figure~\ref{fig:m4_2_bisection_test}). Only the first five rows are shown; the full table is available online.}
\label{tab:m4_2_params}
\begin{tabular}{cccccc}
\hline
Star ID  & Evolutionary & T$_{\rm eff, sp}$ & log~$g$ & $v_t$ & T$_{\rm eff, ph}$      \\
& phase & (K)  & (cgs) & (km/s) & (K)   \\
\hline\hline
788   & AGB & 4877 $\pm$ 52 & 1.71 & 1.56 $\pm$ 0.07 & 4937 \\
3590  & AGB & 4929 $\pm$ 36 & 1.84 & 1.68 $\pm$ 0.06 & 4975 \\
10092 & AGB & 4944 $\pm$ 29 & 1.90 & 1.45 $\pm$ 0.04 & 5051 \\
11285 & AGB & 5137 $\pm$ 69 & 2.08 & 1.73 $\pm$ 0.19 & 5154 \\
13609 & AGB & 5131 $\pm$ 67 & 2.05 & 1.21 $\pm$ 0.10 & 5166 \\
\vdots & \vdots & \vdots & \vdots & \vdots & \vdots \\
\hline
\end{tabular}
\end{table*}

%----------------------------------------------------------

\subsection{Chemical abundance determination}
\label{m4_2_abund_method}

With our improved stellar parameters, we adopted the method of \citetalias{maclean2017} for the determination of chemical abundances. This is mostly the same as the method previously used for this sample (\citetalias{maclean2016}), but with an updated line list (that includes Mg and Al) and non-LTE corrections from more recent sources where available. In brief, the equivalent width (EW) method was used in combination with the {\sc ares} \citep[v2]{ares}, {\sc iraf} {\it onedspec}, and {\sc moog} \citep[June 2014 release]{moog} packages, with $\upalpha$-enhanced (+0.4~dex) 1D model atmospheres interpolated from the \citet{atlas9odfnew} grid. Although the M\,4 spectral data is unchanged from \citetalias{maclean2016}, for consistency all Na\,{\sc i} and O\,{\sc i} EWs were remeasured (with little change), while Mg\,{\sc i}, and Al\,{\sc i} EWs are new, since these abundances were not determined in \citetalias{maclean2016}. 

All absorption lines measured are known to suffer from non-LTE effects \citep{bergemann2014}. Abundances of O, Na, and Al were corrected for these non-LTE effects by interpolation of the grids from \citet[O]{amarsi2016_oxygen}, \citet[Na]{lind2011}, and \citet[Al]{nordlander2017}. Mg was not corrected for non-LTE because it is known (and confirmed in this study) to be homogeneous in M\,4. More detail of this method, and our adopted line list, can be found in \citetalias{maclean2017}.

As in \citetalias{maclean2017}, we were unable to correct our derived Fe abundances for non-LTE effects on a line-by-line basis due to the large number of Fe\,{\sc i} lines in the stellar spectrum. We have therefore performed a test on a representative subset of three RGB and three AGB stars from M\,4, using corrections interpolated from the \citet{amarsi2016_iron} grid for five Fe\,{\sc i} lines\footnote{4788.8\AA, 4839.5\AA, 5701.6\AA, 5753.1\AA~and 7748.3\AA} and two Fe\,{\sc ii} lines\footnote{6516.1\AA~and 7711.7\AA}. For our sample of M\,4 stars the non-LTE effects on Fe\,{\sc i} and Fe\,{\sc ii} are negligible considering our uncertainty in individual abundances (discussed in \S\ref{m4_2_abundance_results}), thus we do not apply them to our final abundances. The O, Na, Al non-LTE corrections for our sample are largely systematic with minimal star-to-star scatter. However, the average corrections for the three species are slightly different ($\Delta_{\rm corr} \sim 0.03$ to 0.06) for the AGB and RGB. We summarise the results of our Fe non-LTE test along with the non-LTE corrections of Na and Al abundances in Table~\ref{tab:m4_2_nlte}.

\begin{table}
\centering
\caption{Summary of average non-LTE corrections for each chemical species.}
\label{tab:m4_2_nlte}
\begin{tabular}{ccc}
\hline 
Species & \multicolumn{2}{c}{Average non-LTE Correction} \\
 & AGB & RGB \\
\hline\hline
Fe\,{\sc i} & $+0.00 \pm 0.03$ & $-0.01 \pm 0.07$ \\
Fe\,{\sc ii} & $-0.01 \pm 0.00$ & $-0.01 \pm 0.01$ \\
O\,{\sc i}  & $-0.16 \pm 0.04$ & $-0.10 \pm 0.02$ \\
Na\,{\sc i} & $-0.11 \pm 0.03$ & $-0.14 \pm 0.03$ \\
Al\,{\sc i} & $-0.13 \pm 0.03$ & $-0.10 \pm 0.03$ \\
\hline
\end{tabular}
\end{table}

%----------------------------------------------------------

\subsection{Abundance results}
\label{m4_2_abundance_results}

Chemical abundances using the new stellar parameters from this study are presented in Table~\ref{tab:m4_2_abunds}. Individual uncertainties cited in these tables are based only on the line-to-line scatter of each abundance. Using the 1$\sigma$ uncertainties of each stellar parameter ($\pm 65$~K in T$_{\rm eff}$, $\pm 0.2$ in log~$g$, $\pm 0.1$~km/s in $v_t$), an atmospheric sensitivity analysis was performed on a representative sub-sample and the results are summarised in Table~\ref{tab:m4_2_atmos}. The uncertainty in abundances due to atmospheric uncertainties is $\leq \pm 0.05$ for all species, except for Fe\,{\sc ii} and O\,{\sc i} which are $\pm 0.10$ and $\pm 0.13$, respectively.

The use of elemental ratios with respect to Fe can be problematic, especially in globular clusters that are homogeneous in Fe abundance at the level of uncertainty in the relevant studies \citep[i.e. when not using differential analysis methods such as in][]{yong2013}. In these cases, dividing star-to-star elemental abundances by Fe abundance adds noise from the imperfect measurement of [Fe/H] and thereby degrades the signal in star-to-star abundance distributions (see \citetalias{campbell2017} for a detailed analysis). Throughout this paper we present all abundances in the form ${\rm log_{\epsilon}(X)}$\footnote{${\rm log_{\epsilon}(X) = log_{10}(N_{X}/N_{H}) + 12.0}$, where N$_X$ represents the number density of atoms of element $X$.}, which eliminates many systematic offsets that may exist in [X/Fe] and [X/H] ratios -- for example adopted solar abundances, and the sensitivity of Fe\,{\sc i} to T$_{\rm eff}$.

A detailed comparison to recent high-resolution spectroscopic studies of M\,4 is not only warranted, but crucial for this cluster. We reserve this analysis and discussion for {\S}\ref{m4_2_lit_comp}, except for a comparison with our previous results (\citetalias{maclean2016}), which is presented in Table~\ref{tab:m4_2_ML16_comp}. The only change of note is in log~$g$. In \citetalias{maclean2016} we assumed a mass of 0.8~M$_{\odot}$ for all stars, while here we assumed a mass of 0.7~M$_{\odot}$ for our AGB sample, which accounts for $-0.10$~dex of the $-0.15$ difference in log~$g$ values for the AGB stars. No other significant changes occurred in the re-analysis, with T$_{\rm eff}$, ${\rm log_{\epsilon}(Fe\,\textsc{i})}$, ${\rm log_{\epsilon}(O)}$, and ${\rm log_{\epsilon}(Na)}$ showing very little change. The scatter is indicative of our parameter uncertainties\footnote{An exception is the scatter in $v_t$ differences, which has little effect on elemental abundances -- see Table~\ref{tab:m4_2_atmos}.} and estimated total abundance errors (Table~\ref{tab:m4_2_uncerts}).

\begin{table*}
\centering
\caption{Chemical abundances for each star in our M\,4 sample. Abundance uncertainties reflect line-to-line scatter (1$\sigma$), and do not take atmospheric sensitivities into account (see Table~\ref{tab:m4_2_atmos}). The last four lines show the cluster average abundances (for the AGB and RGB) with standard error of the mean, and standard deviation to indicate observed scatter. O, Na, and Al abundances were corrected for non-LTE effects. Only the first five rows are shown; the full table is available online.}
\label{tab:m4_2_abunds}
\begin{tabular}{cccccccc}
\hline
ID  & Type & ${\rm log_{\epsilon}(Fe\,\textsc{i})}$ & ${\rm log_{\epsilon}(Fe\,\textsc{ii})}$ & ${\rm log_{\epsilon}(O)}$ & ${\rm log_{\epsilon}(Na)}$ & ${\rm log_{\epsilon}(Mg)}$ & ${\rm log_{\epsilon}(Al)}$ \\
\hline\hline
788   & AGB  & $6.24 \pm 0.08$ & $6.24 \pm 0.03$ & $8.26 \pm 0.05$ & $4.93 \pm 0.01$ & $6.79 \pm 0.03$ & $5.56 \pm 0.02$ \\
3590  & AGB  & $6.27 \pm 0.06$ & $6.29 \pm 0.04$ & $8.07 \pm 0.01$ & $5.15 \pm 0.03$ & $6.73 \pm 0.04$ & $5.67 \pm 0.03$ \\
10092 & AGB  & $6.33 \pm 0.04$ & $6.35 \pm 0.01$ & $8.29 \pm 0.06$ & $4.95 \pm 0.02$ & $6.72 \pm 0.03$ & $5.53 \pm 0.03$ \\
11285 & AGB  & $6.32 \pm 0.07$ & $6.31 \pm 0.04$ & $8.10 \pm 0.04$ & $5.19 \pm 0.02$ & $6.80 \pm 0.02$ & $5.71 \pm 0.06$ \\
13609 & AGB  & $6.32 \pm 0.09$ & $6.38 \pm 0.06$ & $8.12 \pm 0.04$ & $5.02 \pm 0.11$ & $6.76 \pm 0.05$ & $5.57 \pm 0.05$ \\
\vdots & \vdots & \vdots & \vdots & \vdots & \vdots & \vdots & \vdots \\
\hline
Mean & AGB & $6.30 \pm 0.01$ & $6.31 \pm 0.01$ & $8.18 \pm 0.02$ & $5.11 \pm 0.03$ & $6.76 \pm 0.01$ & $5.62 \pm 0.02$ \\
$\sigma$ & & 0.03 & 0.04 & 0.09 & 0.12 & 0.03 & 0.08 \\
Mean & RGB & $6.33 \pm 0.01$ & $6.34 \pm 0.01$ & $8.10 \pm 0.01$ & $5.33 \pm 0.02$ & $6.78 \pm 0.01$ & $5.76 \pm 0.01$ \\
$\sigma$ &  & 0.05 & 0.06 & 0.12 & 0.19 & 0.05 & 0.09 \\
\hline
\end{tabular}
\end{table*}

\begin{table}
\centering
\caption{Typical abundance uncertainties due to the (1$\sigma$) atmospheric sensitivities of a representative sub-sample of three RGB and two AGB stars in our M\,4 data set. Parameter variations (in parentheses) are the adopted uncertainties in the respective parameters. Note the direction of signs.}
\label{tab:m4_2_atmos}
\begin{tabular}{lccccc}
\hline
 & ${\Delta}T_{\rm eff}$          & ${\Delta}$log~$g$    & ${\Delta}v_t$         & \bf Total       \\% & ($\sigma$)          \\
 & ($\pm 65$~K) & ($\pm$0.2) & ($\pm$0.1) \\
\hline\hline
${\rm log_{\epsilon}(Fe\,\textsc{i})}$  & $\pm$0.05    & $\pm$0.00       & $\mp$0.02      & \bf $\pm$0.05 \\% & (0.01) \\
${\rm log_{\epsilon}(Fe\,\textsc{ii})}$  & $\mp$0.05    & $\pm$0.09       & $\mp$0.02      & \bf $\pm$0.10 \\% & (0.03) \\
${\rm log_{\epsilon}(O)}$     		   & $\mp$0.10    & $\pm$0.08       & $\mp$0.01       & \bf $\pm$0.13 \\% & (0.02) \\
${\rm log_{\epsilon}(Na)}$   		   & $\pm$0.05    & $\mp$0.01       & $\mp$0.02       & \bf $\pm$0.04 \\% & (0.02) \\
${\rm log_{\epsilon}(Mg)}$    		   & $\pm$0.03    & $\pm$0.00       & $\mp$0.01       & \bf $\pm$0.03 \\% & (0.03) \\
${\rm log_{\epsilon}(Al)}$    		   & $\pm$0.04    & $\pm$0.00       & $\pm$0.00       & \bf $\pm$0.04 \\% & (0.01) \\
\hline
\end{tabular}
\end{table}

\begin{table}
\centering
\caption{Summary of typical abundance uncertainties (1$\sigma$) from each source identified in the text, and the total uncertainties (when added in quadrature). The first column are the average line-to-line uncertainties of all stars, values in the second column are the total uncertainties from atmospheric sensitivities (Table~\ref{tab:m4_2_atmos}), and the third column represents the typical uncertainties in non-LTE corrections, as reported in the relevant sources (see {\S}\ref{m4_2_abund_method}). Note that individual Fe abundances were not corrected for non-LTE (see text for details).}
\label{tab:m4_2_uncerts}
\begin{tabular}{ccccc}
\hline
Species & Line-to-Line & Atmospheric & non-LTE & \textbf{Total} \\
\hline\hline
Fe\,{\sc i}  & $\pm 0.09$ & $\pm 0.05$ & -          & \bf $\pm$0.10 \\
Fe\,{\sc ii} & $\pm 0.04$ & $\pm 0.10$ & -          & \bf $\pm$0.11 \\
O            & $\pm 0.05$ & $\pm 0.13$ & $\pm 0.05$ & \bf $\pm$0.15 \\
Na           & $\pm 0.04$ & $\pm 0.04$ & $\pm 0.04$ & \bf $\pm$0.07 \\
Mg           & $\pm 0.04$ & $\pm 0.03$ & -          & \bf $\pm$0.05 \\
Al           & $\pm 0.03$ & $\pm 0.04$ & $\pm 0.06$ & \bf $\pm$0.08 \\
\hline
\end{tabular}
\end{table}

\begin{table}
\centering
\caption{The average differences in parameters and abundances between this study and \protect{\citet[\citetalias{maclean2016}]{maclean2016}}. Uncertainties are standard deviations, and indicate the scatter between the studies, if the offsets were removed. The significant change in log~$g$ values is discussed in the text. Note that abundances from Fe\,{\sc ii} lines were not published in \citetalias{maclean2016}.}
\label{tab:m4_2_ML16_comp}
\begin{tabular}{lcc}
\hline
Parameter & \multicolumn{2}{c}{This study $-$ \citetalias{maclean2016}} \\
 & (AGB) & (RGB) \\
\hline\hline
$\Delta T_{\rm eff}$         & $-21$ $\pm$ 44     & $-20$ $\pm$ 57     \\
$\Delta$log~$g$              & $-0.15$ $\pm$ 0.12 & $-0.04$ $\pm$ 0.10 \\
$\Delta v_t$                 & $+0.12$ $\pm$ 0.18 & $+0.12$ $\pm$ 0.15 \\
$\Delta{\rm log_{\epsilon}(Fe\,\textsc{i})}$  & $-0.02$ $\pm$ 0.04 & $-0.02$ $\pm$ 0.05 \\
$\Delta{\rm log_{\epsilon}(O)}$            & $+0.06$ $\pm$ 0.07 & $+0.04$ $\pm$ 0.09 \\
$\Delta{\rm log_{\epsilon}(Na)}$        & $-0.03$ $\pm$ 0.06 & $+0.00$ $\pm$ 0.06 \\
\hline 
\end{tabular}
\end{table}

\begin{figure}
\centering
\includegraphics[width=.9\linewidth]{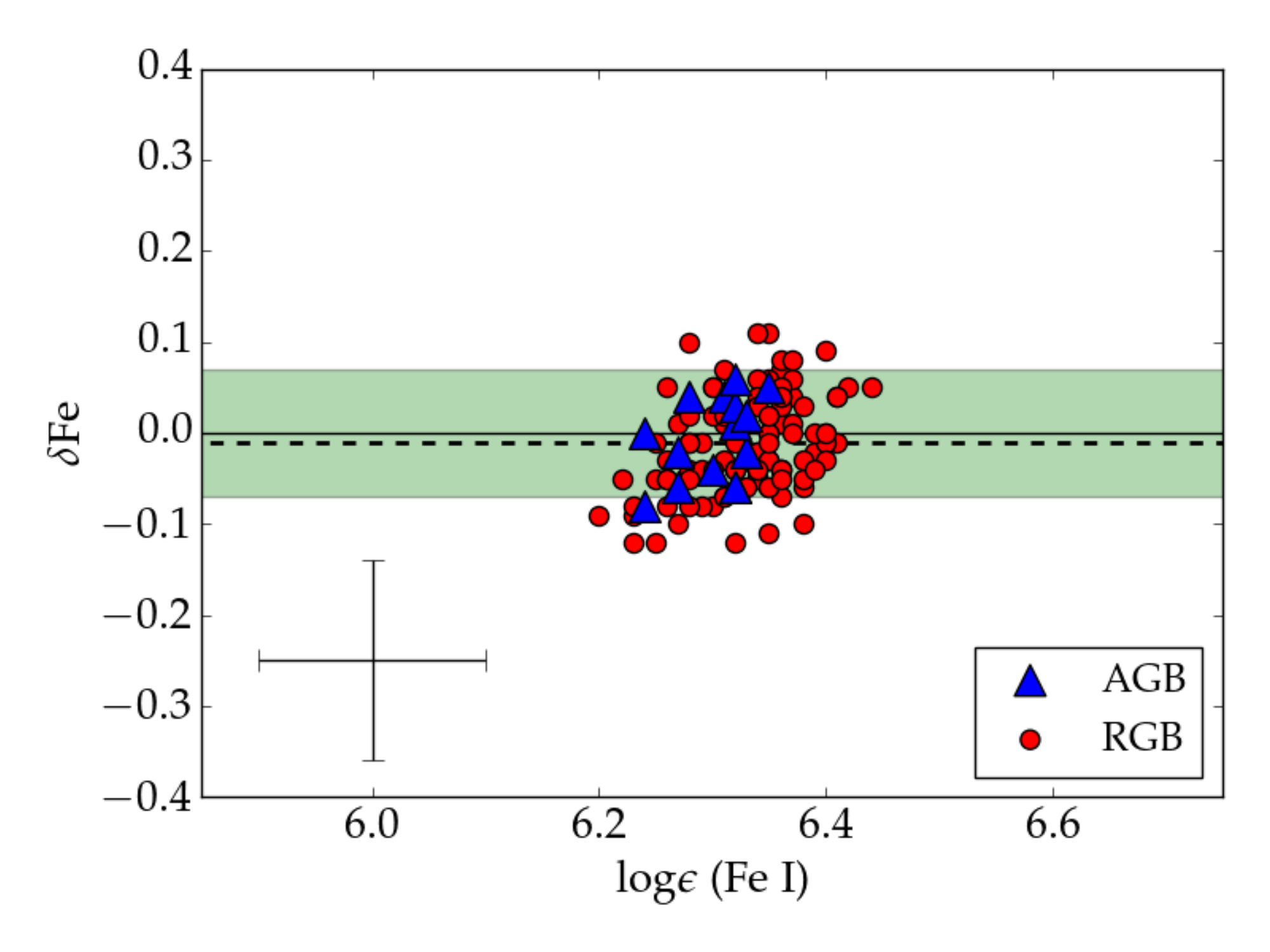}
\caption{Fe abundances for this study. Here, ionisation difference ($\delta$Fe = ${\rm log_{\epsilon}(Fe\,\textsc{i})}$ -- ${\rm log_{\epsilon}(Fe\,\textsc{ii})}$) is plotted against ${\rm log_{\epsilon}(Fe\,\textsc{i})}$ abundance to highlight departures from LTE in Fe\,{\sc i}, and the similarity between the Fe abundances of the AGB and RGB. The error bar indicates typical 1$\sigma$ total uncertainties in individual abundances (i.e. the line-to-line uncertainties and the 1$\sigma$ atmospheric sensitivity uncertainties added in quadrature), while the black dashed line represents the sample average $\delta$Fe value of $-0.01$. The shaded green region indicates the non-LTE uncertainties quoted in \protect{\citet[$\pm 0.05$~dex]{amarsi2016_iron}}, around the expected $\delta$Fe value ($+0.00$~dex, solid black line) from our non-LTE test (see \S\ref{m4_2_abund_method}).}
\label{fig:m4_2_dfe}
\end{figure}

Abundances from Fe\,{\sc ii} lines were not published in \citetalias{maclean2016}, but are included here as part of our re-analysis. In Figure~\ref{fig:m4_2_dfe} we plot ${\rm log_{\epsilon}(Fe\,\textsc{i})}$ against $\delta$Fe (ionisation balance; $\delta$Fe = ${\rm log_{\epsilon}(Fe\,\textsc{i})}$ -- ${\rm log_{\epsilon}(Fe\,\textsc{ii})}$). Our non-LTE test (see \S\ref{m4_2_abund_method}) predicted a theoretical $\delta$Fe value of $0.00 \pm 0.07$, while our observed sample has an average $\delta$Fe of $-0.01 \pm 0.05$. This high level of agreement is strong evidence that our {\sc phobos v2} spectroscopic method is reliable, and that our stellar parameters are accurate.

\begin{figure}
\centering
\includegraphics[width=0.9\linewidth]{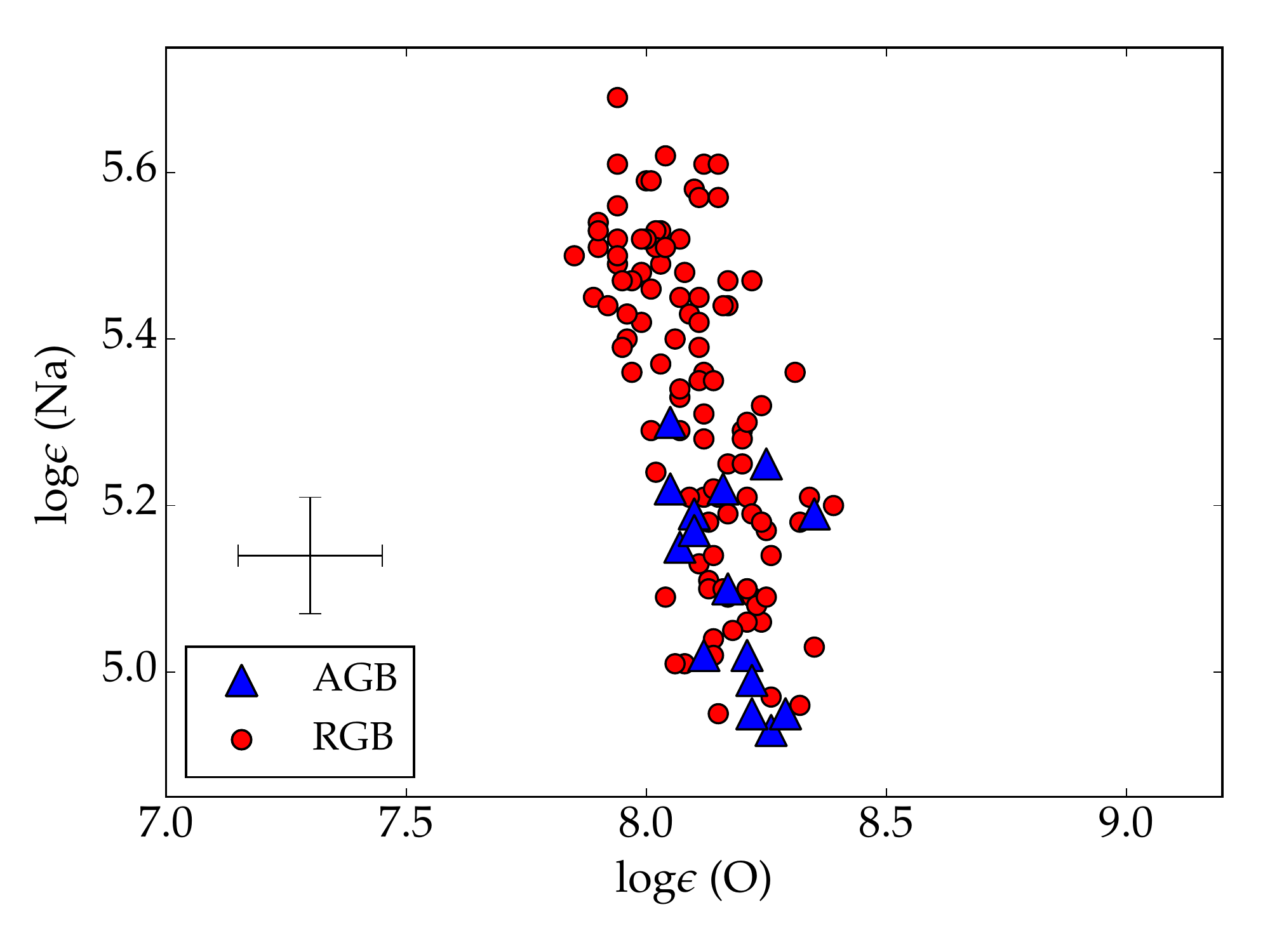}
\caption{O and Na abundances for our M\,4 sample. The error bar indicates typical 1$\sigma$ total uncertainties in individual abundances (i.e. the line-to-line uncertainties and the 1$\sigma$ atmospheric sensitivity uncertainties added in quadrature).}
\label{fig:m4_2_nao}
\end{figure}

As in \citetalias{maclean2016}, M\,4 shows a significant spread in Na abundance among RGB stars ($\sigma = \pm 0.19$~dex; see Figure~\ref{fig:m4_2_nao}). However, considering the uncertainty in O abundance we cannot resolve the Na-O anti-correlation that has been reported elsewhere \citep[e.g.][]{marino2008}. In fact, given the total uncertainty in our O abundances of $\pm 0.15$~dex (Table~\ref{tab:m4_2_uncerts}) -- compared to the O spread on the RGB of $\pm 0.12$~dex (Table~\ref{tab:m4_2_abunds}) -- we cannot say that M\,4 actually shows heterogeneity in O abundance, formally it appears to be homogeneous. This uncertainty in ${\rm log_{\epsilon}(O)}$ comes from the large sensitivity of the 777nm triplet to T$_{\rm eff}$ and log~$g$, and is typically smaller for other O lines that we could not observe with HERMES/AAT. Na, on the other hand, shows a significant star-to-star scatter in both the RGB, and (to a smaller degree; $\sigma = \pm 0.12$~dex) the AGB.

\begin{figure}
\centering
\includegraphics[width=0.9\linewidth]{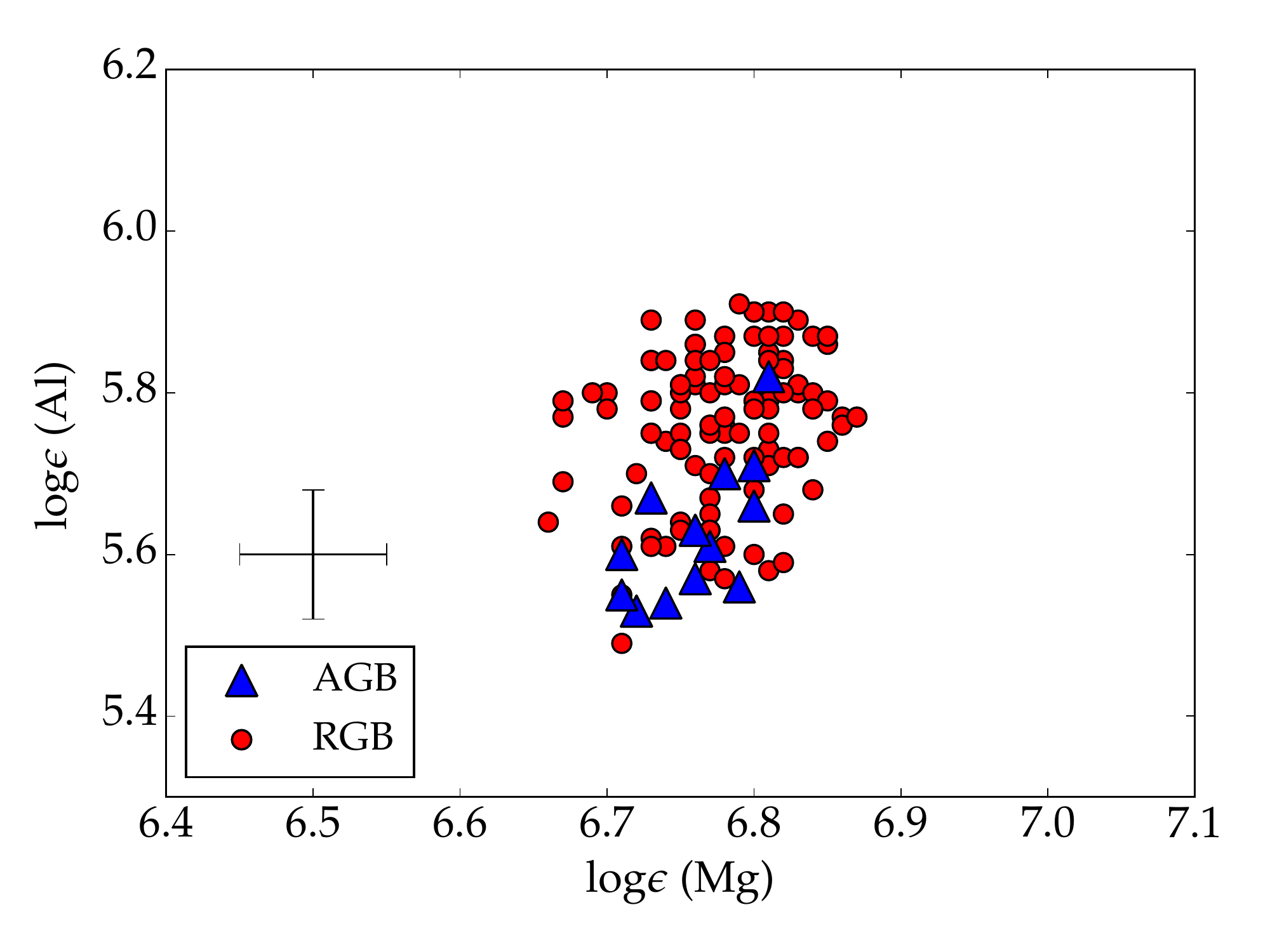}
\caption{Same as Figure~\ref{fig:m4_2_nao}, but for Mg and Al abundances.}
\label{fig:m4_2_mgal}
\end{figure}

We find a correlation between Na and Al abundance, but no evidence of a Mg-Al anti-correlation (Figure~\ref{fig:m4_2_mgal}), in agreement with previous results (e.g. \citetalias{marino2008}). A clear outlier is the star AGB18573 which appears to have a low Na abundance but a high Al abundance. We have not been able to provide an explanation for this anomalous star, however, it was reported by \citetalias{marino2017} to be similarly Na-poor and Al-rich. We find Mg to be homogeneous in M\,4 ($\sigma = \pm 0.05$~dex on the RGB), while Al is difficult to classify because the star-to-star scatter ($\sigma = \pm 0.09$ and $\pm 0.08$~dex on the RGB and AGB, respectively) is similar to our total uncertainties in the abundance ($\pm 0.08$~dex). We note however, that for the AGB, the 1$\sigma$ spread in Al abundance reduces to $\pm 0.06$~dex when the Al-rich outlier AGB18573 is discounted, and can be seen in Figure~\ref{fig:m4_2_mgal} to have a smaller spread than our RGB sample.

As in \citetalias{maclean2016}, the average Na, O, and Al abundances of AGB stars in M\,4 are clearly different to that of the RGB, being heavily weighted toward SP1-like abundances. Our Fe and Mg abundances are constant, and the average RGB and AGB abundances agree. These results are consistent with our claim in \citetalias{maclean2016} that M\,4 may not contain SP2 AGB stars ($\mathscr{F} = 100\%$). Due to the spread in AGB Na abundances, and our abundance uncertainties, we conclude that $\mathscr{F} \gtrsim 65\%$ -- i.e. less than 20\% of AGB stars, or 3 out of 15, have SP2-like abundances. This compares with 55\% on the RGB. This value is considerably higher than that expected from stellar evolutionary theory ($\mathscr{F} = 0\%$) for a cluster with a HB extending only to T$_{\rm eff} \simeq 9000$~K.

%%%%%%%%%%%%%%%%%%%%%%%%%%%%%%%%%%%%%%%%%%%%%%%%%%%%%%%%%%%%%%

\section{Cyanogen band strengths from low-resolution spectra}
\label{m4_2_cn}

As a further observational check of the relative abundance distributions of M\,4's AGB and RGB, we determined CN band-strengths for a sample of M\,4 stars. The bimodality of CN band strengths in M\,4 is well established \citep{norris1981m4,ivans1999}, and can be used to identify to which subpopulation (SP1 or SP2) a star belongs because CN band strengths have been shown to correlate with Na abundance\footnote{CN band strengths are primarily indicative of atmospheric N abundance, which correlates with Na} \citep{cottrell1981,campbell2012,smith2015}.

In addition to our sample of high-resolution spectra, low-resolution spectra of M\,4 stars were collected in September 2009 \citep{campbell2010} using the AAOmega/2dF multi-object spectrograph on the Anglo-Australian Telescope \citep[$R \simeq 3000$;][]{aaomega1,aaomega2,aaomega3}. We used the 1700B grating which gave a spectral coverage from 3755~\AA~to~4437~\AA, while the signal-to-noise ratio for all targets was $\gtrsim 20$. The software package {\sc 2dfdr} \citep[v3.211]{2dfdr} was used to reduce the data in preparation for analysis. This is new and unpublished data, and is included to provide an additional avenue for the investigation of M\,4 abundance distributions. A total of 7 AGB and 19 RGB stars were observed with AAOmega; all but two of which (stars 25133 and 17999) were included in our HERMES target list.

\begin{figure}
\centering
\includegraphics[width=0.9\linewidth]{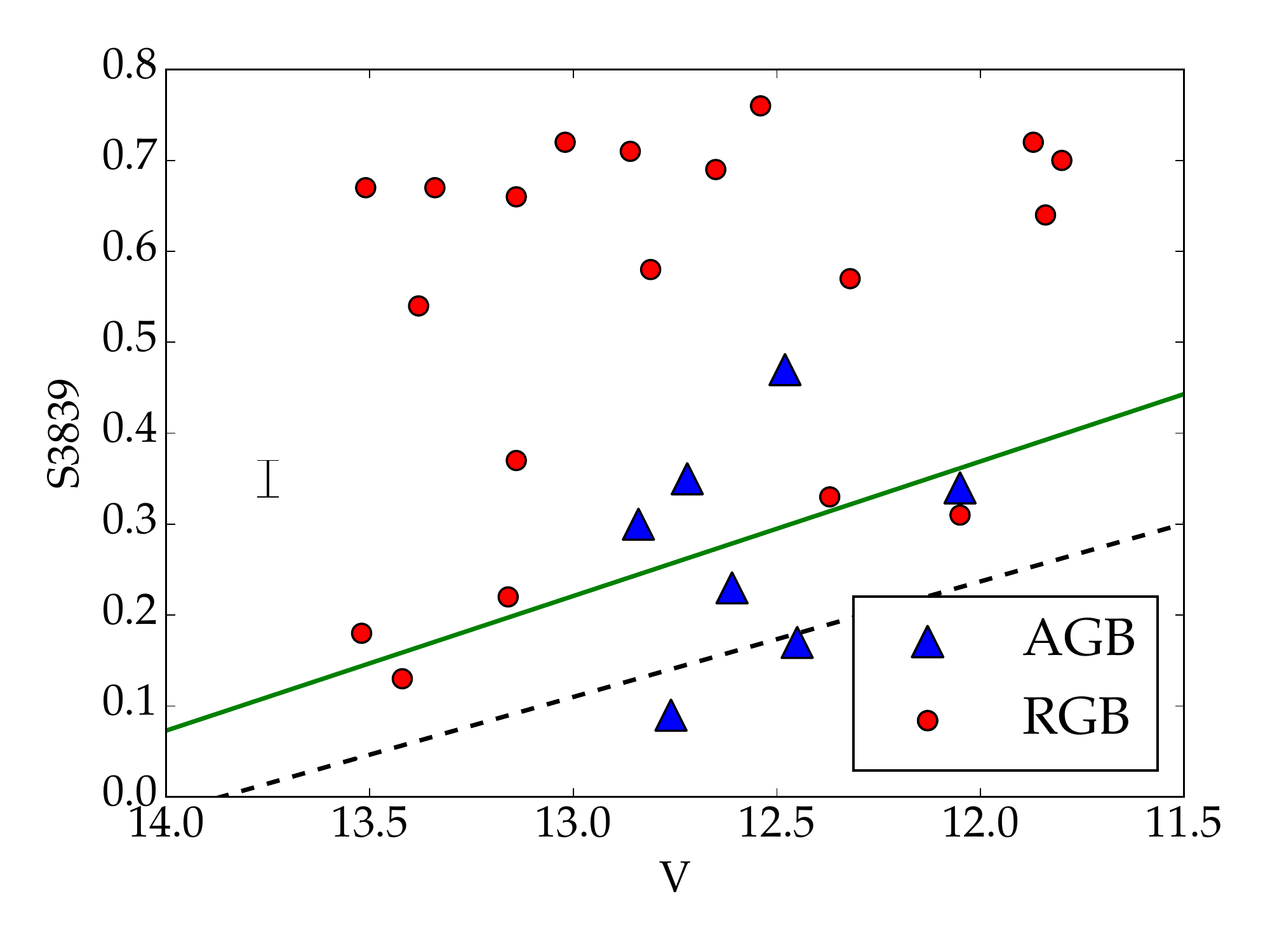}
\caption{S3839 CN index values versus $V$-band magnitudes for our M\,4 low-resolution sample. The green trend-line is a linear best fit for the five RGB stars with the lowest CN band strengths ($S3839 = -0.148V + 2.145$), while the dashed trend-line is the baseline from Figure 3 of \protect{\citet[$S3839 = -0.127V + 1.761$]{norris1981m4}}. The typical S3839 uncertainty is represented on the left.}
\label{fig:m4_2_cn}
\end{figure}

To quantify the CN band strengths we use the S3839 CN index from \citet{norris1981m4} which compares a spectral segment where the CN molecule absorbs light with a neighbouring pseudo-continuum:
\begin{equation}
S3839 = -2.5~{\rm log} \frac{\int^{3883}_{3846}I_{\lambda}d{\lambda}}{\int^{3916}_{3883}I_{\lambda}d{\lambda}}.
\end{equation}
IRAF was used to measure the integrated fluxes of our low-resolution spectra. Target data, S3839 values, and $\delta$S3839 excess values are given in Table~\ref{tab:m4_2_cn}. CN band strengths are presented in Figure~\ref{fig:m4_2_cn}. 

Even without adjusting for the trend with V band magnitude \citep[called the baseline in][]{norris1981m4,ivans1999}, it can be seen that the RGB stars display a significant spread in S3839 values, and that our AGB sample are heavily weighted to low S3839 index values. The green fiducial line in Figure~\ref{fig:m4_2_cn} was used to empirically correct for the trend between V band magnitude and S3839 value ($\delta$S3839 excess is the vertical distance of each star to the green fiducial), and as a reference we include the baseline used by \citet{norris1981m4} which is qualitatively similar.

We adopt the characteristic S3839 uncertainty of $\pm 0.02$ from \cite{campbell2012}, which was based on the typical differences between S3839 measurements from two separate observations of the same star in the GC NGC\,1851 (the spectra of which were obtained during the same observing program and with the same technical specifications as the M\,4 spectra used in this study), and a typical $\delta$S3839 uncertainty of 0.08~dex due to assumptions in determining the trend with V band magnitude. We discuss our CN results further in the next section, in comparison with previous CN studies on M\,4.

\begin{table}
\centering
\caption{S3839 CN index values for the low-resolution M\,4 sample, along with $V$-band magnitudes and $\delta$S3839 excess values. The last four lines show the cluster average abundances (for the AGB and RGB) with standard error of the mean, and standard deviation to indicate observed scatter. Note that all but two (stars 25133 and 17999) of the low-resolution targets were also observed with HERMES in high-resolution. $V$-band magnitudes and IDs are from \protect{\citet{momany2003}}.}
\label{tab:m4_2_cn}
\begin{tabular}{ccccc}
\hline
ID  & Type & $V$ & S3839 & $\delta$S3839 \\
\hline\hline
3590  & AGB  & 12.48  & $0.47$  & $0.17$   \\
10092 & AGB  & 12.61  & $0.23$  & $-0.04$  \\
11285 & AGB  & 12.84  & $0.30$  & $0.06$   \\
13609 & AGB  & 12.76  & $0.09$  & $-0.17$  \\
20089 & AGB  & 12.72  & $0.35$  & $0.05$   \\
25133 & AGB  & 12.45  & $0.17$  & $-0.13$  \\
46676 & AGB  & 12.05  & $0.34$  & $-0.02$   \\
1029  & RGB  & 13.14  & $0.66$  & $0.46$   \\
3114  & RGB  & 13.38  & $0.54$  & $0.37$   \\
4361  & RGB  & 13.51  & $0.67$  & $0.52$   \\
4806  & RGB  & 13.16  & $0.22$  & $0.03$   \\
4938  & RGB  & 12.86  & $0.71$  & $0.46$   \\
6978  & RGB  & 13.34  & $0.67$  & $0.50$   \\
7298  & RGB  & 13.42  & $0.13$  & $-0.03$  \\
8803  & RGB  & 11.87  & $0.72$  & $0.34$   \\
9040  & RGB  & 12.32  & $0.57$  & $0.25$   \\
10801 & RGB  & 12.54  & $0.76$  & $0.45$   \\
10928 & RGB  & 11.80  & $0.70$  & $0.30$   \\
12387 & RGB  & 13.14  & $0.37$  & $0.17$   \\
13170 & RGB  & 13.52  & $0.18$  & $0.04$   \\
14037 & RGB  & 12.05  & $0.31$  & $-0.05$  \\
14350 & RGB  & 12.65  & $0.69$  & $0.42$   \\
14377 & RGB  & 12.81  & $0.58$  & $0.33$   \\
15010 & RGB  & 12.37  & $0.33$  & $0.01$   \\
17999 & RGB  & 11.84  & $0.64$  & $0.24$   \\
23196 & RGB  & 13.02  & $0.72$  & $0.50$  \\
\hline
Mean & AGB & - & $0.28 \pm 0.03$ & $-0.01 \pm 0.03$  \\
$\sigma$ & &  & 0.13 & 0.12  \\
Mean & RGB & - & $0.51 \pm 0.02$ & $0.23 \pm 0.02$  \\
$\sigma$ &  &  & 0.21 & 0.25  \\
\hline
\end{tabular}
\end{table}

\begin{figure}
\centering
\includegraphics[width=0.9\linewidth]{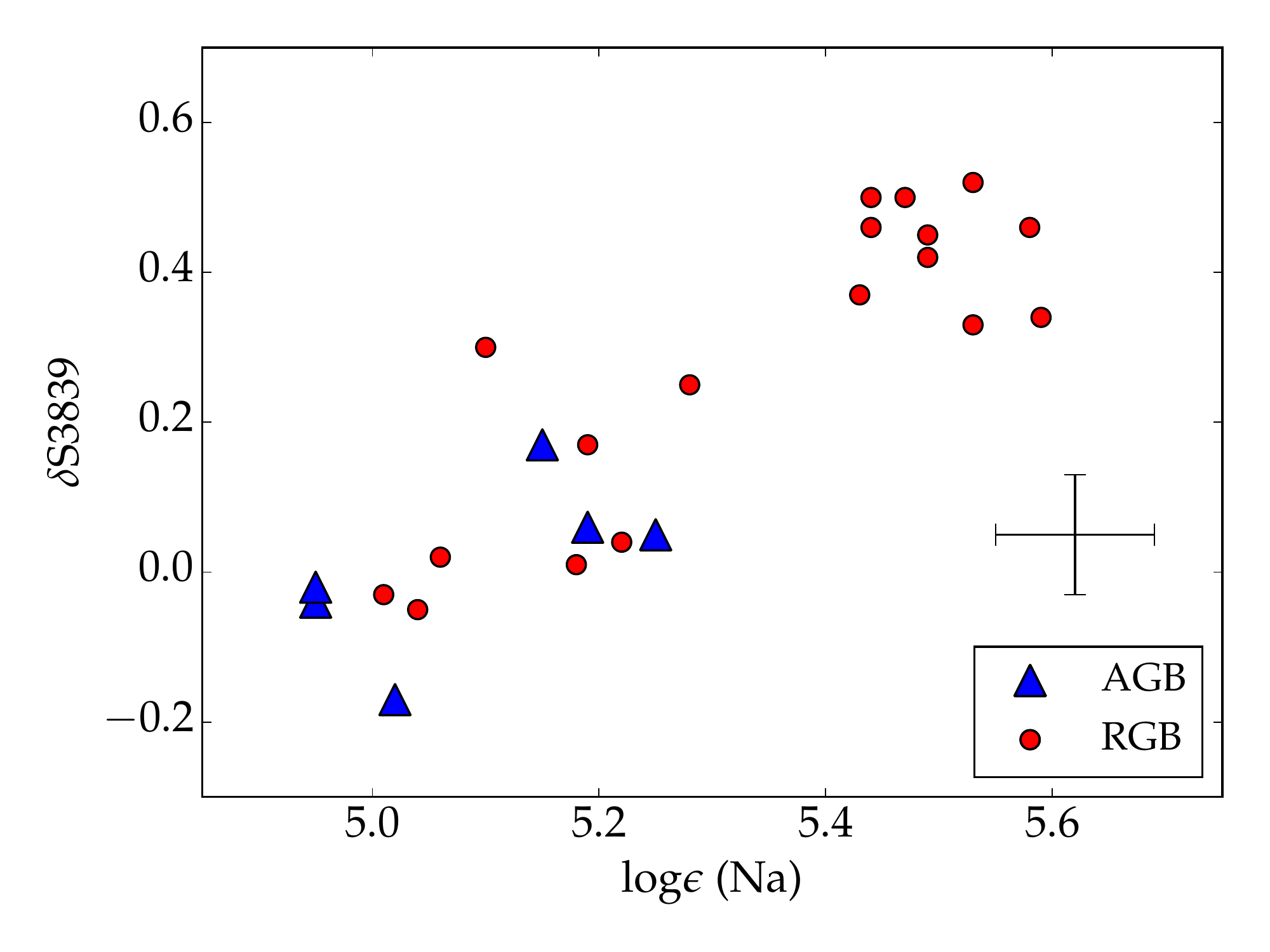}
\caption{$\delta$S3839 (excess CN index values) versus Na abundances for stars in both our M\,4 HERMES and AAOmoga samples. $\delta$S3839 values are the distance a star is above the green trend-line in Figure~\ref{fig:m4_2_cn}. The error bar represents typical uncertainties. Note that for two stars (25133 and 17999) only low-resolution spectra were observed, and they are therefore not included in this plot.}
\label{fig:m4_2_nacn}
\end{figure}

%%%%%%%%%%%%%%%%%%%%%%%%%%%%%%%%%%%%%%%%%%%%%%%%%%%%%%%%%%%%%%

\section{Literature comparison of AGB abundances}
\label{m4_2_lit_comp}

After determining reliable elemental abundances and CN band strengths, we compiled and compared spectroscopic results from the literature in order to investigate the conflicting conclusions regarding M\,4's AGB abundances.

While \citetalias{maclean2016} was the first study that systematically targeted the AGB of M\,4, AGB stars had been included previously in several spectroscopic studies of the cluster: \citet{norris1981m4}, \citet{suntzeff1991}, and \citet[hereafter \citetalias{ivans1999}]{ivans1999}. CN band strengths and abundances from these three studies were compiled and merged into the data set of \citet[hereafter \citetalias{sb2005}]{sb2005} who reported on six AGB stars (two of which they classified as CN-strong, one as CN-intermediate, and the remaining three as CN-weak). \citetalias{ivans1999} reported that their AGB abundances show less evidence of H-burning than their RGB sample, and described their AGB results as ``puzzling''.

Soon after the publication of \citetalias{maclean2016}, \citet{lardo2017} disputed our conclusion by utilising a pseudo-CMD with the photometric index C$_{UBI}$ = ($U - B$) $-$ ($B - I$), which has been used to separate the RGB (and the AGB more recently) subpopulations of GCs \citep[e.g.][]{monelli2013,garciahernandez2015}. They demonstrated that the spread in C$_{UBI}$ for their sample of AGB stars is statistically similar to that of the RGB. \citetalias{marino2017} performed a similar study using both the C$_{UBI}$ index and the combination of Hubble Space Telescope (HST) filters C$_{F275W,F336W,F438W} = (m_{F275W} - m_{F336W}) - (m_{F336W} - m_{F438W}$). They came to a similar conclusion as \citet{lardo2017} -- that photometric data of M\,4 stars suggest that both SP1 and SP2 stars ascend the AGB. Evidence for this lies in the broadness of the branches in the pseudo-CMDs, for which a double sequence (or a single sequence that is broader than expected from observational errors) is understood to indicate a spread in the abundances of H-burning products (primarily He, N, and C; see \citealt{milone201247tuc}). Both photometric investigations of the AGB of M\,4 found that the broadness of the branch is consistent with a heterogeneity in He, N, and C of similar magnitude as the RGB of M\,4, in contradiction to the conclusions of \citetalias{maclean2016}.

Although C$_{UBI}$ has been used to infer most of these results, the broadness of the $UBI$ filter pass-bands means that they incorporate a multitude of atomic lines and molecular bands, which makes abundance information that has been inferred from photometric bands difficult to interpret, and can only be used to infer the collective differences that may be the result of a range of spectroscopic features. In an era where medium to high-resolution spectroscopic data is available, these spectra provide a much more definitive answer to the discussion of subpopulations. We therefore focus on spectroscopic data in this investigation.

In response to the unexpected findings of \citetalias{maclean2016}, two high-resolution spectroscopic studies -- both using VLT/FLAMES spectra -- have been performed on M\,4 AGB stars: \citetalias{marino2017} and \citet[\citetalias{wang2017}]{wang2017}. \citetalias{marino2017} determined the abundances of a range of species (most relevant to this comparison are the abundances of O, Na, Mg, Al, and Fe) for a sample of 17 AGB stars, but did not re-observe or redetermine abundances for RGB stars. They reported that their AGB sample showed similar [Na/Fe] and [O/Fe] values to a sample of RGB abundances from \citetalias{marino2008} -- on average their AGB sample had [Na/Fe] values only 0.08~dex lower than the RGB sample -- thereby challenging the conclusion of \citetalias{maclean2016} by reporting the discovery of both SP1 and SP2-like AGB stars in M\,4. 

\citetalias{wang2017} observed a sample of 19 AGB and 68 RGB stars in M\,4, and determined Fe and Na abundances for each star. They reported that their AGB sample shows, on average, lower [Na/H] values than their RGB sample (by 0.14~dex). This was in broad agreement with \citetalias{maclean2016}, however they reported a larger spread in Na abundances on the AGB -- $\sigma = 0.17$~dex compared to $0.14$~dex in \citetalias{maclean2016}; however their uncertainties in [Na/H] are larger than those determined for our Na abundances ($\pm 0.16$~dex compared to $\pm 0.11$~dex). They also noted a smaller difference in maximum [Na/H] between the RGB and AGB ($\Delta$[Na/H]$_{max}$ = 0.26~dex compared to 0.40~dex in \citetalias{maclean2016}). Curiously, the [Na/H] results of \citetalias{wang2017} also agreed well with an overlapping sub-sample of \citetalias{marino2017}, confusing the situation further since the conclusions of \citetalias{marino2017} and \citetalias{maclean2016} are in contradiction.

In summary, for our comparison we have collated: 
\renewcommand\labelenumi{(\roman{enumi})} 
\renewcommand\theenumi\labelenumi
\begin{enumerate}
\item the O, Na, Mg, Al, and Fe abundances from \citetalias{ivans1999}, 
\item the CN band strengths from \citetalias{sb2005}, 
\item the O, Na, Mg, Al, and Fe abundances from \citetalias{marino2017} and \citetalias{marino2008}, 
\item the Na and Fe abundances from \citetalias{wang2017}, 
\item the O, Na, Mg, Al, and Fe abundances from this study, and 
\item CN band strengths from this study.
\end{enumerate}

The evolutionary-phase designation of targets in \citetalias{ivans1999} was questioned in \citetalias{sb2005}, who reclassified several of the \citetalias{ivans1999} AGB targets. Star 4633 was determined by \citetalias{sb2005} and \citet{suntzeff1991} to be on the RGB, and here we adopt this classification. Targets 2519, 4201, 1701, and 4414 are listed in \citetalias{sb2005} as `uncertain', and we did not include them in our comparison for this reason (we note that their exclusion does not affect the result). For our analysis of the CN band strengths from \citetalias{sb2005}, we redetermined $\delta$S3839 excess values using the green fiducial from Figure~\ref{fig:m4_2_cn} to ensure consistency with the CN results of this study.

The studies of \citetalias{marino2008}, \citetalias{marino2017}, and \citetalias{wang2017} included many of the same stars in M\,4 as \citetalias{maclean2016}, and a direct comparison of the adopted stellar parameters and reported abundances is possible for the overlapping samples. For our comparisons, we use the ${\rm log_{\epsilon}(X)}$ notation in order to avoid including systematic offsets such as solar abundance choice and dividing abundances by Fe abundance. Differences between the values determined in this study and those published in \citetalias{marino2008}, \citetalias{marino2017}, and \citetalias{wang2017} are summarised in Table~\ref{tab:m4_2_overlap}. 

The AGB stellar parameters adopted in this study are largely similar to those in \citetalias{marino2017}, while the RGB sample in \citetalias{marino2008} has, on average, higher log~$g$ values by 0.25~dex than our RGB sample, which is likely connected to their Fe\,{\sc i} abundances which are systematically larger by 0.09~dex\footnote{Ionisation balance was forced in \citetalias{marino2008}, which is controlled primarily by log~$g$.}. There are significant offsets between our abundances and those in \citetalias{marino2008} and \citetalias{marino2017} (up to an average difference of 0.25~dex), however the scatter around these offsets -- typically considered a better indication of the agreement between abundance analysis studies -- is consistent with the uncertainties quoted in this study. A detailed investigation of the differences in Na abundance between our work and \citetalias{marino2017} (AGB EWs were kindly provided by A. F. Marino via priv. comm.) revealed that all offsets were able to be accounted for by quantifiable differences in stellar parameters, non-LTE corrections, choice of atmospheric models, atomic line data, and EWs. The measured EWs for lines in common (the 568nm doublet) were quite similar, with typical differences of the order of 5 m\AA, corresponding to $\Delta{\rm log_{\epsilon}(Na)} \sim 0.09$~dex. 

Comparing our work with that of \citetalias{wang2017}, we note that while the adopted T$_{\rm eff}$ values are quite different ($\sim 100$~K difference), the abundances agree more closely than with \citetalias{marino2008}/\citetalias{marino2017}. There is still a notable offset in AGB Na abundance ($\Delta{\rm log_{\epsilon}(Na)}$ = 0.14~dex), however the large uncertainties quoted in \citetalias{wang2017} ($\pm 0.16$~dex) make it difficult to determine its significance.

We were unable to identify overlapping sample stars with \citetalias{sb2005} and \citetalias{ivans1999}, and therefore could not directly compare the CN band strengths and elemental abundances from these studies in the same manner.

\begin{table*}
\centering
\caption{The average star-to-star differences in parameters and abundances between the published results of \protect{\citet[\citetalias{marino2017}, AGB only]{marino2017}}, \protect{\citet[\citetalias{marino2008}, RGB only]{marino2008}}, \protect{\citet[\citetalias{wang2017}]{wang2017}}, and those of this study. Uncertainties are standard deviations, and indicate the scatter between the studies, if the offsets were removed. While significant offsets exist between our work those of \citetalias{marino2017}, \citetalias{marino2008} and \citetalias{wang2017}, the scatter around the offsets are consistent with the uncertainties quoted in this study (see text for discussion).}
\label{tab:m4_2_overlap}
\begin{tabular}{lcccc}
\hline
Parameter & \citetalias{marino2017} - this study  & \citetalias{marino2008} - this study & \multicolumn{2}{c}{\citetalias{wang2017} - this study} \\
 & (AGB) & (RGB) & (AGB) & (RGB) \\
\hline\hline
$\Delta T_{\rm eff}$                          & $-30$ $\pm$ 64     & $-37$ $\pm$ 61     & $-94$ $\pm$ 57     & $-113$ $\pm$ 88    \\
$\Delta$log~$g$                               & $+0.06$ $\pm$ 0.21 & $+0.25$ $\pm$ 0.13 & $-0.06$ $\pm$ 0.03 & $+0.00$ $\pm$ 0.06 \\
$\Delta v_t$                                  & $+0.15$ $\pm$ 0.17 & $-0.07$ $\pm$ 0.13 & $-0.10$ $\pm$ 0.21 & $-0.17$ $\pm$ 0.20 \\
$\Delta{\rm log_{\epsilon}(Fe\,\textsc{i})}$  & $-0.02$ $\pm$ 0.06 & $+0.09$ $\pm$ 0.07 & $+0.05$ $\pm$ 0.09 & $+0.07$ $\pm$ 0.11 \\
$\Delta{\rm log_{\epsilon}(Fe\,\textsc{ii})}$ & $+0.03$ $\pm$ 0.06 & ~~\,-              & $-0.01$ $\pm$ 0.06 & $+0.03$ $\pm$ 0.10 \\
$\Delta{\rm log_{\epsilon}(O)}$               & $-0.10$ $\pm$ 0.07 & $-0.03$ $\pm$ 0.12 & ~~\,-              & ~~\,-              \\
$\Delta{\rm log_{\epsilon}(Na)}$              & $+0.19$ $\pm$ 0.06 & $+0.21$ $\pm$ 0.09 & $+0.14$ $\pm$ 0.09 & $+0.06$ $\pm$ 0.11 \\
$\Delta{\rm log_{\epsilon}(Mg)}$              & $+0.10$ $\pm$ 0.06 & $+0.22$ $\pm$ 0.08 & ~~\,-              & ~~\,-              \\
$\Delta{\rm log_{\epsilon}(Al)}$              & $+0.13$ $\pm$ 0.04 & $+0.18$ $\pm$ 0.08 & ~~\,-              & ~~\,-              \\
\hline 
\end{tabular}
\end{table*}

In order to facilitate comparisons both between the AGB and RGB, and each individual study, we present kernel density estimation (KDE) histograms of the O, Na, Mg, Al, and Fe abundances in Figures~\ref{fig:m4_2_lit_fei}, \ref{fig:m4_2_lit_o}, \ref{fig:m4_2_lit_na}, \ref{fig:m4_2_lit_mg}, and \ref{fig:m4_2_lit_al}, respectively, and KDEs of CN band strengths in Figure~\ref{fig:m4_2_lit_cn}. The published abundance uncertainties in each study were adopted, and used for the smoothing bandwidths applied to the KDE histograms. We now discuss each element individually.

\subsubsection*{Iron}

The ${\rm log_{\epsilon}(Fe\,\textsc{i})}$ values as published in \citetalias{marino2008}, \citetalias{marino2017}, \citetalias{wang2017}, and the Fe abundances determined in this study (\S\ref{m4_2_abundance_results}) are presented in Figure~{\ref{fig:m4_2_lit_fei}. In the cases of this study, \citetalias{wang2017}, and \citetalias{ivans1999}, the respective samples of RGB and AGB stars were observed simultaneously and analysed in a consistent manner, and the reported Fe abundances agree very well internally, with average differences between the AGB and RGB no larger than 0.04~dex. 

In contrast, \citetalias{marino2017} did not observe an RGB sample at the same time as their AGB sample was observed, nor did they re-analyse the results of \citetalias{marino2008} (in which a sample of 105 RGB stars was observed and analysed spectroscopically). Instead, they compared their AGB results directly with their RGB abundances from \citetalias{marino2008}. A significant difference in [Fe/H] of 0.14 can be seen between their AGB and RGB samples, larger than the total [Fe/H] uncertainty quoted in either publication. This difference can cause significant problems if elements are scaled by Fe abundance as it implicitly assumes that all other elemental abundances are offset by the same amount. As discussed earlier, we have chosen not to scale abundances with Fe in this study. The reason that the Fe abundances do not agree for these samples is likely to be changes in the adopted spectroscopic method (that, for example, produce systematic offsets in T$_{\rm eff}$ or log~$g$), however we cannot determine the true cause with the available data. 

The Fe abundances from neutral lines are very consistent between these studies, except for the disagreement between \citetalias{marino2008} and \citetalias{marino2017}. The average abundance for all five studies is [Fe/H] $= -1.14 \pm 0.07$ (assuming a solar Fe abundance of 7.50).

\begin{figure*}
\centering
\includegraphics[width=0.9\linewidth]{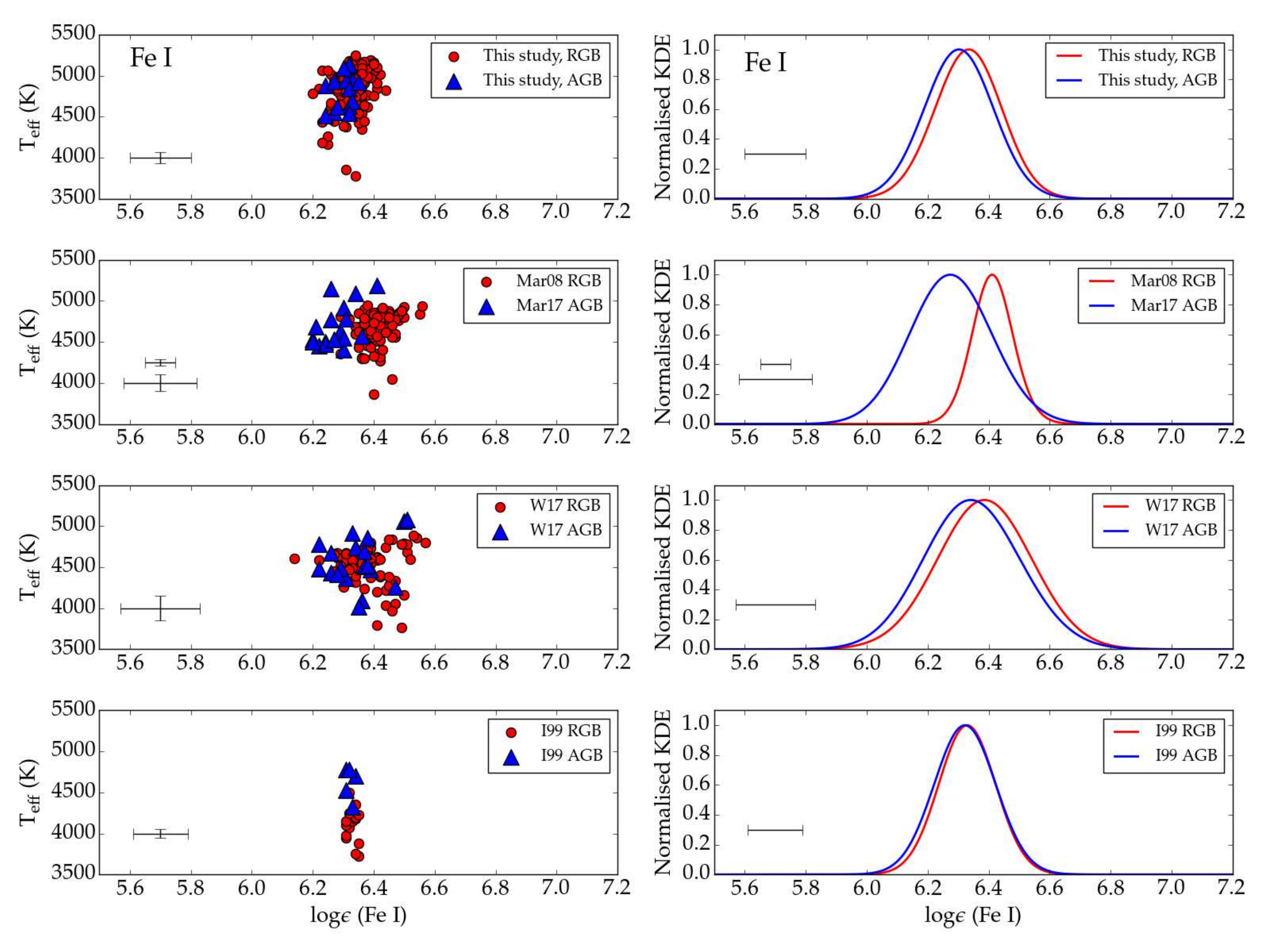} 
\caption{Abundances determined from Fe\,{\sc i} absorption lines from this study, \citetalias{marino2017}, \citetalias{marino2008}, \citetalias{wang2017}, and \citetalias{ivans1999} are presented in the left-hand panels, with kernel density estimations (KDEs) of these data presented in the right-hand panels. Typical abundance errors are shown, as published in the relevant studies (in the \citetalias{marino2017}/\citetalias{marino2008} panel the top error bars are those of the RGB sample in \citetalias{marino2008}), and were used as the bandwidths of the KDEs in the right-hand panels.}
\label{fig:m4_2_lit_fei}
\end{figure*}

\subsubsection*{Oxygen}

The O abundances of this study, \citetalias{marino2017}, \citetalias{marino2008}, and \citetalias{ivans1999} are presented in Figure~\ref{fig:m4_2_lit_o}. Both our re-analysed AGB sample and that of \citetalias{ivans1999} show, on average, slightly higher O abundances than the respective RGB samples ($\Delta{\rm log_{\epsilon}(O)} = 0.08$ for both studies), while the AGB abundances of \citetalias{marino2017} are slightly \textit{lower} than the RGB values from \citetalias{marino2008} ($\Delta{\rm log_{\epsilon}(O)} = -0.08$). The moderate systematic offsets between studies (up to 0.14~dex) can be largely accounted for by line-list differences (in this study we used the 777nm triplet, while \citetalias{marino2008}, \citetalias{marino2017} and \citetalias{ivans1999} used the 630nm forbidden line), however these offsets are still smaller than our uncertainty in ${\rm log_{\epsilon}(O)}$. 

In our work, the difference between the branches ($\Delta{\rm log_{\epsilon}(O)} = 0.08$) is smaller than the total uncertainty in our O abundances ($\pm 0.15$, see Table~\ref{tab:m4_2_uncerts}), and the scatter in our RGB O abundances ($\pm 0.12$). We therefore do not make any conclusions about the AGB of M\,4 from these data. Similarly for the results of \citetalias{marino2017} and \citetalias{ivans1999}, the differences between the O abundances of the giant branches are of the order of the uncertainties ($\pm 0.12$ and $\pm 0.08$, respectively), and are therefore too small to claim any significant variation.

These O abundances shed little light on the nature of AGB stars in M\,4 due to the large uncertainties and relatively small spread in values. Most notable are the O abundances of our work and that of \citetalias{marino2017}, whose scatter in ${\rm log_{\epsilon}(O)}$ ($\pm 0.12$ for \citetalias{marino2017}) is of the order of the total reported uncertainty. Furthermore, we detect no bimodality in O abundance, and it is possible that the bimodality seen in the RGB abundances of \citetalias{marino2008} is an artefact of the very small uncertainty of $\pm 0.04$, which is less than half the magnitude of the O uncertainty in \citetalias{marino2017}, which utilised the same method and absorption lines. This casts doubt on the confidence with which a Na-O anti-correlation can be claimed, and it cannot be confirmed that a heterogeneity in O abundance exists within M\,4 giant stars (\citealt{carretta2009vii} similarly reported a formal homogeneity in O for M\,4).

\begin{figure*}
\centering
\includegraphics[width=0.9\linewidth]{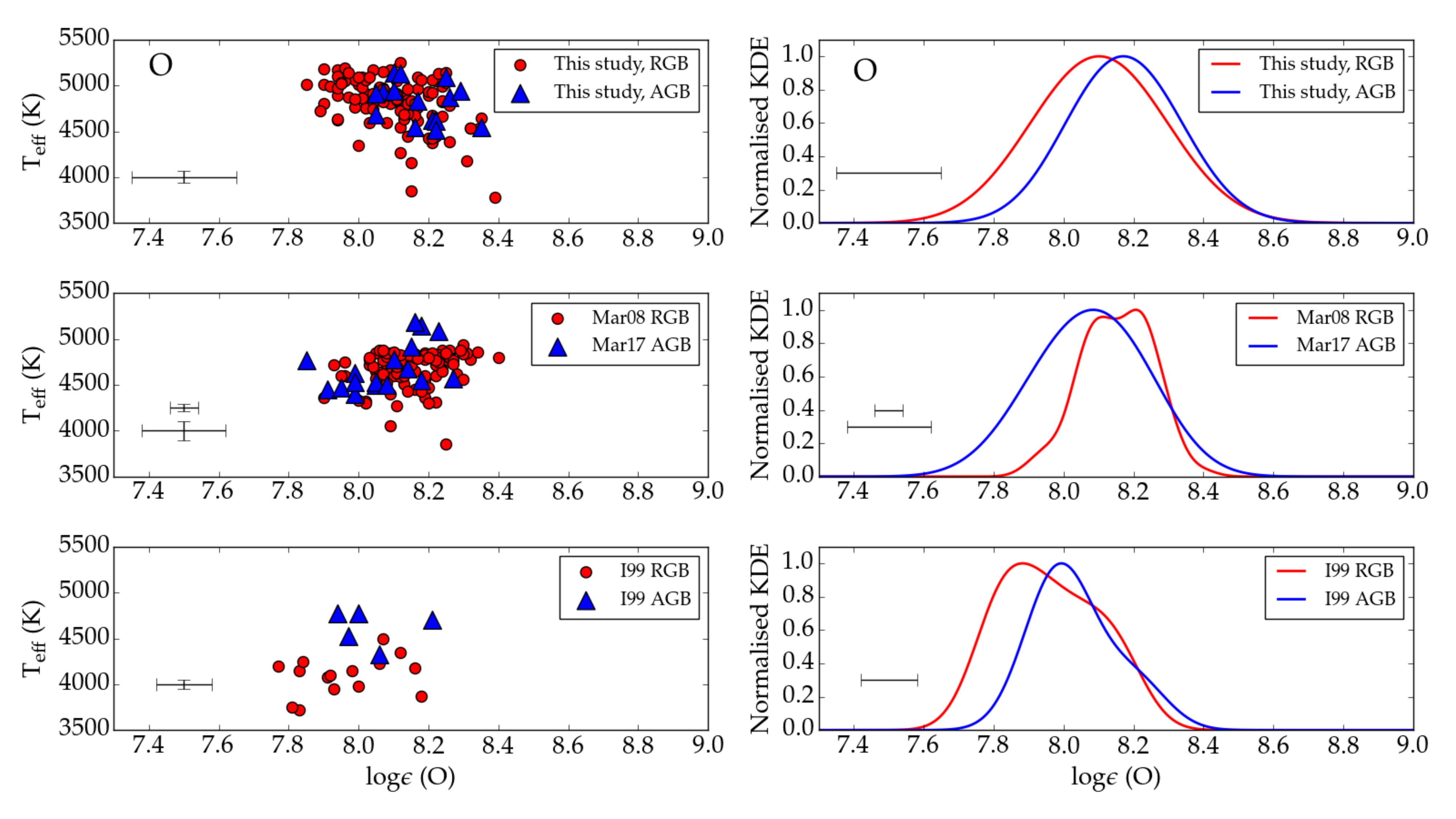} 
\caption{Same as Figure~\ref{fig:m4_2_lit_fei}, but for the abundances determined from O\,{\sc i} absorption lines from this study, \citetalias{marino2017}, \citetalias{marino2008}, and \citetalias{ivans1999}.}
\label{fig:m4_2_lit_o}
\end{figure*}

\subsubsection*{Sodium}

The Na abundances reported by \citetalias{marino2017}, \citetalias{marino2008}, \citetalias{wang2017}, \citetalias{ivans1999}, and this study are presented in Figure~\ref{fig:m4_2_lit_na}. A significant spread larger than the uncertainties exists within all abundance samples, with many showing strong evidence of bimodality.

In all AGB studies of M\,4, there is an apparent absence on the AGB of the most Na-rich stars, when compared to the corresponding sample of RGB stars. The various data sets are surprisingly similar, with only one AGB star having ${\rm log_{\epsilon}(Na)} > 5.5$ (in the sample of \citetalias{marino2017}); while in all RGB samples, the largest density of ${\rm log_{\epsilon}(Na)}$ values is between 5.5 and 5.7. The RGB and AGB of \citetalias{wang2017} overlap to a larger extent than those of the other studies, but the lack of the most Na-rich stars on the AGB is clear (as noted by \citetalias{wang2017}). The differences between the giant branches in this work, and that of \citetalias{marino2008}/\citetalias{marino2017}, \citetalias{wang2017}, and \citetalias{ivans1999} are $\Delta{\rm log_{\epsilon}(Na)} = -0.22$, $-0.21$, $-0.14$, and $-0.20$, respectively (these values are all larger than the respective uncertainties in Na abundance, except for that of \citetalias{wang2017}).

It is important to note that in all cases there is also evidence of heterogeneity in the Na abundances of M\,4's AGB population (in this study we found a spread of $\sigma = 0.12$~dex, compared to a total Na uncertainty of $\pm 0.07$~dex). This may indicate that stars that have some Na enrichment (i.e. SP2 stars) are indeed present on the AGB, but that there is a limiting factor that is preventing stars with the highest Na abundances from either evolving to the AGB, or appearing as Na-rich on the AGB as they would have on the RGB. We also note (especially among our abundances, and those of \citetalias{marino2017}) that some AGB stars in M\,4 appear to have lower Na abundances than the most Na-poor RGB stars of the cluster. This suggests that there may be a systematic offset in Na abundance between the two giant branches. We explore this possibility in Section \ref{m4_2_atmos_tests}. 

Finally, we note that the Na abundance uncertainty of \citetalias{wang2017} ($\pm 0.16$~dex) appears to be overestimated, most likely due to the selection of stellar parameters which resulted in an uncertainty in T$_{\rm eff}$ of $\pm 150$~K. The uncertainty in Na abundance in the study of \citetalias{ivans1999} ($\pm 0.04$~dex) appears to be underestimated -- the structure seen in the \citetalias{ivans1999} KDE is unlikely to be real, but is more likely an artefact of both small uncertainties and a small sample size -- however we chose to adopt the published uncertainties.

\begin{figure*}
\centering
\includegraphics[width=0.9\linewidth]{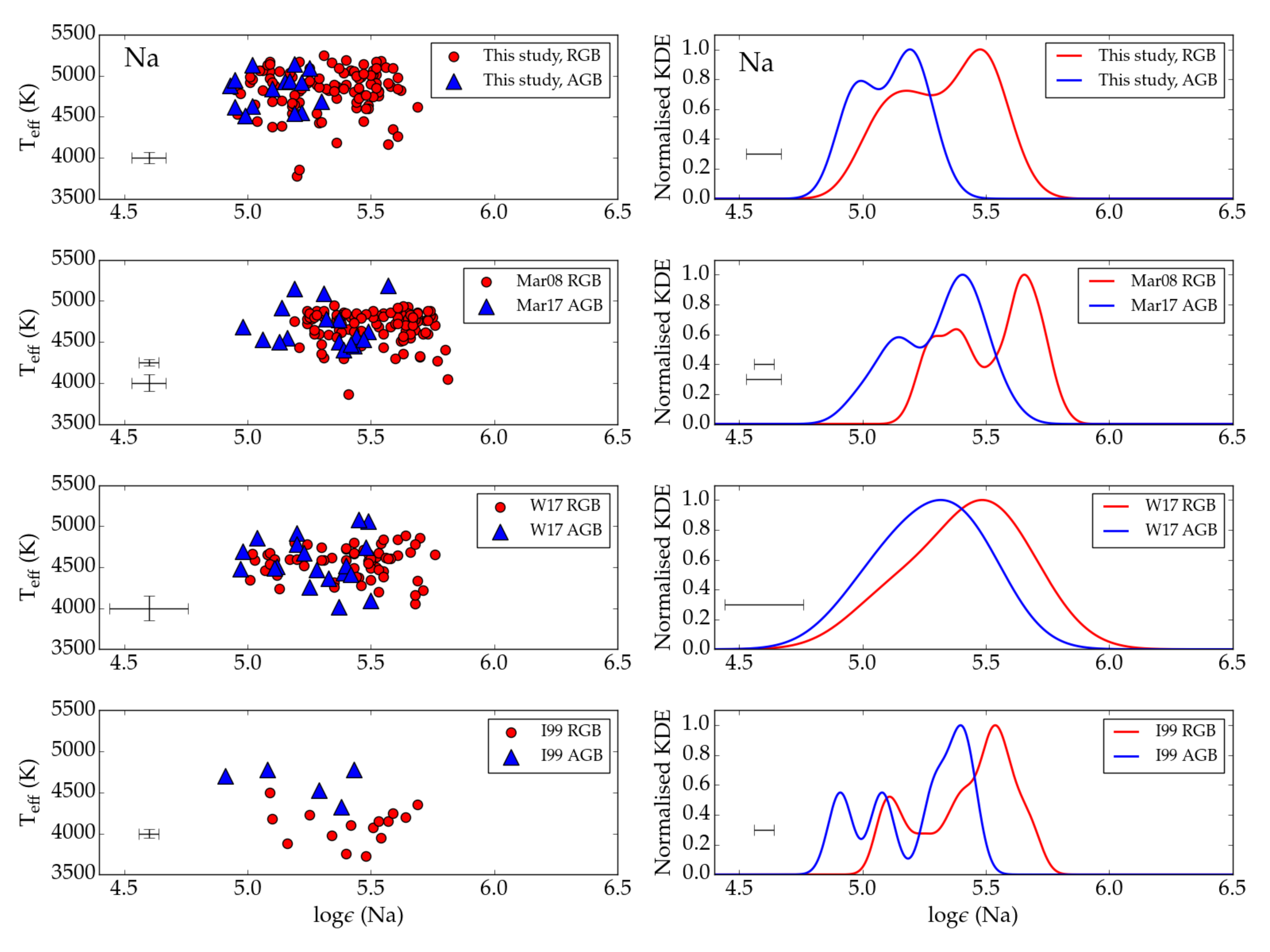} 
\caption{Same as Figure~\ref{fig:m4_2_lit_fei}, but for the abundances determined from Na\,{\sc i} absorption lines from this study, \citetalias{marino2017}, \citetalias{marino2008}, \citetalias{wang2017}, and \citetalias{ivans1999}.}
\label{fig:m4_2_lit_na}
\end{figure*}

\subsubsection*{Magnesium}

Mg abundances from our work, \citetalias{marino2017}, \citetalias{marino2008}, and \citetalias{ivans1999} are presented in Figure~\ref{fig:m4_2_lit_mg}. Previous studies have concluded that M\,4 is homogeneous in Mg, and we find this for all samples included here. 

Due to the homogeneity of Mg, we do not expect any significant difference between the ${\rm log_{\epsilon}(Mg)}$ values of AGB and RGB stars in the cluster. While this is the case with the results of this study and those of \citetalias{ivans1999} ($\Delta{\rm log_{\epsilon}(Mg)} = -0.02$ and $0.00$, respectively), the abundances of \citetalias{marino2008} and \citetalias{marino2017} indicate that AGB stars in M\,4 present as significantly more Mg-poor than the RGB ($\Delta{\rm log_{\epsilon}(Mg)} = -0.14$). We consider this to be unlikely, and it may be related to the discrepancy in Fe abundance between the two studies.

\begin{figure*}
\centering
\includegraphics[width=0.9\linewidth]{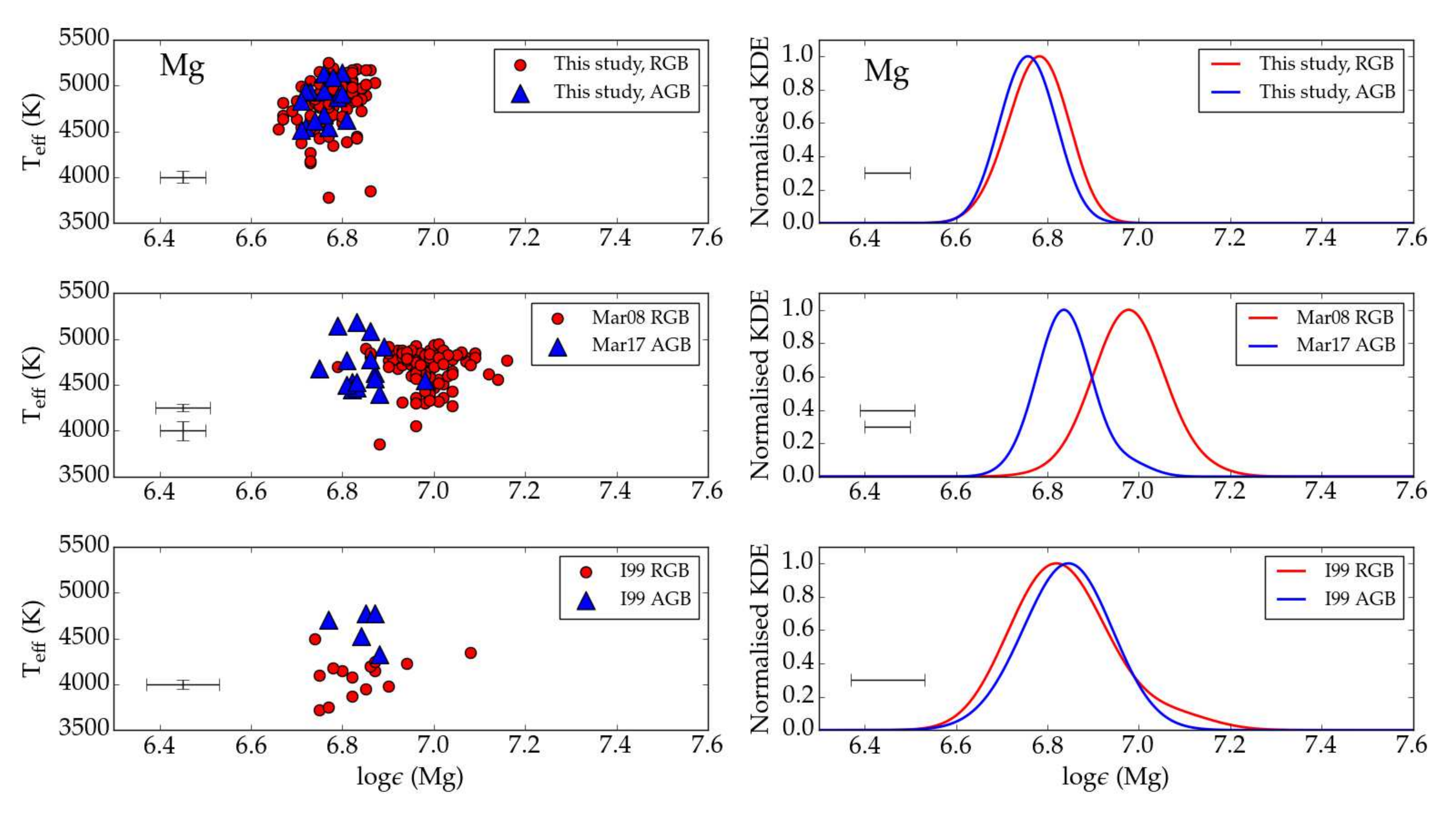} 
\caption{Same as Figure~\ref{fig:m4_2_lit_fei}, but for the abundances determined from Mg\,{\sc i} absorption lines from this study, \citetalias{marino2017}, \citetalias{marino2008}, and \citetalias{ivans1999}.}
\label{fig:m4_2_lit_mg}
\end{figure*}

\subsubsection*{Aluminium}

Figure~\ref{fig:m4_2_lit_al} presents the Al abundances of this study, \citetalias{marino2017}, \citetalias{marino2008}, and \citetalias{ivans1999}. The spread in RGB ${\rm log_{\epsilon}(Al)}$ values, while significant in each sample at the 1$\sigma$ level ($\pm 0.09$, $\pm 0.12$, and $\pm 0.12$ for our work, \citetalias{marino2008}, and \citetalias{ivans1999}, respectively) is quite small and there is no evidence of bimodality. The spread in AGB Al abundances, however, is even smaller than for each of the respective RGB samples, and shows potentially homogeneous abundances (except for the single Al-rich outlier in this study and \citetalias{marino2017}; 2MASS ID 16234085-2631215).

The similarity between the Al abundances of this study, \citetalias{marino2017}, and \citetalias{ivans1999} is noteworthy, with the AGB samples in all cases being significantly offset to lower values ($\Delta{\rm log_{\epsilon}(Al)} = -0.14$, $-0.18$, and $-0.18$, respectively), indicating that M\,4 stars on the AGB are more Al-poor, on average, than those on the RGB. 

While the Al abundance uncertainty reported in \citetalias{ivans1999} ($\pm 0.03$~dex) appears to be underestimated (as with Na), we adopt this value for our comparison while noting that the structure in the bottom right panel of Figure~\ref{fig:m4_2_lit_al} is likely an artefact of this underestimation.

\begin{figure*}
\centering
\includegraphics[width=0.9\linewidth]{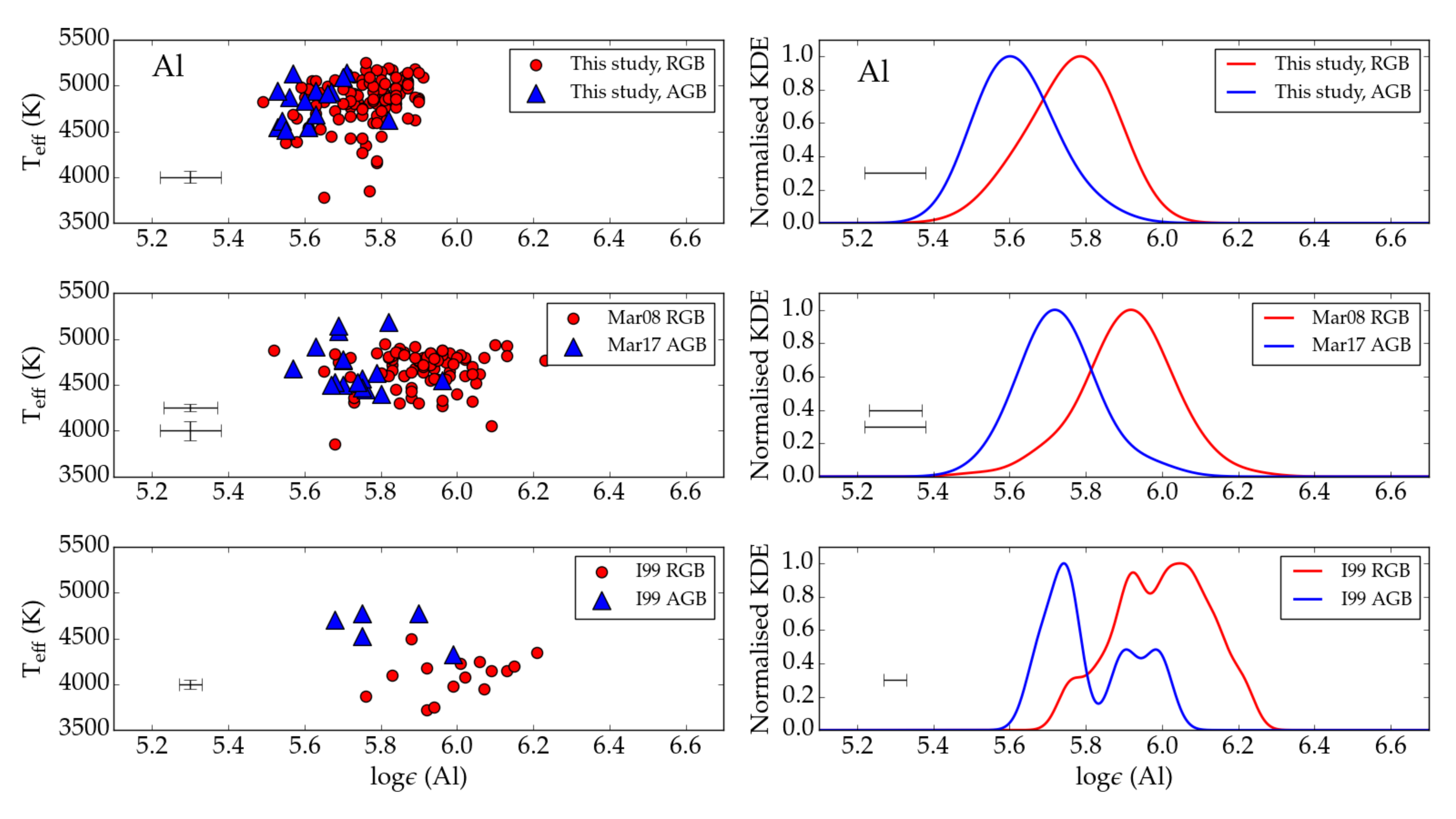} 
\caption{Same as Figure~\ref{fig:m4_2_lit_fei}, but for the abundances determined from Al\,{\sc i} absorption lines from this study, \citetalias{marino2017}, \citetalias{marino2008}, and \citetalias{ivans1999}.}
\label{fig:m4_2_lit_al}
\end{figure*}

\subsubsection*{Cyanogen}

In Figure~\ref{fig:m4_2_lit_cn}, the compiled CN band strengths of \citetalias{sb2005} and the results from this study (from \S\ref{m4_2_cn}) are presented. A clear bimodality in $\delta$S3839 values is visible in the RGB samples of both studies (albeit with a larger spread of $\pm 0.25$ in the results from this study, compared to $\pm 0.19$ in \citetalias{sb2005}), which has been noted in previous CN studies of M\,4 \citep{norris1981m4,suntzeff1991}.

Both studies strongly suggest an extreme paucity of CN-strong AGB stars in the cluster: ${\Delta}{\delta}$S3839 $= -0.20$ and $-0.14$ for this study and \citetalias{sb2005}, respectively. In both AGB samples, however, there is a significant spread in $\delta$S3839 values ($\pm 0.12$ and $0.11$, respectively), with an apparent bimodality in the AGB sample of \citetalias{sb2005} (although there are only 6 stars in this sample). This striking similarity between the independently observed and analysed CN results provides significant weight to our Na and Al abundance results, along with the strong correlation between $\delta$S3839 and ${\rm log_{\epsilon}(Na)}$ values (see Figure~\ref{fig:m4_2_nacn}).

\begin{figure*}
\centering
\includegraphics[width=0.9\linewidth]{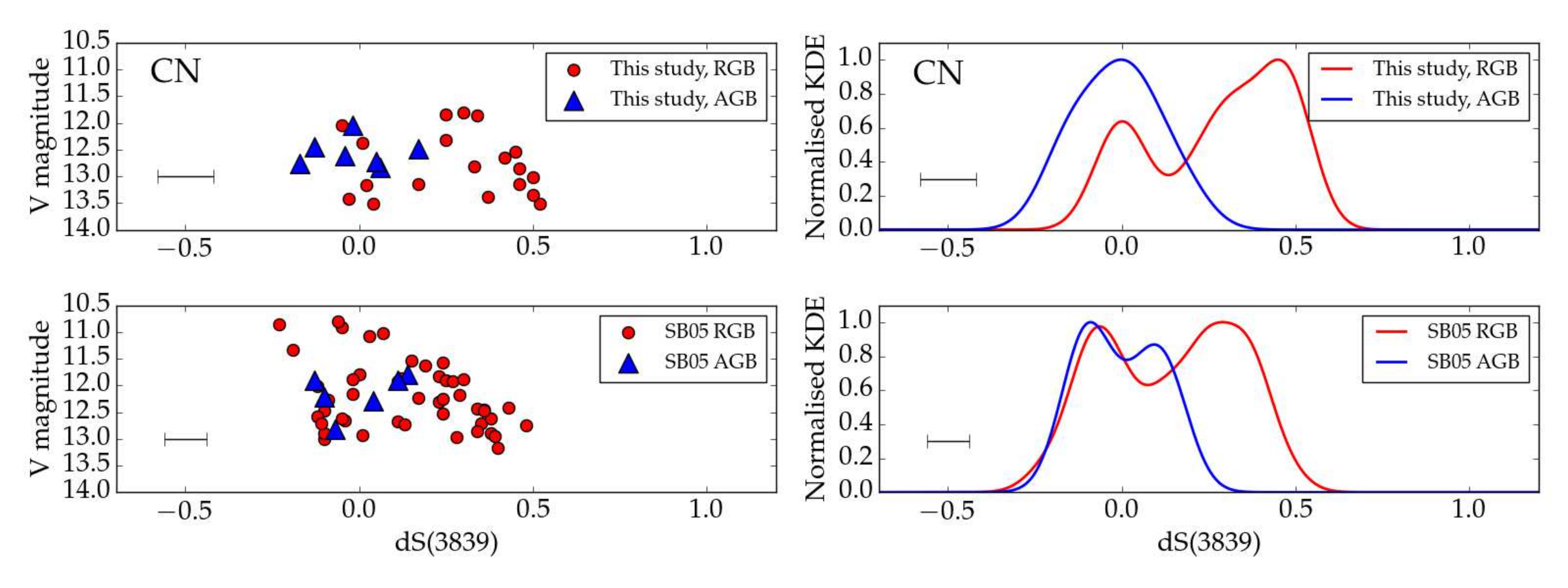} 
\caption{Same as Figure~\ref{fig:m4_2_lit_fei}, but for the CN band strengths ($\delta$S3839 values) from this study and \citetalias{sb2005}.}
\label{fig:m4_2_lit_cn}
\end{figure*}

\subsubsection*{Comparison Summary}

In summary, we have identified four main conclusions from the literature comparison:
\renewcommand\labelenumi{(\roman{enumi})}
\renewcommand\theenumi\labelenumi
\begin{enumerate} 
\item there is no systematic offset between the Mg and Fe abundances of AGB and RGB stars in M\,4,
\item the AGB of M\,4 is systematically offset to lower values in Na and Al abundances, and CN band strength compared to the RGB,
\item no conclusions can be drawn concerning differences in the O abundances of AGB or RGB stars in M\,4, and 
\item due to (iii) there may be no Na-O anti-correlation in M\,4.
\end{enumerate}

Three of the most common diagnostic tools of multiple populations in M\,4 -- Na abundances, Al abundances, and CN band strengths -- consistently indicate a significant difference between the light-elemental distributions of AGB and RGB stars in this globular cluster, with an apparent deficit of AGB stars enhanced in H-burning products. The only exception to this are the O abundances, from which no conclusion can be consistently drawn. Indeed, we detect little evidence of a spread in O abundance for M\,4. Thus, taken at face value, most of the results presented in this section show that, in general, the AGB stars in M\,4 contain less H-burning products than RGB stars in the cluster. It is possible that the stars currently on the AGB have experienced less of the `self-pollution' that M\,4 (and other Galactic GCs) is thought to have experienced early in its life \citep{dorazi2010}.

We can see only two possible explanations for the results presented here:
\renewcommand\labelenumi{(\roman{enumi})}
\renewcommand\theenumi\labelenumi
\begin{enumerate}
\item The most Na-enhanced -- and by correlation, He-enriched \citep{dantona2002,chantereau2016} -- stars in M\,4 are not evolving to the AGB, but are becoming AGB-manqu\'{e} stars, evolving directly from the HB to the WD phase.
\item Systematic errors are affecting both the high-resolution spectroscopic method of abundance determination \textit{and} the calculation of S3839 index values of AGB stars across several studies, consistently resulting in AGB samples appearing more Na-poor, Al-poor, and CN-weak than they are in reality.  
\end{enumerate}

We investigate i) in \S\ref{m4_2_monstar} with 1D stellar evolution models, and ii) in \S\ref{m4_2_atmos_tests} by conducting tests on the impact of using a range of different atmospheric models for the determination of elemental abundances.

%%%%%%%%%%%%%%%%%%%%%%%%%%%%%%%%%%%%%%%%%%%%%%%%%%%%%%%%%%%%%%

\section{Expectations from theoretical stellar evolution models}
\label{m4_2_monstar}

In the stellar evolutionary models of \citet{dorman1993}, at the approximate metallicity of M\,4 ([Fe/H] $\sim -1.15$), stars with zero-age HB (ZAHB) effective temperatures of $15,000 \lesssim$ T$_{\rm eff} \lesssim 19,000$~K have short early-AGB lives and evolve to the white dwarf cooling phase without fully ascending the AGB. These stars may not be detectable on the AGB due to the short time-scale of this phase of evolution. Stars with T$_{\rm eff} \gtrsim 19,000$~K at the ZAHB become AGB-manqu\'{e} stars and never join the AGB. If applied to M\,4, this implies that \textit{all} stars in M\,4 should evolve to and ascend the AGB. This is because the hottest HB stars in the cluster have T$_{\rm eff} \sim 9500$~K \citep{villanova2012}.

The spectroscopic abundances of M\,4's AGB population, as presented in this study, appear to suggest that the most Na-rich stars \citep[these stars populate the blue-HB due to the correlation between He and Na abundance;][]{marino2011,chantereau2016} either do not evolve to the early-AGB, or spend a very short amount of time in this phase\footnote{Such that no such Na-rich stars are in the AGB phase at the present time.}.

To establish a precise, quantitative theoretical expectation of M\,4's AGB abundances we have calculated a range of theoretical stellar model tracks for M\,4 stars. We have done this in order to determine the likelihood of the blue HB stars in the cluster avoiding the AGB, thereby intrinsically creating the abundance distributions observed in this study -- where the most Na-rich stars are present on the HB, but missing on the AGB. The stellar models were calculated using the Monash University stellar structure code {\sc MONSTAR} \citep{lattanzio86,campbell08} with \citet{spruit2015} overshooting in the core helium-burning phase, as described in \citet{constantino2017}. The code has been updated with low temperature opacity tables which follow variations in C, N and O \citep{marigo09,constantino14}. The \citet{reimers75} mass loss prescription was used for the RGB.

Our aim was to determine the optimal parameters for M\,4 stars that allowed us to most accurately match the observed bimodal HB, and to identify whether these stars evolve to the AGB. We then sought to determine the approximate HB T$_{\rm eff}$ required for M\,4 stars to avoid the AGB phase. At a given age and metallicity, the HB T$_{\rm eff}$ of a star is a function of both initial mass\footnote{Since the core mass at the onset of helium burning is relatively fixed at $\sim 0.475 M_{\odot}$ due to the degenerate equation of state \citep{sweigart1978}, the amount of leftover envelope after the core helium flash directly influences the HB T$_{\rm eff}$.} and helium mass fraction -- a higher Y value decreases the time on the main sequence, so for a coeval cluster with a helium abundance variation, a star enhanced in He will have a lower initial mass, and therefore have a higher HB T$_{\rm eff}$.

We began by identifying the most important observational and theoretical constraints that affect HB morphology, and created a range of parameters over which to test. We tested three parameters: helium enrichment ($\Delta$Y), cluster age, and RGB mass loss rate. Cluster metallicity also has an effect on HB morphology, however, this value is well constrained for M\,4 -- therefore we assumed [Fe/H]$= -1.15$ for all evolutionary models. Published estimates of these constraints from the literature, and the values adopted for our evolutionary models, are summarised in Table~\ref{tab:m4_2_model_constraints}. 

For the helium mass-fraction of SP1 stars in M\,4 we adopted Y = 0.245 \citep{valcarce2014}, and for SP2 stars we adopted Y = 0.275 (so $\Delta$Y = 0.03, see Table~\ref{tab:m4_2_model_constraints}). For C, N and O abundances, we adopted the values reported by \citet{villanova2012} for the N-poor (SP1) and N-rich (SP2) populations\footnote{SP1: [C/Fe] = $-0.20$, [N/Fe] = $+0.16$, [O/Fe] = $+0.42$.\\ SP2: [C/Fe] = $-0.36$, [N/Fe] = $+0.80$, [O/Fe] = $+0.25$.}. We calculated models over a range of ages (determined primarily by initial mass and Y, for which the dependence was controlled) and RGB mass loss rates. We compared the maximum T$_{\rm eff}$ reached on the HB -- our primary observational constraint -- with observed values reported in the literature, as determined by \citet[maximum red-HB T$_{\rm eff} = 6250$~K]{marino2011} and \citet[maximum blue-HB T$_{\rm eff} = 9500$~K]{villanova2012}. A summary of our model tracks is presented in Table~\ref{tab:m4_2_monstar}.

\begin{table}
\centering
\caption{A summary of M\,4 observational constraints for helium enrichment ($\Delta$Y), age, and RGB mass loss parameter (Reimers $\eta$). The values adopted for use in our theoretical models are listed in the last row.}
\label{tab:m4_2_model_constraints}
\begin{tabular}{cccc}
\hline
Reference & $\Delta$Y & Age & Reimers \\
 & & (Gyr) & $\eta$  \\
\hline \hline 
H02$^1$ & - & 12.70 $\pm$ 0.70 & - \\
MF09$^2$ & - & 12.65 $\pm$ 0.64 & - \\ 
V12$^3$ & 0.04 & - & - \\
Val14$^4$ & $\lesssim 0.01$ & - & -  \\
MZ15$^5$ & - & 11.81 $\pm$ 0.66 & 0.40 $\pm$ 0.08   \\
N15$^6$ & 0.02 & - & - \\
\hline
Adopted & 0.03 & 12.45 $\pm$ 0.7 & 0.40 $\pm$ 0.08 \\
\hline
\multicolumn{4}{l}{\footnotesize{$^1$\protect\shortcite{hansen2002}; $^2$\protect\shortcite{marinfranch2009}}} \\
\multicolumn{4}{l}{\footnotesize{$^3$\protect\shortcite{villanova2012}; $^4$\protect\shortcite{valcarce2014}}} \\
\multicolumn{4}{l}{\footnotesize{$^5$\protect\shortcite{mcdonald2015}; $^6$\protect\shortcite{nardiello2015}}} \\
\end{tabular}
\end{table}

\begin{table}
\centering
\caption{A summary of theoretical stellar models calculated for M\,4. The last column indicates the highest T$_{\rm eff}$ that was reached in the HB phase of each model track, our primary observational constraint. The first ten models listed are representative of SP1 stars, with a T$_{\rm eff}$ constraint on the red-HB from \protect{\citet{marino2011}}. The next ten models are representative of SP2 stars, with a T$_{\rm eff}$ constraint on the blue-HB from \protect{\citet{villanova2012}}. The final nine models are tests using extreme values of RGB mass loss, age, and helium enrichment, to explore AGB-manqu\'{e} evolution. In Figure~\ref{fig:m4_2_monstar} we show tracks of the three models in bold text, which we found to best match the red-HB (Y = 0.245), the blue-HB (Y = 0.275), and also the lowest HB T$_{\rm eff}$ required to produce an AGB-manqu\'{e} star (Y = 0.325).}
\label{tab:m4_2_monstar}
\begin{tabular}{ccccc}
\hline
Y & Age (RGB-tip) & Initial & Reimers & HB Max \\
 & (Gyr) & Mass (M$_{\odot}$) & $\eta$ & T$_{\rm eff}$ (K) \\
\hline \hline
\multicolumn{5}{c}{\textit{SP1 (Observed HB Max T$_{\rm eff} = 6250$~K)}} \\
0.245 & 11.83         & 0.839     & 0.32 & 5375          \\
0.245 & 11.83         & 0.839     & 0.40 & 5540          \\
0.245 & 11.83         & 0.839     & 0.47 & 6030          \\
0.245 & 12.45         & 0.827     & 0.32 & 5425          \\
0.245 & 12.45         & 0.827     & 0.40 & 5675          \\
\textbf{0.245} & \textbf{12.45}         & \textbf{0.827}     & \textbf{0.44} & \textbf{6120} \\
0.245 & 12.45         & 0.827     & 0.47 & 6960          \\
0.245 & 13.24         & 0.813     & 0.32 & 5510          \\
0.245 & 13.24         & 0.813     & 0.40 & 6050          \\
0.245 & 13.24         & 0.813     & 0.47 & 8250          \\
\hline
\multicolumn{5}{c}{\textit{SP2 (Observed HB Max T$_{\rm eff} = 9500$~K)}} \\
0.275 & 11.79         & 0.796     & 0.32 & 5720          \\
0.275 & 11.79         & 0.796     & 0.40 & 7250          \\
0.275 & 11.79         & 0.796     & 0.47 & 9370          \\
0.275 & 12.40         & 0.785     & 0.32 & 6025          \\
0.275 & 12.40         & 0.785     & 0.40 & 8150          \\
\textbf{0.275} & \textbf{12.40}         & \textbf{0.785}     & \textbf{0.44} & \textbf{9400} \\
0.275 & 12.40         & 0.785     & 0.47 & 10390         \\
0.275 & 13.22         & 0.771     & 0.32 & 6950          \\
0.275 & 13.22         & 0.771     & 0.40 & 9380          \\
0.275 & 13.22         & 0.771     & 0.47 & 11870         \\
\hline
\multicolumn{5}{c}{\textit{Tests of extreme $\eta$ values}} \\
0.275 & 12.40         & 0.785     & 0.55 & 13530         \\
0.275 & 12.40         & 0.785     & 0.58 & 15200         \\
0.275 & 12.40         & 0.785     & 0.60 & 17000         \\
%\hline
\multicolumn{5}{c}{\textit{Tests of extreme ages}} \\
0.275 & 14.59         & 0.750     & 0.44 & 13070         \\
0.275 & 15.70         & 0.735     & 0.44 & 16000         \\
%\hline
\multicolumn{5}{c}{\textit{Tests of extreme Y values}} \\
0.295 & 12.40         & 0.757     & 0.44 & 11840         \\
0.315 & 12.40         & 0.729     & 0.44 & 14500         \\
\textbf{0.325} & \textbf{12.40}         & \textbf{0.715}     & \textbf{0.44} & \textbf{15500} \\
0.350 & 12.40         & 0.680     & 0.44 & 19200        \\
\hline
\end{tabular}
\end{table}

We found that in order to match the HB morphology of M\,4, based on spectroscopic HB T$_{\rm eff}$ values and helium mass-fractions in the literature, we required a Reimers mass loss rate of $\eta = 0.44 \pm 0.04$ and initial masses of $0.827 \pm 0.013$ and $0.785 \pm 0.013$ M$_{\odot}$ for SP1 and SP2, respectively; which gave a cluster age of $12.4 \pm 0.6$~Gyr. Uncertainties given here are the ranges in each value for which the HB morphology was able to be reproduced.

In Figure~\ref{fig:m4_2_monstar} we present model tracks with the mean mass loss rates and initial masses required to match the HB of M\,4 (according to the maximum T$_{\rm eff}$ reached on the HB), which are indicated in bold text in Table~\ref{tab:m4_2_monstar}. Included for reference are the stellar parameters (reported T$_{\rm eff}$ and photometric log~$g$) of HB stars determined by \citet{marino2011} and \citet{villanova2012}, and AGB stars determined with {\sc phobos v2} in this study. As an example of an AGB-manqu\'{e} star, we also included a stellar model with a very large helium enhancement (Y = 0.32 and $\Delta$Y = 0.08, see Table~\ref{tab:m4_2_monstar}), for which we adopted the mean age and mass loss rate that we determined for M\,4 (12.4~Gyr, $\eta = 0.44$).

\begin{figure*}
\centering
\includegraphics[width=0.9\linewidth]{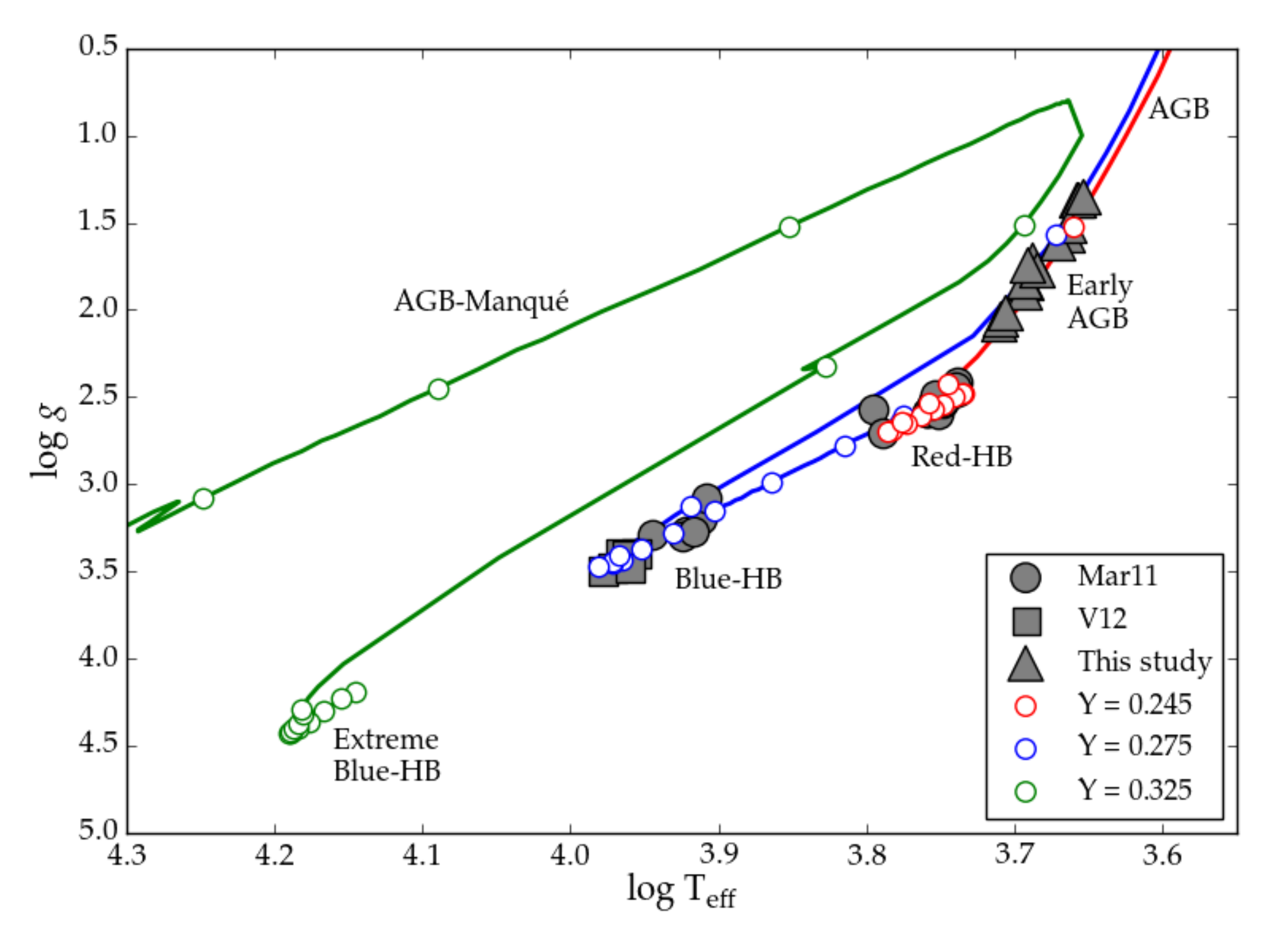} 
\caption{Evolutionary tracks of the three models found to best match the red-HB (red track, Y = 0.245), the blue-HB (blue track, Y = 0.275), and the lowest HB T$_{\rm eff}$ required to produce an AGB-manqu\'{e} star (green track, Y = 0.325) -- see Table~\ref{tab:m4_2_monstar} and text for model details. While each model was evolved from the beginning of the main sequence, we show the evolution of each model from the ZAHB. Points along the evolutionary tracks are separated in age by 10 Myr to give an indication of time spent in each phase, and hence the likelihood of observing stars in each phase. Also included are the T$_{\rm eff}$ and log~$g$ values for our AGB sample (from \S\ref{m4_2_reanalysis}), and the T$_{\rm eff}$ values of HB stars from \protect{\citet[Mar11]{marino2011}} and \protect{\citet[V12]{villanova2012}} for which we redetermined log~$g$ photometrically \protect{\citep[using the empirical relation from][so that all observations are on the same log~$g$ scale]{alonso1999}}. Also note that the blue-HB model begins on the red-HB before quickly moving to canonical blue-HB temperatures, possibly indicating that some red-HB stars may in fact be SP2 stars that are still in the early HB phase. While \protect{\citet{marino2011}} did not report on any Na-rich stars on the red-HB, they did find a larger spread of Na abundances among red-HB stars than blue-HB stars.}
\label{fig:m4_2_monstar}
\end{figure*}

All stellar models whose maximum T$_{\rm eff}$ on the HB closely matched the values in the literature (6250~K for the red-HB, and 9500~K for the blue-HB) evolved to the AGB. In fact, all models with a maximum HB T$_{\rm eff} \lesssim 15,500$~K spend enough time on the early-AGB to potentially be observed. This provides a very strong prediction that every star in M\,4 should evolve to (at least) the early-AGB, and that the light elemental abundance distribution of the AGB should match that of the HB and RGB. Furthermore, we find that only HB stars with a maximum T$_{\rm eff} \gtrsim 15,500$~K are likely to avoid the AGB, or have short enough AGB-lifetimes to avoid detection -- this agrees well with the HB models of \citet{dorman1993}. We note that there is a difference of 6000~K in T$_{\rm eff}$ between the observed blue end of M\,4's HB and the values required for the evolution of AGB-manqu\'{e} stars. Comparing to the reported uncertainty in T$_{\rm eff}$ of $\pm 50$~K in \citet{marino2011} and \citet{villanova2012}, this is a very large difference. This shows that there is a very clear expectation that \textit{all} stars on the M\,4 blue-HB should become AGB stars. 

In chemical space, this implies that the Na, Al, and CN distributions should be identical on the AGB and RGB. Given the abundance results from multiple spectroscopic studies (see \S\ref{m4_2_lit_comp}), which indicate that these abundance distributions are \textit{not} identical, there is a clear discordance between the observations of M\,4 stars and theoretical expectations. 

In the next section, we investigate various uncertainties and assumptions that may affect the abundances of AGB stars in M\,4, to see whether aspects of the spectroscopic method may be responsible for the contradictory results found thus far.

%%%%%%%%%%%%%%%%%%%%%%%%%%%%%%%%%%%%%%%%%%%%%%%%%%%%%%%%%%%%%%

\section{Atmospheric model tests}
\label{m4_2_atmos_tests}

\subsection{Stellar parameter test}

Determining precise effective temperatures for stars can be difficult -- random and systematic errors are often of the order of $100-200$~K \citep[e.g.][also see Table~\ref{tab:m4_2_teff_diffs}, Figure~\ref{fig:m4_2_bisection_test}, and Table~\ref{tab:m4_2_overlap}]{ramirez2005,wang2017}. While the random errors in our work that are associated with uncertainties in atmospheric parameters are presented in Table~\ref{tab:m4_2_atmos}, we conducted an additional test of stellar parameters, in an effort to investigate the effects of systematic errors in T$_{\rm eff}$ on our sample of M\,4 stars.

We redetermined LTE Na and Fe abundances for our M\,4 stellar sample using three different empirical colour-T$_{\rm eff}$ relations (see \S\ref{m4_2_atmos_method}), chosen to maximise the systematic differences between the estimated effective temperatures. These relations are the $B-V$ relation from \citet{alonso1999}, the $B-V$ relation from \citet{casagrande2010}, and the $V-K$ relation from \citet[note that some stars do not have reliable 2MASS magnitudes and were therefore not included here]{ramirez2005}. The average differences between the T$_{\rm eff}$ values determined from these relations and those adopted for our final T$_{\rm eff, sp}$ results in \S~\ref{m4_2_reanalysis} are $1 \pm 67$~K, $-83 \pm 105$~K and $129 \pm 109$~K, respectively. The star-to-star differences are presented in Figure~\ref{fig:m4_2_1d_teff}, showing individual T$_{\rm eff}$ differences of up to 500K and a total 1$\sigma$ scatter of 127~K for the entire sample. Values of log~$g$ and $v_t$ were determined using the same method as in \S\ref{m4_2_atmos_method}. 

\begin{figure}
\centering
\includegraphics[width=\linewidth]{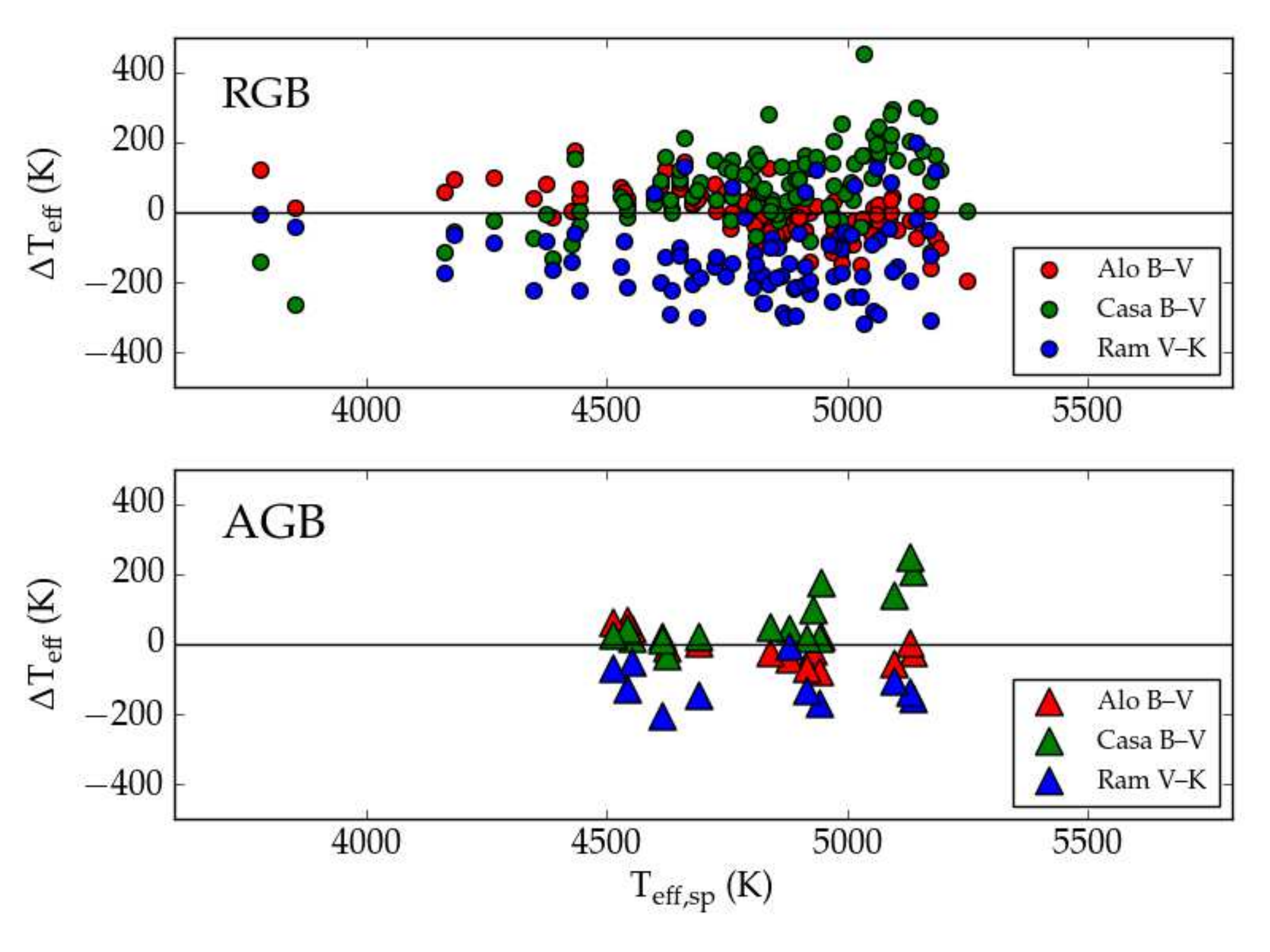} 
\caption{The star-to-star differences in T$_{\rm eff}$ between our T$_{\rm eff, sp}$ values and those of three empirical colour-T$_{\rm eff}$ relations: \protect{\citet[$B-V$]{alonso1999}}, \protect{\citet[$B-V$]{casagrande2010}}, and \protect{\citet[$V-K$]{ramirez2005}}, where $\Delta$T$_{\rm eff}$ = T$_{\rm eff, relation} - $T$_{\rm eff, sp}$. The top panel shows our sample of RGB stars in M\,4, while the bottom panel presents our AGB sample.}
\label{fig:m4_2_1d_teff}
\end{figure}

The LTE Fe and Na abundances determined using the stellar parameters from these three relations (the line-list and method are the same as in \S\ref{m4_2_reanalysis}) are presented in Figures~\ref{fig:m4_2_1d_fe}~and~\ref{fig:m4_2_1d_na}. Systematic differences in T$_{\rm eff}$ have a large effect on the spread and distribution of Fe abundances, with the \citet{casagrande2010} and \citet{ramirez2005} relations producing significant trends between T$_{\rm eff}$ and ${\rm log_{\epsilon}(Fe\,\textsc{i})}$ (also see \citetalias{campbell2017}). Our adopted T$_{\rm eff, sp}$ values (included in the bottom panels for comparison) produce the tightest distribution of Fe abundances ($\sigma = 0.05$).

In contrast to the effect on Fe abundance, large systematic variations in T$_{\rm eff}$ appear to have little impact on the distribution of Na abundances, despite some stars' T$_{\rm eff}$ varying by up to nearly 500~K between the three empirical relations and those adopted in this study. As seen in Figure~\ref{fig:m4_2_1d_na}, the Na-poor nature of our AGB sample is present irrespective of the T$_{\rm eff}$ scale adopted. This demonstrates that conservative systematic changes in stellar atmospheric parameters have virtually no bearing on our results, and that ${\rm log_{\epsilon}(Na)}$ is much more robust to sample-wide T$_{\rm eff}$ variations than ${\rm log_{\epsilon}(Fe\,\textsc{i})}$ (which we also found to be the case for NGC\,6752; see \citetalias{campbell2017}). Next, we investigated the effect of including helium enhancement in atmospheric models.

\begin{figure}
\centering
\includegraphics[width=\linewidth]{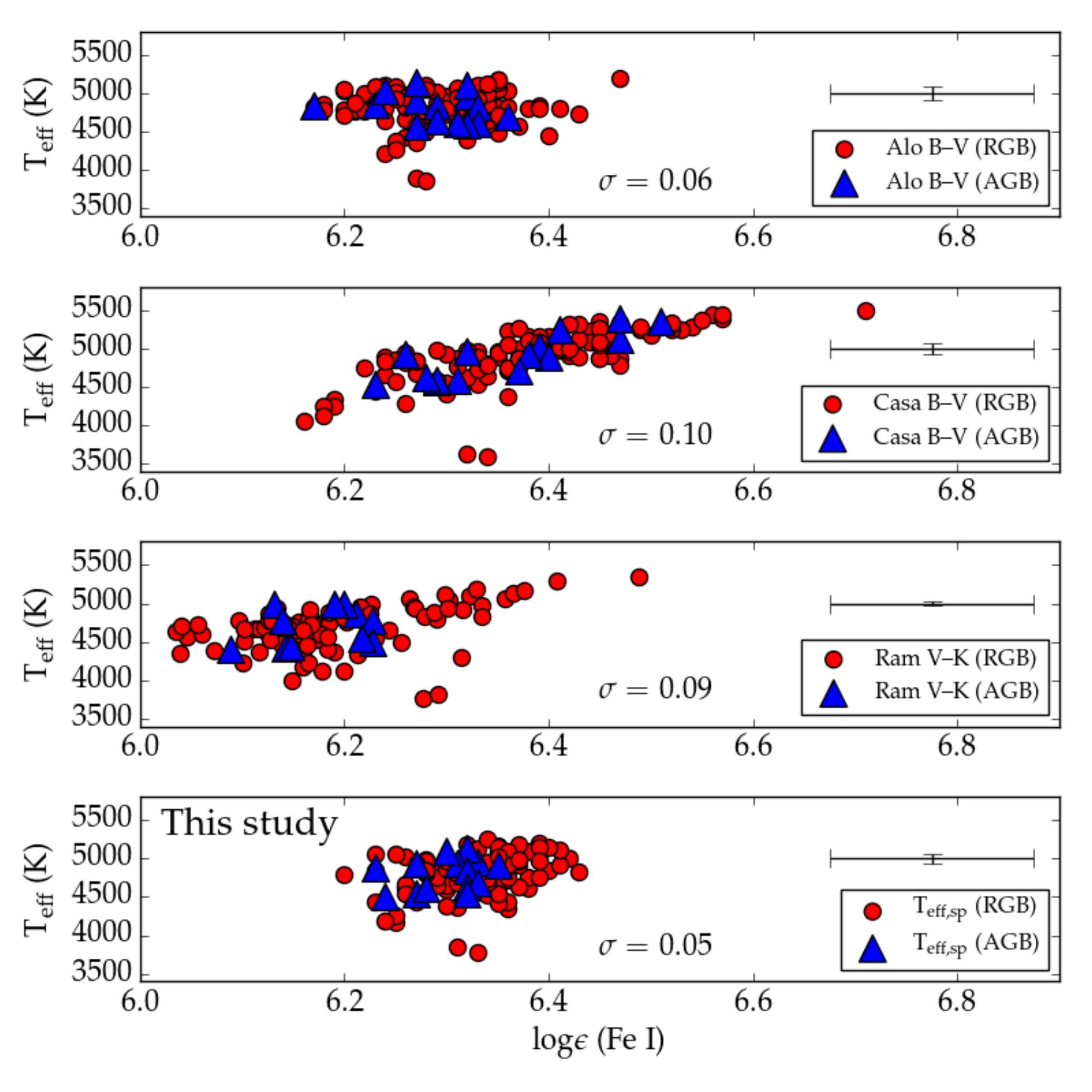} 
\caption{Fe abundances plotted against T$_{\rm eff}$, as determined using three different empirical colour-T$_{\rm eff}$ relations (top three panels) -- \protect{\citet[$B-V$]{alonso1999}}, \protect{\citet[$B-V$]{casagrande2010}}, and \protect{\citet[$V-K$]{ramirez2005}} -- and our spectroscopic stellar parameters (T$_{\rm eff, sp}$, using {\sc phobos}; bottom panel). The total uncertainty in ${\rm log_{\epsilon}(Fe\,\textsc{i})}$ is indicated (see Table~\ref{tab:m4_2_uncerts}), along with the relevant quoted uncertainties in T$_{\rm eff}$ for each relation.}
\label{fig:m4_2_1d_fe}
\end{figure}

\begin{figure}
\centering
\includegraphics[width=\linewidth]{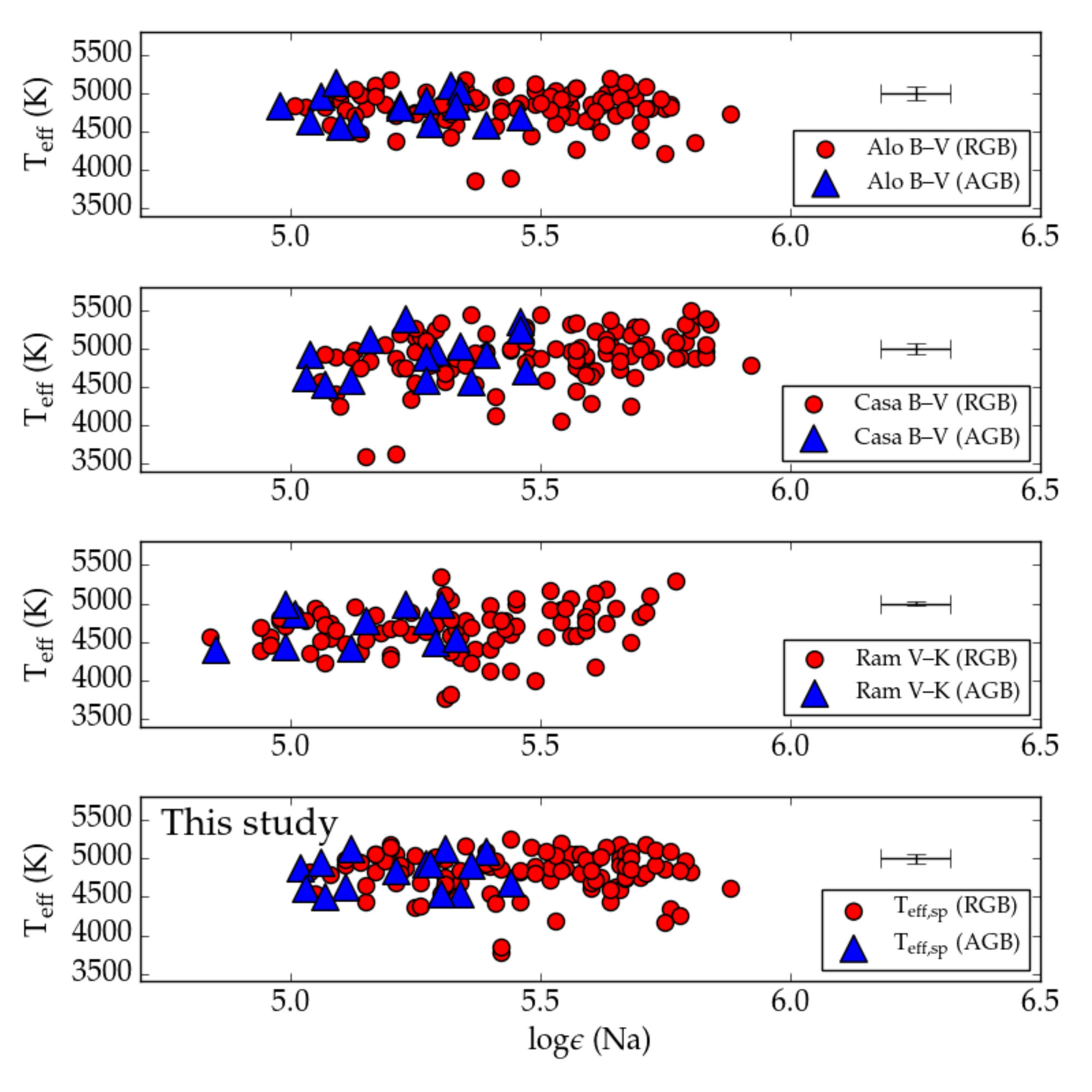} 
\caption{Same as Figure~\ref{fig:m4_2_1d_fe}, but for Na abundance.}
\label{fig:m4_2_1d_na}
\end{figure}

%------------------------------------------------------------------

\subsection{Helium enriched model test}
\label{m4_2_he_rich}

The KURUCZ/{\sc ATLAS9} atmospheric models used in the determination of abundances in this study adopt the solar abundances of \citet{GS98} -- with a helium mass fraction of Y $= 0.248$, which is similar to the primordial value assumed for SP1 stars in M\,4 \citep[Y $\sim 0.245$;][]{valcarce2014}. It is accepted that some GC stars are significantly enriched in helium \citep[by more than $\Delta$Y $= 0.15$ in some clusters, for example NGC2808;][]{dantona05}. \cite{villanova2012} determined helium abundances for a sample of blue HB stars in M\,4 (assumed to represent the most He-rich stars in the cluster), and found $\Delta$Y to be of the order of 0.03-0.04, while \citet{valcarce2014} and \citet{nardiello2015} determined $\Delta$Y values of $\lesssim 0.01$ and 0.02, respectively. 

Here we investigate the effects of including a He-enhancement in the atmospheric models used in chemical abundance determination. We redetermined the LTE abundances of O, Na, Mg, Al, and Fe for a sub-sample of M\,4 stars using a representative helium rich model available in the {\sc ATLAS9} database. Few He enhanced models have been computed for the {\sc ATLAS9} grid, so we conducted this test using the model with parameters closest to our M\,4 sample: [Fe/H]$=-1.5$, T$_{\rm eff} = 5000$~K, log $g = 1.5$, $v_t = 2.0$~km/s, and $\Delta$Y$ = +0.1$ (Y $= 0.352$). Due to the restriction of model selection, only a small subset of stars in our sample have stellar parameters similar to this model; therefore only a representative test was possible. 

For a sub-sample of four AGB and eight RGB stars (which cover the entire range of Na abundance as determined with {\sc phobos v2}), we determined LTE abundances using: i) the He enhanced model (`Y-enh/$\upalpha$-norm') which has scaled solar abundances for all other species, ii) a model with scaled solar abundances and Y $= 0.248$ (`Y-norm/$\upalpha$-norm'), and iii) a model with Y $= 0.248$ and an $\upalpha$-element element enhancement of +0.4~dex (`Y-norm/$\upalpha$-enh'). With these three models\footnote{There are no `Y-enh/$\upalpha$-enh' models in the {\sc ATLAS9} database.}, we were able to quantify the effect of increased He on elemental abundances while controlling for $\upalpha$-enhancement ($\upalpha$-enhanced atmospheric models were adopted for our abundance determination in \S\ref{m4_2_abund_method}). All three models had the same values of [Fe/H], T$_{\rm eff}$, log~$g$, and $v_t$ to ensure a consistent comparison. The results of this test are summarised in Table~\ref{tab:m4_2_he_rich}.

The differences between the abundances determined using the two Y $= 0.248$ models (`Y-norm/$\upalpha$-enh' and `Y-norm/$\upalpha$-norm'; see column three of Table~\ref{tab:m4_2_he_rich}), were constant throughout the sub-sample of 12 stars. Therefore we found that the effects of an $\upalpha$-enhancement are small and entirely systematic, with offsets $\leq 0.04$~dex for all species. 

Similarly, the effects of helium enhancement were systematic -- that is, an offset across the test sample -- for every species except Na, for which there was a 0.04~dex ($\sigma = 0.01$) range in abundance differences. This was smaller than our total uncertainty in ${\rm log_{\epsilon}(Na)}$ of 0.07~dex. As seen in Figure~\ref{fig:m4_2_he_rich}, which presents the quantitative effect of He-enhancement on Na abundance for the 12 stars in our sub-sample, the relative increase in Na abundance when the `Y-enh/$\upalpha$-norm' model is used positively correlates with Na abundance. Notably, the maximum change in Na abundance (0.08~dex) is of the order of our uncertainties ($\pm 0.07$~dex), and is significantly smaller than the mean difference in abundance between the RGB and AGB ($\Delta{\rm log_{\epsilon}(Na)}$ = 0.22~dex in our work).

We conclude that using helium enhanced 1D atmospheric models for the determination of chemical abundances of helium enriched stars in M\,4 would not alter the findings of this study for the following reasons:
\renewcommand\labelenumi{(\roman{enumi})}
\renewcommand\theenumi\labelenumi
\begin{enumerate}
\item The `Y-enh/$\upalpha$-norm' model affects the Na abundance of AGB stars in the same direction and magnitude as RGB stars of similar parameters and Na abundance, so distributions are not altered.
\item A helium enhancement of $\Delta$Y = +0.1~dex alters ${\rm log_{\epsilon}(Na)}$ by $\lesssim 0.07$~dex, which is smaller than our uncertainty in ${\rm log_{\epsilon}(Na)}$. Therefore, a helium enhancement more appropriate to M\,4 (0.01 $< \Delta$Y $<$ 0.04) would most likely not produce a measurable change in Na abundance.
\item A helium enhancement preferentially spreads out the high-Na stars to even higher values, making the AGB stars even more representative of SP1 RGB stars.
\end{enumerate}

\begin{table*}
\centering
\caption{The average differences in elemental abundance, for a representative sub-sample of M\,4 stars, when three {\sc ATLAS9} atmospheric models of varying composition -- helium-enhanced/$\upalpha$-normal ('Y-enh/$\upalpha$-norm'), helium-normal/$\upalpha$-normal ('Y-norm/$\upalpha$-norm'), and helium-normal/$\upalpha$-enhanced ('Y-norm/$\upalpha$-enh') -- were used, in combination with our standard spectroscopic method of abundance determination. All three models had the following stellar parameters: [Fe/H]$=-1.5$, T$_{\rm eff} = 5000$~K, log $g = 1.5$, $v_t = 2.0$~km/s, while the T$_{\rm eff}$ of each star in our sub-sample (8 RGB, and 4 AGB stars) was between 4900 < T$_{\rm eff}$ < 5100~K. Errors are the standard deviation of abundance difference over our 12 star sub-sample.}
\label{tab:m4_2_he_rich}
\begin{tabular}{ccc}
\hline
Species  & \multicolumn{2}{c}{$\Delta{\rm log_{\epsilon}(X)}$} \\
 & (Y-enh/$\upalpha$-norm -- Y-norm/$\upalpha$-enh) & (Y-norm/$\upalpha$-enh -- Y-norm/$\upalpha$-norm) \\
\hline\hline
Fe\,{\sc i}  & $+0.029 \pm 0.003$ & $-0.023 \pm 0.001$ \\
Fe\,{\sc ii} & $-0.013 \pm 0.004$ & $+0.040 \pm 0.001$ \\
O    & $+0.092 \pm 0.002$ & $+0.010 \pm 0.006$ \\
Na   & $+0.050 \pm 0.012$ & $-0.020 \pm 0.002$ \\
Mg   & $+0.041 \pm 0.002$ & $-0.017 \pm 0.001$ \\
Al   & $+0.024 \pm 0.002$ & $-0.011 \pm 0.000$ \\
\hline
\end{tabular}
\end{table*}
	
\begin{figure}
\centering
\includegraphics[width=0.9\linewidth]{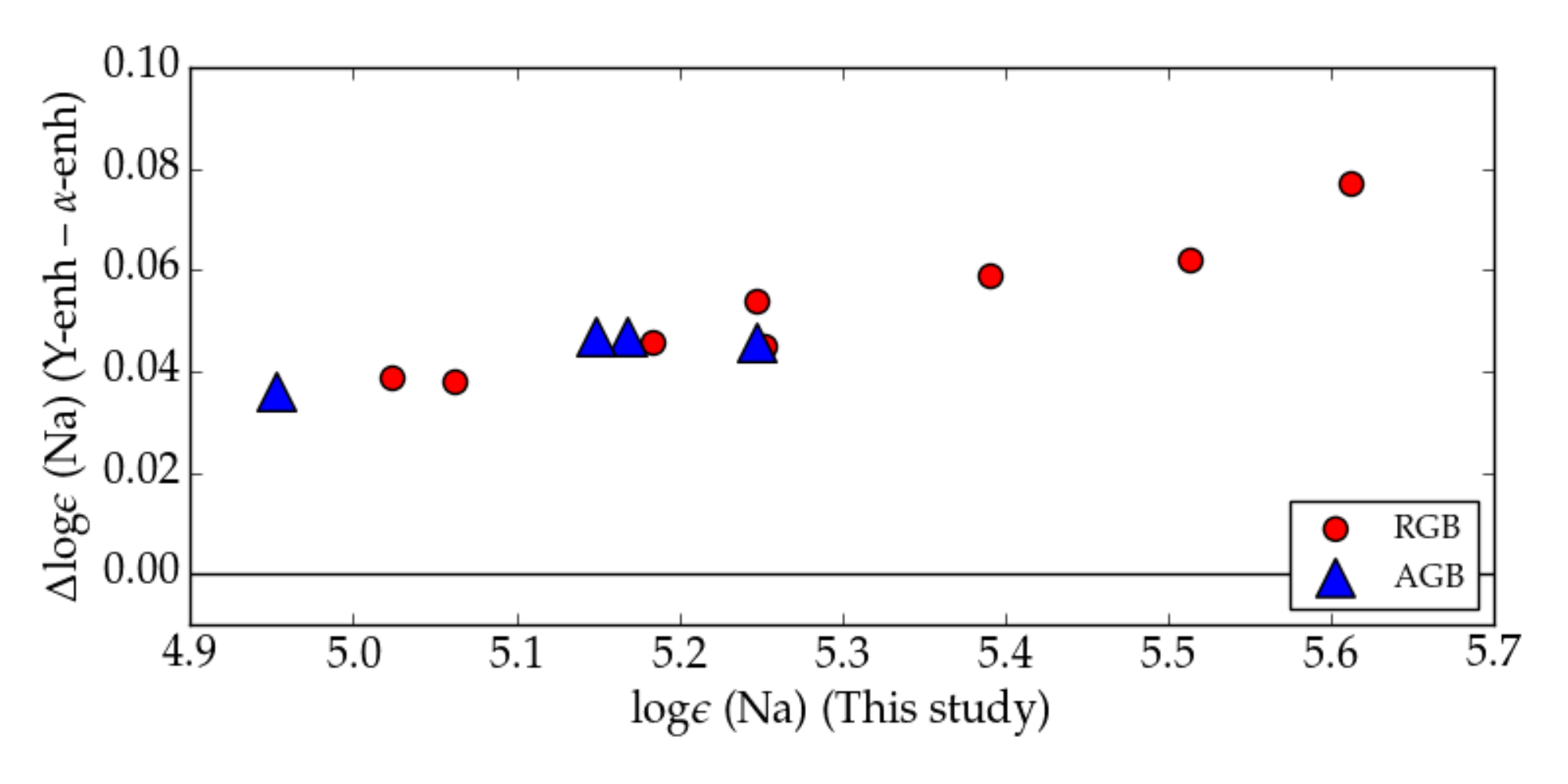} 
\caption{The star-to-star differences in Na abundance, for a representative sub-sample of 12 M\,4 stars, when two {\sc ATLAS9} atmospheric models of varying composition -- helium-enhanced and $\upalpha$-enhanced -- were used, in combination with our standard spectroscopic method of abundance determination. Na abundances on the x-axis are those adopted as the final abundances in this study.}
\label{fig:m4_2_he_rich}
\end{figure}

%------------------------------------------------------------------

\subsection{MARCS and $\langle$3D$\rangle$~{\sc Stagger}-grid test}
\label{m4_2_marcs_stagger}

In this study, and our previous GC investigations using AAT/HERMES spectra (\citetalias{maclean2016} and \citetalias{maclean2017}), we have exclusively employed the {\sc ATLAS9} grid of stellar atmospheric models. As a further test of the effects of using different model atmospheres on abundance determination, we investigate the effect on chemical abundance when two other sets of atmospheric models are employed: the 1D MARCS grid, and the mean-3D {\sc Stagger}-grid. Moreover, we do this with a totally independent abundance determination code, providing a further test of the robustness of our results. 

We determined non-LTE ${\rm log_{\epsilon}(Na)}$ and ${\rm log_{\epsilon}(O)}$ values for our entire M\,4 sample using the 3D non-LTE {\sc balder} code (\citealt{amarsi2018_balder}; based on the {\sc multi3d} code, \citealt{multi3d}). This method was very different to that used in Section~\ref{m4_2_abund_method}. Synthetic equivalent widths were calculated across a grid of Na and O abundances (in steps of 0.2~dex) by direct integration across the line, and then interpolated onto our spectroscopic stellar parameters (determined with {\sc phobos v2}). Abundances were evaluated by interpolating [X/Fe] values (a constant value of [Fe/H] $= -1.17$ was adopted) as a function of synthetic equivalent width onto our measured equivalent widths (from \S\ref{m4_2_abund_method}) for each star. Calculations were based on the Na model atom from \citet{lind2011}, and the O model atom from \citet{amarsi2018_oxygen}.

This abundance determination was done twice for our entire M\,4 stellar sample, with different grids of atmospheric models: i) the spherical 1D MARCS model atmospheres of scaled-solar chemical composition and $v_t = 2.0$~km/s \citep{gustafsson08}, and ii) the spatially- and temporally-averaged mean 3D ($\langle$3D$\rangle$) model atmospheres of the {\sc Stagger}-grid \citep{magic13}. For the latter analysis (based on $\langle$3D$\rangle$~model atmospheres), several stars in our sample, including all AGB stars, required extrapolation in T$_{\rm eff}$ or log~$g$, as they lie outside the parameter space of the {\sc Stagger}-grid.

\begin{table*}
\centering
\caption{Na and O abundances for each star in our M\,4 sample, determined using the {\sc balder} code with i) the 1D MARCS, and ii) the $\langle$3D$\rangle$~{\sc Stagger}-grid of stellar atmospheric models (see \S\ref{m4_2_marcs_stagger} for details). Abundance uncertainties reflect line-to-line scatter (1$\sigma$), and do not take atmospheric sensitivities into account. The last four lines show the cluster average abundances (for the AGB and RGB) with standard error of the mean, and standard deviation to indicate observed scatter. The final column indicates, for each star, whether extrapolation in the stellar parameters was required for the analysis based on the $\langle3D\rangle$~{\sc Stagger}-grid. Note that the stellar parameters from Table~\ref{tab:m4_2_params} were used for all abundance determinations. Only the first five rows in the top panel are shown; the full table is available online.}
\label{tab:m4_2_marcs_stagger}
\begin{tabular}{ccccccc}
\hline
ID  & Type & \multicolumn{2}{c}{1D MARCS} & \multicolumn{2}{c}{$\langle$3D$\rangle$~{\sc Stagger}} & $\langle$3D$\rangle$~extrapolation \\
  & & ${\rm log_{\epsilon}(Na)}$ & ${\rm log_{\epsilon}(O)}$ & ${\rm log_{\epsilon}(Na)}$ & ${\rm log_{\epsilon}(O)}$ & required? \\
\hline\hline
788   & AGB & $4.85 \pm 0.03$ & $8.20 \pm 0.06$ & $4.85 \pm 0.03$ & $8.37 \pm 0.06$ & Yes \\
3590  & AGB & $5.06 \pm 0.07$ & $8.04 \pm 0.00$ & $5.08 \pm 0.06$ & $8.20 \pm 0.01$ & Yes \\
10092 & AGB & $4.88 \pm 0.06$ & $8.28 \pm 0.03$ & $4.89 \pm 0.05$ & $8.42 \pm 0.03$ & Yes \\
11285 & AGB & $5.12 \pm 0.03$ & $8.09 \pm 0.04$ & $5.13 \pm 0.02$ & $8.19 \pm 0.05$ & Yes \\
13609 & AGB & $4.95 \pm 0.06$ & $8.10 \pm 0.03$ & $4.96 \pm 0.08$ & $8.20 \pm 0.02$ & Yes \\
\vdots & \vdots & \vdots & \vdots & \vdots & \vdots & \vdots \\
\hline
Mean & AGB & $5.00 \pm 0.03$ & $8.15 \pm 0.02$ & $5.03 \pm 0.03$ & $8.34 \pm 0.03$  \\
$\sigma$ & & 0.12 & 0.09 & 0.12 & 0.13 \\
Mean & RGB & $5.17 \pm 0.02$ & $8.07 \pm 0.01$ & $5.21 \pm 0.02$ & $8.20 \pm 0.01$  \\
$\sigma$ &  & 0.15 & 0.11 & 0.16 & 0.15  \\
\hline
\end{tabular}
\end{table*}

Abundances determined using the {\sc balder} code in combination with the MARCS grid are presented in Figure~\ref{fig:m4_2_marcs}, along with the star-to-star differences in non-LTE Na abundances between those from Section~\ref{m4_2_abundance_results} and those from this test. The results of this test are also presented in Table~\ref{tab:m4_2_marcs_stagger}. Comparing the top panel in Figure~\ref{fig:m4_2_marcs} with Figure~\ref{fig:m4_2_nao} shows that the spread and distribution of O and Na abundances using the MARCS grid and the {\sc balder} code are similar to those determined with {\sc phobos v2}. The bottom panel, however, indicates that significant changes to the absolute Na abundances occurred. The differences between the models and methods are correlated with Na abundance, and not evolutionary status. 

For Na, the average difference (in the sense of {\sc balder} -- {\sc phobos}, see Figure~\ref{fig:m4_2_marcs}) for stars with ${\rm log_{\epsilon}(Na)} > 5.25$ (indicated by the dashed line in Figure~\ref{fig:m4_2_marcs}) was $\Delta{\rm log_{\epsilon}(Na)} = -0.19 \pm 0.06$, which includes only one AGB star. For stars with ${\rm log_{\epsilon}(Na)} < 5.25$, $\Delta{\rm log_{\epsilon}(Na)} = -0.11 \pm 0.04$ for the AGB, and $\Delta{\rm log_{\epsilon}(Na)} = -0.12 \pm 0.04$ for the RGB. This acts to reduce the 1$\sigma$ spread in RGB Na abundance by 0.04~dex (to $\pm 0.15$~dex, see Table~\ref{tab:m4_2_abunds}), but does not alter the spread in AGB Na abundances. It also reduces the average difference in AGB and RGB Na abundance to $\Delta{\rm log_{\epsilon}(Na)} = -0.17$ from $-0.22$.

For O, the average difference was $\Delta{\rm log_{\epsilon}(O)} = -0.03 \pm 0.02$ for both samples, indicating no significant difference between the O abundances determined with the two methods. We again find M\,4 to be homogeneous in O. It is interesting to note that Na abundance was more sensitive than O abundance to the differences in method and atmospheric models examined in this test.

\begin{figure}
\centering
\includegraphics[width=0.9\linewidth]{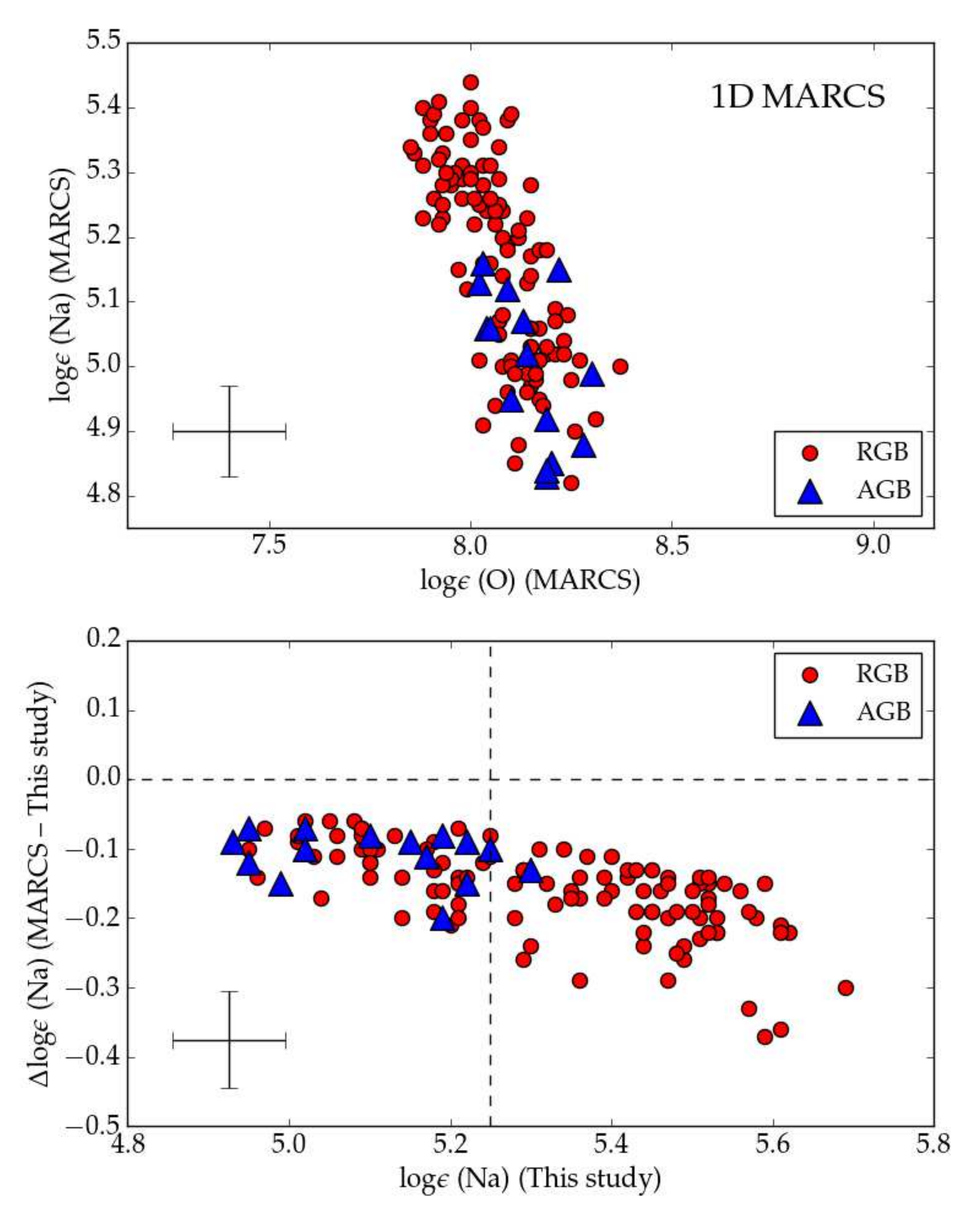} 
\caption{\textbf{Top panel:} Na and O abundances for each star in our M\,4 sample, determined using the {\sc balder} code with the 1D MARCS grid of stellar atmospheric models (see \S\ref{m4_2_marcs_stagger} for details). Error bars indicate our total abundance uncertainties (Table~\ref{tab:m4_2_uncerts}). \textbf{Bottom panel:} The star-to-star differences in Na abundance as determined using i) the {\sc balder} code with the 1D MARCS grid, and ii) {\sc phobos v2} with the 1D {\sc ATLAS9} grid of atmospheric models. Error bars indicate our total uncertainty in Na abundance. The dashed vertical line is at ${\rm log_{\epsilon}(Na)} = 5.25$, see \S\ref{m4_2_marcs_stagger} for details. Note that the stellar parameters from Table~\ref{tab:m4_2_params} were used for all determinations.}
\label{fig:m4_2_marcs}
\end{figure}

Abundances determined using the $\langle$3D$\rangle$~{\sc Stagger}-grid are presented in Figure~\ref{fig:m4_2_stagger}, along with the star-to-star differences in non-LTE Na abundance between the two sets of model atmospheres (the MARCS and $\langle$3D$\rangle$~{\sc Stagger}-grid), to indicate the impact of utilising atmospheric profiles computed in $\langle$3D$\rangle$~compared to 1D. The results of this test are included in Table~\ref{tab:m4_2_marcs_stagger}, and stars that required extrapolation outside of the {\sc Stagger}-grid are indicated.

As with the MARCS grid results, use of the $\langle$3D$\rangle$~{\sc Stagger}-grid for Na abundance determination gives a similar distribution to our {\sc phobos v2} abundances (Figure~\ref{fig:m4_2_nao}). The bottom panel of Figure~\ref{fig:m4_2_stagger} indicates that the Na abundances determined with the MARCS and $\langle$3D$\rangle$~{\sc Stagger}-grid were very similar, where the average difference was $\Delta{\rm log_{\epsilon}(Na)} = -0.03 \pm 0.02$ for AGB stars, and $\Delta{\rm log_{\epsilon}(Na)} = -0.05 \pm 0.01$ for RGB stars (excluding the two brightest stars in our sample, see caption of Figure~\ref{fig:m4_2_stagger}). The O abundances were impacted to a much higher degree; however this was mostly due to the extrapolation that was required for several stars (all AGB stars and several RGB stars required extrapolation, particularly those with high O abundances in Figure~\ref{fig:m4_2_stagger}). The average difference in O abundance between the MARCS and {\sc Stagger}-grid was $\Delta{\rm log_{\epsilon}(O)} = -0.20 \pm 0.08$ for AGB stars, and $\Delta{\rm log_{\epsilon}(O)} = +0.13 \pm 0.08$ for all RGB stars ($\Delta{\rm log_{\epsilon}(O)} = +0.09 \pm 0.04$ excluding those that required extrapolation).

\begin{figure}
\centering
\includegraphics[width=0.9\linewidth]{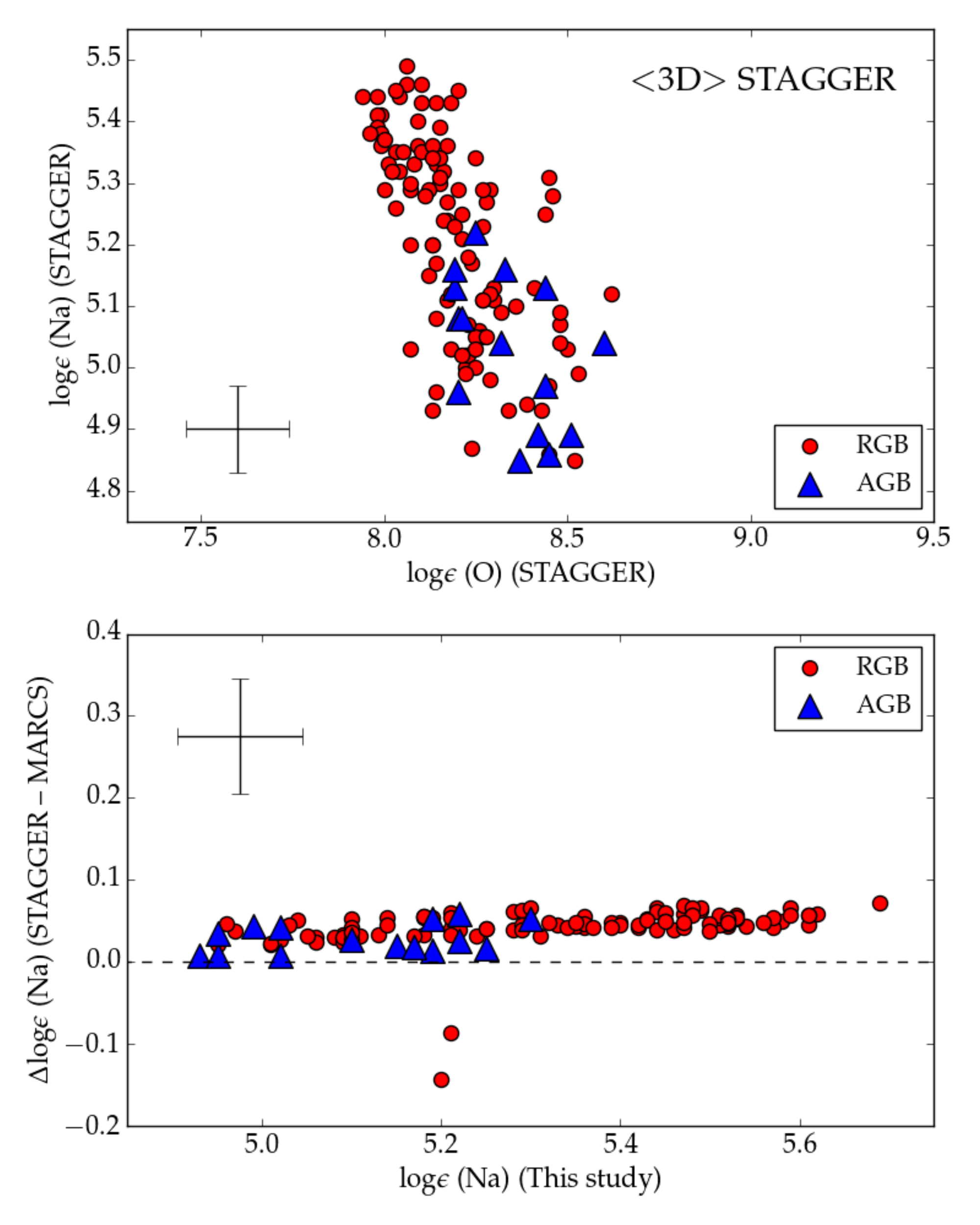} 
\caption{\textbf{Top panel:} Na and O abundances for each star in our M\,4 sample, determined using the non-LTE {\sc balder} code with the $\langle$3D$\rangle$~{\sc Stagger}-grid of stellar atmospheric models (see \S\ref{m4_2_marcs_stagger} for details). Error bars indicate our total abundance uncertainties (Table~\ref{tab:m4_2_uncerts}). \textbf{Bottom panel:} The star-to-star differences in Na abundance as determined using the non-LTE {\sc balder} code with i) the 1D MARCS, and ii) the $\langle$3D$\rangle$~{\sc Stagger}-grid of atmospheric models. The two outlying stars with negative differences are the two brightest stars in our sample, and were outside the {\sc Stagger}-grid by $\sim$1.0~dex in log~$g$. Error bars indicate our total uncertainty in Na abundance.}
\label{fig:m4_2_stagger}
\end{figure}

Comparing the bottom panels of Figures~\ref{fig:m4_2_marcs} and \ref{fig:m4_2_stagger}, we can see that the largest difference is between the {\sc balder} code and {\sc phobos v2}, rather than between the MARCS and {\sc Stagger}-grid stellar atmospheric models. With the tests performed, however, we cannot disentangle the effects of using the {\sc ATLAS9} vs MARCS grids from the effects of using the {\sc balder} vs {\sc phobos v2} codes. For low-Na stars (including all AGB stars) there is essentially an offset when the {\sc balder} code is used, and it compresses the range in Na by $\sim 0.08$~dex in high-Na RGB stars. This is independent of the choice of atmospheric model.

While the abundances of some stars are significantly different when this alternative method is employed, the overall result is unchanged, with M\,4 displaying an SP2 AGB deficit using both the MARCS and $\langle$3D$\rangle$~{\sc Stagger}-grids, and different abundance determination methods. It is interesting to note that the extrapolation of $\langle$3D$\rangle$~{\sc Stagger}-grid models had a large effect on O abundance, but almost no affect on Na abundance.

%--------------------------------------------------------------------

\subsection{Full-3D {\sc Stagger}-grid results}
\label{m4_2_full_3d}

In addition to using the $\langle$3D$\rangle$~{\sc Stagger}-grid, we also conducted a test using atmospheric models from the full-3D {\sc Stagger}-grid. This grid cannot be interpolated in T$_{\rm eff}$ and log~$g$ to provide star-specific models (as with 1D grids), so only a representative test was possible. We chose three models from the {\sc Stagger}-grid, which are approximately representative of i) an upper-RGB star (T$_{\rm eff} = 4500$~K, log~$g = 2.0$), ii) a lower-RGB star (T$_{\rm eff} = 5000$~K, log~$g = 3.0$), and iii) an early-AGB star (T$_{\rm eff} = 5000$~K, log~$g = 2.0$). For each model, we determined non-LTE stellar spectra in the region of the 568nm Na doublet feature at two representative Na abundances: [Na/Fe] = 0.0~dex (${\rm log_{\epsilon}(Na)} \sim 5.24$), and [Na/Fe] = 0.5~dex (${\rm log_{\epsilon}(Na)} \sim 5.74$).

We then computed non-LTE spectra in the same region using 1D atmopsheric models using the same stellar parameters and microphysics \citep[those used for 1D comparisons in][]{magic13}, and with a range of abundances between -1.0 < [Na/Fe] < +1.2, and microturbulence values between 1.0 < $v_t$ < 2.0~km/s. We quantified the corrections that should be applied to 1D Na abundances in order to account for 3D effects by comparing abundances between the 1D- and 3D-computed spectra at a given EW (corresponding to [Na/Fe] = 0.0 and 0.5~dex in the 3D regime). 

The choice of microturbulence is vital to this test, due to the sensitivity of the corrections to $v_t$, which can be difficult to determine accurately \citep{gratton1996}. We therefore interpolated the corrections based on representative $v_t$ values for stars with T$_{\rm eff}$ and log~$g$ similar to the three adopted {\sc Stagger}-grid models. All spectra were determined using the non-LTE {\sc balder} code, as in \S\ref{m4_2_marcs_stagger}.

Full-3D abundance corrections, as determined with the {\sc Stagger}-grid for the three representative atmospheric models, are presented in Table~\ref{tab:m4_2_full_3d}. 

\begin{table*}
\centering
\caption{Corrections to 1D non-LTE Na abundances in order to account for 3D non-LTE effects (`1D--3D') for three different sets of stellar parameters, representative of i) an upper-RGB star, ii) a lower-RGB star, and iii) an early-AGB star, respectively, and for two different Na abundances. These corrections were determined using the {\sc balder} code with the 1D MARCS grid, and full-3D {\sc Stagger}-grid of atmospheric models. Corrections were interpolated in $v_t$ based on the typical microturbulence values of representative stars in our M\,4 sample.}
\label{tab:m4_2_full_3d}
\begin{tabular}{lccccc}
\hline
Evolutionary & \multicolumn{2}{c}{Model parameters} & & \multicolumn{2}{c}{3D--1D correction} \\
Phase & T$_{\rm eff}$  &  log~$g$ & $v_t$ & [Na/Fe] = 0.0 & [Na/Fe] = 0.5  \\
 & (K) & (cgs) & (km/s) & (dex) & (dex) \\
\hline \hline
Upper-RGB & 4500  & 2.0 & $1.5$ & $0.06$ & $0.12$ \\
Lower-RGB & 5000  & 3.0 & $1.2$ & $0.01$ & $0.02$ \\
Early-AGB & 5000  & 2.0 & $1.6$ & $0.03$ & $0.04$ \\
\hline
\end{tabular}
\end{table*}

We found that in 3D, Na abundances are quite insensitive to changes in surface gravity -- a difference of $\Delta$log~$g$ = 1.0 only changes the Na correction by 0.02~dex, far below our total uncertainty in ${\rm log_{\epsilon}(Na)}$ ($\pm 0.07$~dex). Na corrections are more sensitive to changes in effective temperature, where $\Delta$T$_{\rm eff}$ = 500~K alters the correction by $\leq 0.08$~dex. It is important to note that significant confounding variables were unable to be accounted for in this test, including molecule (e.g. CH, NH) rearrangement due to CN processing and `deep mixing' on the upper-RGB, and differences in electron number densities due to the intrinsic Na and Al abundance variations.

The primary effect of these corrections is that the 3D non-LTE distribution of RGB Na abundances would likely extend toward higher values, thus exacerbating the difference to the AGB stars. We conclude that Na-rich stars are not likely to be incorrectly identified as being Na-poor due to 3D non-LTE effects on the lines, and that using full-3D atmospheric models for our entire sample of stars would be unlikely to alter our primary result for M\,4. 

Moreover, all of the tests here suggest that while the RGB Na dispersion can be altered with different methods and atmospheric models, the AGB stars all remain Na-poor. We found that AGB stars change in ${\rm log_{\epsilon}(Na)}$ in the same direction and the same approximate magnitude as RGB stars with comparable Na abundance -- we could not identify any way of systematically shifting the ${\rm log_{\epsilon}(Na)}$ values of AGB stars differently to those of RGB stars. In effect, these tests retain the relative Na distributions of the AGB and RGB that we found in Section~\ref{m4_2_abundance_results}.

%%%%%%%%%%%%%%%%%%%%%%%%%%%%%%%%%%%%%%%%%%%%%%%%%%%%%%%%%%%%%%

\section{Summary}
\label{m4_2_summary}

In light of conflicting results in several spectroscopic studies targeting the AGB of M\,4, we sought to i) present robust abundances for a sample of AGB and RGB stars in M\,4, ii) compare these abundances to those in the recent literature to investigate whether the results agree or disagree, and iii) attempt to predict and explain the abundance distributions of AGB stars in M\,4.

In Section~\ref{m4_2_reanalysis}, we analysed a sample of 15 AGB and 106 RGB stellar spectra in M\,4, observed with HERMES/AAT, and originally published in \citetalias{maclean2016}. We redetermined O, Na, and Fe abundances, and additionally report new Mg and Al abundances for each star. In this study, we were especially careful in our determination of stellar parameters (particularly T$_{\rm eff}$), and developed our spectroscopic code {\sc phobos v2} to avoid a reliance on photometric estimates of T$_{\rm eff}$. We found M\,4 to be heterogeneous in Na and Al, while our total uncertainties in O, Fe, and Mg abundances were larger than the spread in the respective values -- therefore we report that M\,4 is homogeneous in these species, within uncertainties. Furthermore, we found the atmospheres of our AGB sample to be lower in Na and Al, on average, compared to those of our RGB sample ($\Delta{\rm log_{\epsilon}(Na)} = -0.22$ and $\Delta{\rm log_{\epsilon}(Al)} = -0.14$), and with a smaller star-to-star spread in these abundances.

In Section~\ref{m4_2_cn}, we presented new CN band strengths for a sample of 7 AGB and 19 RGB stars in M\,4 based on independent low-resolution spectra. We identified the bimodality in CN band-strengths that was first observed by \citet{norris1981m4}, and found $\delta$S3839 to correlate with our ${\rm log_{\epsilon}(Na)}$ values from \S\ref{m4_2_reanalysis}. We found the average AGB band-strength to be weaker than that of our RGB sample ($\Delta\delta$S3839 $=0.24$), and with a smaller spread in values -- similar to our Na and Al results.

In Section~\ref{m4_2_lit_comp}, we compiled spectroscopic results from the literature. We used values from \citetalias{ivans1999} (O, Na, Mg, Al, and Fe abundances), \citetalias{sb2005} (CN band-strengths), \citetalias{marino2008} and \citetalias{marino2017} (O, Na, Mg, Al, and Fe abundances), and \citetalias{wang2017} (Na and Fe abundances). We compared the AGB and RGB distributions of ${\rm log_{\epsilon}(X)}$ and $\delta$S3839 values from these six studies to this study (as determined in \S\ref{m4_2_reanalysis} and \S\ref{m4_2_cn}). We found that all Fe abundance distributions agree well (both between studies, and between the giant branches within each study), except for \citetalias{marino2017} whose separately determined AGB and RGB abundances did not agree. We found a similar result for Mg. The uncertainties in the O abundances prevented us from drawing any conclusions for this element other than a formal homogeneity within M\,4 stars.

A bimodality is visible in the Na abundances of \citetalias{ivans1999}, \citetalias{marino2008}/\citetalias{marino2017}, and our work (but not \citetalias{wang2017}, however this is most likely due to their large uncertainties). In the abundances of every study, the AGB samples have notably lower ${\rm log_{\epsilon}(Na)}$ values, but with a bimodality still present (except \citetalias{wang2017}). The Al abundances all show a similar offset between the AGB and RGB, however no bimodality could be identified, except in the results of \citetalias{ivans1999} (this may be an artefact of underestimated errors and a small sample size). The CN band-strengths from \citetalias{sb2005} and this study both show bimodality, while both AGB samples show an extreme paucity of CN-strong members.

In Section~\ref{m4_2_monstar}, we calculated a series of theoretical stellar evolutionary models with the {\sc MONSTAR} code, using observational constraints on M\,4 stars from the literature. This was done in order to establish a precise, quantitative theoretical expectation of the abundances of AGB stars in M\,4. We found that in order to match the HB morphology, as determined spectroscopically by \citet{marino2011} and \citet{villanova2012}, and using a helium enhancement for SP2 stars of $\Delta$Y = 0.03, we required a Reimers mass loss rate of $\eta = 0.44 \pm 0.03$ and initial masses of $0.827 \pm 0.013$ and $0.785 \pm 0.013$ M$_{\odot}$ for SP1 and SP2, respectively; which gave a cluster age of $12.4 \pm 0.6$~Gyr. All stellar models whose HB T$_{\rm eff}$ matched the observed values ascended the AGB, indicating that all post-HB stars in M\,4, irrespective of Na abundance, should evolve to the AGB. We also demonstrated that at the metallicity of M\,4, only stars that reach a T$_{\rm eff} \gtrsim 15,500$~K on the HB  -- 6000~K hotter than the bluest HB stars in M\,4 -- should have AGB lifetimes short enough to avoid detection, in agreement with the models of \citet{dorman1993}.

Confronted with this discordance between our observational results and the prediction of stellar theory, we investigated the robustness of our spectroscopic abundance determinations. We did this in Section~\ref{m4_2_atmos_tests} by conducted a range of tests using various stellar atmospheric models in order to determine the robustness of our elemental abundance results to uncertainties in atmospheric structure. Specifically, we i) redetermined LTE Na and Fe abundances for our entire M\,4 sample using three different sets of photometric T$_{\rm eff}$ estimates (with individual T$_{\rm eff}$ differences of up to 500~K), ii) determined elemental abundances for a sub-sample of M\,4 stars using a He-enhanced ($\Delta$Y = 0.10) model from the {\sc ATLAS9} grid to estimate the effect of including He variations in atmospheric models, iii) redetermined Na and O abundances independently using the non-LTE {\sc balder} code \citep{amarsi2018_balder} in combination with atmospheric models from the 1D MARCS grid and the $\langle$3D$\rangle$~{\sc Stagger}-grid, and iv) using the full-3D {\sc Stagger}-grid, determined corrections to 1D non-LTE Na abundances to account for 3D non-LTE effects for three sets of stellar parameters. All tests indicated that Na-rich stars (on the AGB or RGB) are unlikely to be misidentified as being Na-poor.

%%%%%%%%%%%%%%%%%%%%%%%%%%%%%%%%%%%%%%%%%%%%%%%%%%%%%%%%%%%%%%

\section{Conclusions}
\label{m4_2_discussion}

A significant strength of the spectroscopic results presented in this study (\S\ref{m4_2_reanalysis}--\ref{m4_2_cn}) lies in the combining of two independent methods of separating the subpopulations in chemical abundance space (using both high- and low-resolution spectra). Both of our independent sets of M\,4 results in this paper, namely (i) the re-analysed high-resolution spectra, with additional chemical abundances (Figure~\ref{fig:m4_2_mgal}), and (ii) the new CN band strengths (Figure~\ref{fig:m4_2_cn}), support the conclusions of \citetalias{maclean2016} that AGB stars in M\,4 are largely representative of SP1 stars -- namely, that there is a significant paucity of SP2 AGB stars, with an SP2 AGB deficit of $\mathscr{F} \gtrsim 65\%$ -- as evidenced by their Na and Al abundances, and CN band-strengths, compared to those of stars on the RGB. This adds M\,4 to the list of GCs that have been reported to contain significant SP2 AGB deficits, alongside NGC\,6752 \citep{campbell2013} and M\,62 \citep{lapenna2015}.

A comparison of these results with those from the literature (\S\ref{m4_2_lit_comp}) indicate that this is unlikely to be an artefact of our method of abundance determination: spectroscopic M\,4 studies that included AGB stars have consistently shown the AGB to be systematically lower in Na abundance, Al abundance, and CN band strength \citep[typically indicative of N abundance;][]{cottrell1981} than the RGB -- in agreement with our original findings in \citetalias{maclean2016}. In stark contrast to this strong observational result, we predicted -- using theoretical evolutionary models representative of M\,4 stars (\S\ref{m4_2_monstar}) -- that the abundance distributions of the AGB and RGB should be \textit{identical} for all species investigated in this study (except for CN due to extra mixing of N to the stellar surface on the RGB). In an attempt to reconcile the models and observations, we found that we were unable to significantly alter our abundance results by utilising a variety of atmospheric models (\S\ref{m4_2_atmos_tests}), including those with systematically offset stellar parameters, those that included a helium enhancement, different grids of 1D atmospheric models, or 3D atmospheric models.

Two recent photometric investigations of M\,4 (\citealt{lardo2017}, and \citetalias{marino2017}) have reported that their data of M\,4 AGB stars are consistent with the AGB containing both SP1 and SP2 stars. The spectroscopic results presented in this study similarly suggest that some proportion of SP2 stars may evolve to the AGB. However the photometric indices C$_{UBI}$ and C$_{F275W,F336W,F438W}$ are unlikely to be precise enough to detect whether or not the most Na-enhanced SP2 stars are missing on the AGB, as suggested by the spectroscopically-determined abundances presented in this paper. We note that our CN index results -- which are analogous to very narrow-band photometry -- agree with the conclusions drawn from high-resolution spectroscopy, and disagree with those drawn from photometric pseudo-CMDs.

Na, Al, and N are all products of hydrogen burning \citep{kippenhahn1990book}, and are three of the species most commonly observed to vary among the stars of globular clusters \citep[other species include He, C, O, and Mg;][]{gratton2004}, both Galactic and extragalactic \citep{gratton2012,extragalacticgcs}. Of these, atmospheric Na and Al abundances are not predicted to change throughout the lives of individual present-day GC stars -- these abundances are typically assumed to be an intrinsic property of the star because low-mass stars do not reach temperatures high enough to activate the Ne-Na or Mg-Al H-burning chains \citep{norris1981m4,iben1984} -- while N is observed to increase on the RGB via `deep mixing' \citep{henkel2017}.

In conclusion, with no viable mechanism to reduce these abundances in-situ between the RGB and AGB, and the prediction that all stars in M\,4 should evolve through to the AGB, we can see few remaining potential explanations for the consistent observations that AGB stars in M\,4 have significantly lower abundances of Na, Al, and N (inferred from CN) than RGB stars in the cluster. Avenues to consider in order to resolve this disparity are diminishing, but include investigating the effect of interstellar extinction on AGB stellar spectra (M\,4 experiences large differential reddening), and exploring differences between the atmospheric structures of AGB and RGB stars. We note, however, that any solution must simultaneously account for the observed disparity in both the elemental abundance \textit{and} CN band strength distributions, which are determined using different spectroscopic methods.

%%%%%%%%%%%%%%%%%%%%%%%%%%%%%%%%%%%%%%%%%%%%%%%%%%%%%%%%%%%

\section*{Acknowledgements}

Based in part on data acquired through the AAO, via programs 09B/15, 14B/27 and 15A/21 (PI Campbell); some of these observations were conducted by Elizabeth Wylie de Boer, Richard Stancliffe, and George Angelou. Part of this work was supported by the DAAD (PPP project 57219117) with funds from the German Federal Ministry of Education and Research (BMBF). Parts of this research were conducted by the Australian Research Council Centre of Excellence for All Sky Astrophysics in 3 Dimensions (ASTRO 3D), through project number CE170100013. BTM acknowledges the financial support of the Research Training Program (RTP) scheme scholarship. SWC acknowledges federal funding from the Australian Research Council though the Future Fellowship grant entitled ``Where are the Convective Boundaries in Stars?'' (FT160100046). AMA acknowledges funds from the Alexander von Humboldt Foundation in the framework of the Sofja Kovalevskaja Award endowed by the Federal Ministry of Education and Research. TN acknowledges funding from the Australian Research Council (grant DP150100250). VD acknowledges support from the AAO distinguished visitor program 2016. LC gratefully acknowledges support from the Australian Research Council (grants DP150100250, FT160100402). This work was supported by computational resources provided by the Australian Government through the National Computational Infrastructure (NCI) under the National Computational Merit Allocation Scheme (project ew6). We thank Yazan Momany for providing M\,4 $UBVI$ photometry, and Benjamin Hendricks for providing the M\,4 reddening map. We also thank Anna Marino and Yue Wang for providing additional data that aided in the literature comparison, and Yue for helpful discussions and suggestions.

%-----------------------------------------------------------

\bibliography{References}

%-----------------------------------------------------------

\label{lastpage}

\end{document}